\documentclass[11pt,a4paper]{article}
\pdfoutput=1 % Required by JHEP style

\usepackage{jheppub}

\usepackage{hyperref}                            % In-document hyperlinks
\usepackage{graphicx}                            % Figures and things
\usepackage{soul}                                % Highlighting, strikethrough, etc
\usepackage{microtype}                           % Better typography?
\usepackage[sharp]{easylist}                     % Simpler nested lists
\usepackage{amsmath}                             % Align environments
\usepackage{amssymb}                             % Symbols like \mathbb
\usepackage{tensor}                              % Tensor subscripts and superscripts
\usepackage{booktabs}                            % Better tables
\usepackage{wrapfig}                             % Wrap text around figures
\usepackage{subcaption}                          % Figures side-by-side
\usepackage[T1]{fontenc}                         % Not sure what this does...
\usepackage{titlesec}                            % Remove the word Chapter/Appendix
\usepackage[font=small]{caption}                 % Smaller caption fonts

\usepackage{color}
    \definecolor{darkgreen}{rgb}{0,0.5,0}
    \definecolor{darkred}{rgb}{0.5,0,0}
    \definecolor{darkblue}{rgb}{0,0,0.6}
    \definecolor{purple}{rgb}{0.4,.2,0.7}

\usepackage{tikz}

\title{Strong Cosmic Censorship and Eigenvalue Repulsions for rotating de Sitter black holes in higher-dimensions}
\author[a]{Alex Davey,}
\author[a]{Oscar J. C. Dias,}
\author[a]{Paul Rodgers,}
\author[b]{Jorge E. Santos}

\affiliation[a]{STAG research centre and Mathematical Sciences, Highfield Campus, University of Southampton, Southampton SO17 1BJ, UK}
\affiliation[b]{DAMTP, University of Cambridge, Wilberforce Road, Cambridge CB3 0WA, UK}

\emailAdd{amd1g13@soton.ac.uk}
\emailAdd{ojcd1r13@soton.ac.uk}
\emailAdd{P.W.Rodgers@soton.ac.uk}
\emailAdd{jss55@cam.ac.uk}

\abstract{It has been established that Christodoulou’s formulation of Strong Cosmic Censorship (SCC) is violated by Reissner-Nordstr\"om$-$de Sitter black holes, but holds in four-dimensional Kerr-de Sitter black holes.
We show that SCC is also respected by equal angular momenta (cohomogeneity-1) Myers-Perry-de Sitter (MP-dS)  in odd $d\geq 5$ spacetime dimensions. This suggests that the preservation of SCC in rotating backgrounds might be a universal property of Einstein gravity and not limited to the  $d=4$ Kerr-dS background.
As required to discuss SCC in de Sitter spacetimes, we also study important aspects of the  scalar field quasinormal mode (QNM) spectra of MP-dS. 
In particular, we find eigenvalue repulsions similar to those recently observed in the QNM spectra of asymptotically flat Kerr-Newman black holes. For axisymmetric modes (\emph{i.e.} with azimuthal quantum number $m=0$) there are three distinct families of QNM (de Sitter, photon sphere and near-horizon).
However, typically, for non-axisymmetric $(m \ne 0)$ QNMs, we find that the entire spectra can be described by just two families of QNM (since several overtone sections of the photon sphere and near-horizon families merge). For completeness, we also study the full scalar field QNM spectra of higher-dimensional Schwarzschild-de Sitter black holes.}

\begin{document}
%=======================================================

\maketitle
\flushbottom

\clearpage
\section{Introduction}

In the initial value formulation of General Relativity \cite{10.1007/BF02392131,Choquet-Bruhat:1969ywq}, we specify initial data on some partial Cauchy hypersurface $\Sigma$, for example a constant time slice or a pair of null hypersurfaces, and evolve it forwards according to the Einstein equations. The boundary of the \emph{maximal Cauchy development} of that initial data, if it exists, is called the Cauchy horizon. The Strong Cosmic Censorship (SCC) conjecture posits that, for generic initial data, the maximal Cauchy development is inextendible beyond the Cauchy horizon as a suitably regular manifold~\cite{penroseSingularitiesSpacetime1978}. A violation of SCC indicates a failure of predictability, since if one \emph{can} extend beyond the Cauchy horizon then this new region would not depend causally on only the initial data specified on $\Sigma$. The question of the precise notion of the level of regularity that one should require has a long history (see~\cite{diasStrongCosmicCensorship2018a} for a review), however the modern formulation of Strong Cosmic Censorship is due to Christodoulou~\cite{christodoulouFormationBlackHoles2008}, which states that the maximal Cauchy development should be inextendible beyond the Cauchy horizon as a \emph{weak solution} of the Einstein equations or the gravitational equations with matter fields.

There is growing evidence that the Strong Cosmic Censorship is respected for asymptotically flat initial data close to Reissner-Nordström~\cite{Poisson:1990eh,Dafermos:2003wr,lukProofLinearInstability2017,Luk:2017jxq} and Kerr~\cite{dafermosTimeTranslationInvarianceScattering2017,Dafermos:2017dbw}. Ultimately, this is due to the well-known infinite blueshift effect near the Cauchy horizon and the associated Price law~\cite{simpsonInternalInstabilityReissnerNordstrom1973,priceNonsphericalPerturbationsRelativistic1972}. However, for positive cosmological constant ($\Lambda > 0$) there is a competing redshift associated with the gravitational potential well of asymptotically de Sitter spacetimes. As a result of the delicate competition between these two effects, the decay of generic linear perturbations depends on the magnitude of the imaginary part of the slowest-decaying quasinormal mode  (QNM) of the system \cite{HV2017,Costa:2016afl}. Indeed, in recent years, a large body of work indicates that initial data close to Reissner-Nordström$-$de Sitter (RN-dS) violates SCC \cite{cardosoQuasinormalModesStrong2018,diasStrongCosmicCensorship2018a,diasStrongCosmicCensorship2019}\footnote{\label{foot1}Note that charged scalar fields (charged matter fields are required in a theory that allows the formation of RN-dS black holes through gravitational collapse) still lead to a violation of SCC in RN-dS \cite{diasStrongCosmicCensorship2019}. Refs. \cite{PhysRevD.98.124025,PhysRevD.98.104007} arrived to the opposite conclusion because these studies did not extend the analysis sufficiently close to extremality where the violation is finally observed (see associated discussion in \cite{diasStrongCosmicCensorship2019}).}$\,$\footnote{Note that it was proposed that a reformulation of SCC which takes into account quantum corrections can rescue SCC in RN-dS~\cite{diasStrongCosmicCensorship2018a,hollandsQuantumInstabilityCauchy2020,PhysRevD.102.085004,PhysRevD.104.025009} and in BTZ  \cite{Dias:2019ery,Emparan:2020znc,Pandya:2020ejc} backgrounds where Christodoulou’s formulation of SCC is violated. Further note that in all our SCC discussions we assume smooth initial data. However, if one allows rough initial data then Christodoulou’s version of SCC is true for linear perturbations of RN-dS black holes since the solution at the Cauchy horizon is, generically, rougher than the initial data \cite{Dafermos:2018tha} (see also extended discussion in \cite{diasStrongCosmicCensorship2018a}).}, while Kerr-dS does not~\cite{diasStrongCosmicCensorship2018}. These analyses were accomplished within linear theory, and it remains an open problem to show that these extend to the full nonlinear theory \cite{Luna:2019olw}.

Perhaps the strongest motivation to study SCC for initial data close to Reissner-Nordström$-$dS is the fact that, in many respects, the RN-dS black hole appears to be a (much simpler) toy model for initial data close to Kerr-dS. Thus it might come as a surprise that SCC is violated for initial data close to RN-dS but not for initial data close to Kerr-dS. However, there is an important distinction between the two, which is best fleshed out if we recall the parallel between the two spacetimes. In particular, \emph{charged} scalar fields of mass $\mu$ and charge $q$ around a Reissner-Nordström-dS black hole have been shown to lead to a violation of SCC for any \emph{finite} value of $q$ \cite{diasStrongCosmicCensorship2019}$^{\ref{foot1}}$. If, however, we take (the rather unphysical limit) $q\to+\infty$, these violations disappear altogether \cite{diasStrongCosmicCensorship2019}. For Kerr-dS, the analog of $q$ is the azimuthal quantum number $m$, which counts the number of nodes of the scalar perturbation along the direction of rotation. However, unlike RN-dS, for Kerr-dS we are forced to consider initial data with arbitrarily large $m$, and it is this data that ends up saving SCC in Kerr-dS.

It remains, however, a mystery as to why the quasinormal mode spectrum of Kerr-dS black holes at large $m$ behaves just so as to save SCC. It could have been that the large $m$ behaviour was such that SCC would still be violated: it just so happens that, after a long calculation, it is not! One might then wonder whether this is a result one can derive for a large class of rotating black holes, perhaps by studying the universal properties of rotating near-horizon geometries. Whatever the mechanism might be, it appears to depend on the details of the near-horizon geometry. However, it has been argued in \cite{Hod:2018lmi,casalsGlimpsesViolationStrong2020} that initial data close to Kerr-Newman-dS black holes will generically violate SCC if the black hole charge is large enough. This shows, to some extent, that not all rotating black holes necessarily preserve SCC and that preserving SCC cannot be a universal property of \emph{all} rotating near-horizon geometries.

In this manuscript, instead of turning on charge, we change yet another dial: the spacetime dimension $d$. As a first step in this direction, it was shown that scalar field perturbations of RN-dS in \(d = 5, 6\)  violate SCC (much alike in the $d=4$ case), with the expectation that this conclusion does not change in even higher dimensions~\cite{liuStrongCosmicCensorship2019}. RN-dS black holes have, however, been shown to be unstable in dimensions $d\geq6$ \cite{Konoplya:2008au,Cardoso:2009cnd,Konoplya:2013sba,Dias:2020ncd}. As a second step, in this manuscript we consider scalar field perturbations of Myers-Perry$-$de Sitter (MP-dS), \emph{i.e.} the higher-dimensional extension of the Kerr-de Sitter solution. For simplicity, we restrict our analysis to odd $d$ spacetime dimensions and to black holes with equal angular momentum. In this case the resulting line element is cohomogeneity-1, \emph{i.e.} it depends non-trivially on only the radial coordinate. We will find that Christodoulou's formulation of SCC holds in cohomogeneity-1 MP-dS, very much like in the $d=4$ Kerr-dS case. The generic considerations of \cite{rahmanFateStrongCosmic2019} further indicate that this result extends to other, perhaps all, MP-dS solutions. Together with \cite{liuStrongCosmicCensorship2019} we thus have strong evidence in favour of the following universal result for perturbations excited by scalar fields: for arbitrary spacetime dimensions in de Sitter, Christodoulou's formulation of SCC holds in dynamically stable, vacuum, rotating black hole solutions of the Einstein equations, but can be violated if charged matter is included. We note, however, that the leading WKB behaviour is spin independent, and thus the result quoted above could indeed also be true for gravitational perturbations.

As stated above, the question of SCC in de Sitter backgrounds is intimately linked to quasinormal modes \cite{BZ1997,BH2008,Dyatlov2012,Dyatlov2015,HV2017,HV2018,Hintz2018}, so we naturally also take the opportunity to discuss some aspects of the QNM spectra of MP-dS. In particular, we want to identify all possible families of QNMs in MP-dS (for a given set of relevant wave quantum numbers) and, ultimately, the family with the slowest decaying QNM at each point in the 2-parameter space of MP-dS. For that we resort to a numerical computation of the QNM spectra in the full parameter space (using pseudospectral collocation methods) but also to analytical analyses (in the appropriate corners of the parameter space) to elucidate the physical origin of each family.

Asymptotically flat, four-dimensional spacetimes (the Kerr-Newman family and its Kerr and Reissner-Nordstr\"om limits) which admit a Cauchy horizon have two distinct families of QNMs. One of them is the  \emph{photon sphere} (PS) family. In the eikonal limit, where the angular momenta harmonic quantum numbers are large ($|m| = l \to \infty$), these PS modes are connected to the behaviour of unstable null geodesics in the equatorial plane~\cite{ferrariNewApproachQuasinormal1984}. This PS family also exists in de Sitter black holes and we verify (by direct comparison with the numerical results) that this correspondence still holds in rotating black holes in higher dimensions. Moreover, there is also a second QNM family $-$ the \emph{near-horizon} (NH) modes $-$ which are generally suppressed (\emph{i.e.} have a fast decay) except very near extremality, where the Cauchy horizon approaches the event horizon, $r_{-} \to r_{+}$. These NH modes have a wavefunction that is highly localised near the event horizon, and a vanishing imaginary part in the extremal limit~\cite{cardosoQuasinormalModesStrong2018,yangBranchingQuasinormalModes2013a,zimmermanDampedZerodampedQuasinormal2016,diasStrongCosmicCensorship2018a}. The NH modes are still present in MP-dS and to capture them in an analytical approximation, we perform a matched asymptotic expansion. We find the eigenfunction near the horizon and then match it with a vanishing wavefunction solution far from the horizon. We highlight the link between the NH modes we find and the near-horizon geometry of the extremal spacetime.

In asymptotically de Sitter spacetimes, unlike in the $\Lambda=0$ case, there is a third family of QNM modes: the \emph{de Sitter} (dS) family. The frequency of these modes approaches the QNM mode frequency of pure de Sitter space in the limit where the mass and the angular momentum of MP-dS vanish.  In $d=4$, dS modes have a weak dependence on the black hole parameters. We will find that that this is no longer true in higher dimensions.

In cohomogeneity-1 Myers-Perry-de Sitter, axisymmetric mode perturbations (\emph{i.e.} with azimuthal quantum number $m = 0$) feature all three families. 
However, for $m \ne 0$, we will find that the PS and NH modes will typically (but not always) merge into a single family. This family generally dominates, \emph{i.e.} has more slowly decaying modes than the dS family. As a result, we will be able to use both our eikonal and near-horizon approximations in tandem to study SCC.
No less interestingly, we also find that in certain regions of the parameter space some of these families will exhibit eigenvalue repulsions, similar to those recently observed in Kerr-Newman~\cite{diasEigenvalueRepulsionsQuasinormal2021}. For example, this occurs between dS and NH modes.

The plan of this paper is as follows. In Section~\ref{sec:myers-perry} we review the main properties of cohomogeneity-1 Myers-Perry-de Sitter black holes. We use separation of variables to study the Klein-Gordon equation for scalar fields in this background, and we describe the numerical scheme used to solve for the scalar field perturbations. In Section~\ref{sec:AnalyticalQNMfamilies}, we derive analytic approximations for the three families of QNM $-$ dS, PS and NH $-$ that can be present in MP-dS for regions of the parameter space that are susceptible to such analytical approximations. In Section~\ref{sec:QNMspectra} we describe important features of the full QNM spectra after comparing our numerical results with the aforementioned analytical approximations. 
Finally, in Section~\ref{sec:scc} we tackle the question of whether or not Strong Cosmic Censorship holds in cohomogeneity-1 MP-dS black holes. For completeness, and because this depends significantly on the spacetime dimension, we study the QNM spectra of higher-dimensional Schwarzschild-de Sitter black holes in Appendix~\ref{sec:SdS_and_small_a}. Some details of the NH modes are referred to Appendix~\ref{sec:explicit_N} and we discuss the numerical convergence of our results in Appendix~\ref{sec:convergence}.

%=======================================
%=======================================
\section{Scalar perturbations of cohomogeneity-1 Myers-Perry$-$de Sitter}\label{sec:myers-perry}
%=======================================
%=======================================

\subsection{Cohomogeneity-1 Myers-Perry$-$de Sitter black holes}

The Myers-Perry black hole is a stationary and axisymmetric spacetime in $d \ge 4$ dimensions, parameterised by a mass parameter \(M\) and angular momentum parameters \(a_{i}\) in each of the \(n = \left\lfloor \frac{d-1}{2} \right\rfloor\) rotational planes~\cite{myersBlackHolesHigher1986}. For general \(a_{i}\) this black hole has the isometry group \(\mathbb{R}\times U(1)^{n}\). However, in the \emph{equal angular momenta} case \(a_{i} = a\) and in odd dimensions (only), the symmetry is enhanced to \(\mathbb{R} \times U(n)\). Consequently, the resulting metric is \mbox{\emph{cohomogeneity-1}}, \emph{i.e.} it depends non-trivially on only the radial coordinate. This is in contrast to Kerr (and even-dimensional Myers-Perry with or without equal angular momenta) which has non-trivial angular dependence. Thus, cohomogeneity-1 Myers-Perry black holes are easier to study than Kerr.

Myers-Perry can be generalised to include a cosmological constant $\Lambda$. This solution was first found in \(d=5\) (the Hawking-Hunter-Taylor black hole~\cite{hawkingRotationAdSCFT1999}), and generalized  later to arbitrary dimensions~\cite{gibbonsGeneralKerrdeSitter2005,gibbonsRotatingBlackHoles2004}, and they retain all the symmetries discussed above in the $\Lambda=0$ case. We will focus on the equal angular momenta Myers-Perry$-$de Sitter spacetime in odd dimensions, \(d = 2N + 3\), where \(N \ge 1\) is an integer, abbreviating it to simply MP-dS when unambiguous. In Boyer-Lindquist-like (BL) coordinates\footnote{These are related to the \emph{unified} Boyer-Lindquist coordinates of~\cite{gibbonsGeneralKerrdeSitter2005,gibbonsRotatingBlackHoles2004} by the transformations $r^{2} \to (r^{2}+a^{2})\left(1 + a^{2}/L^{2}\right)^{-1}$ and $M \to M(1 + a^{2}/L^{2})^{N+2}$.} \(x_{a} = (t, r, \psi, x_{i})\), the metric can be written as~\cite{diasScalarFieldCondensation2010}
\begin{equation}
  ds^2 = -\frac{f(r)}{h(r)} dt^2 + \frac{1}{f(r)} dr^2 + r^{2} h(r) \Big( d\psi + \mathcal{A} - \Omega(r) \, dt \Big)^2 + r^2 d\Sigma^{2}
 \label{eqn:metric_BL}
\end{equation}
where
\begin{equation}
  f(r) = 1 - \frac{r^2}{L^2} - \frac{2 M}{r^{2N}}\left(1+\frac{a^2}{L^2}\right) + \frac{2 M a^2}{r^{2N+2}}, \qquad h(r) = 1 + \frac{2 M a^{2}}{r^{2N+2}}, \qquad \Omega(r) = \frac{2 M a}{r^{2N+2}h(r)},\label{eqn:metric_BL2}
\end{equation}
with $L$ being the de Sitter radius, and we have expressed the sphere \(S^{2N+1}\) as a fibration (parameterised by \(\psi\)) over \(\mathbb{CP}^{N}\), with Fubini-Study metric \(d\Sigma^{2} = \hat{g}_{ij}dx^{i} dx^{j}\), where the latin indices run over the \(\mathbb{CP}^{N}\) coordinates \(1, \dots , 2N\).  The volume element is \(\sqrt{-g} = r^{2N+1}\sqrt{\hat{g}}\). The one-form \(\mathcal{A} = \mathcal{A}_{i} dx^{i}\) is a local potential for the Kähler form \(\mathcal{J}\) on \(\mathbb{CP}^{N}\), \emph{i.e.} \(d\mathcal{A} = 2 \mathcal{J}\). In the \(N = 1\) case, \(\mathbb{CP}^{1}\) is isomorphic to \(S^{2}\), so we can introduce the standard spherical polar coordinates \((x_{1},x_{2}) = (\theta, \phi)\), with
\begin{equation}
  \hat{g} = \frac{1}{4}\left( d\theta^{2} + \sin^{2}\theta \, d\phi^{2} \right), \qquad \mathcal{A} = \frac{1}{2} \cos\theta \, d\phi.
  \label{eqn:CP1_coords}
\end{equation}
See appendix B of~\cite{diasInstabilityHigherdimensionalRotating2010} for an explicit construction of \(\hat{g}_{ij}\) and \(\mathcal{A}\) for \(N > 1\).

We assume that the mass parameter \(M\) is positive, and we can also assume that the angular momentum
\(a \ge 0\) without loss of generality due to the $t-\psi$ symmetry. The positive real roots of $f(r)$ define the horizon radii. From Descartes' rule of signs, we can show that \(f(r)\) has either three or one positive real roots, counted with multiplicity. Since asymptotically de Sitter black holes must have at least two roots, we require that \(f(r)\) has a single zero at each of the three positive real roots  \(r_{-} \le r_{+} \le r_{c}\), which define the Cauchy horizon ($r_-$), event horizon ($r_+$) and cosmological horizon ($r_{c}$). This requirement restricts the parameter space \((M, a, L)\). Since $f(r)$ is negative in the limit \(r \to \infty\), we see that \(f(r)\) is positive for \(r_- < r < r_+\) and negative for \(r_+ < r < r_c\).
\begin{figure}
  \centering
  \begin{subfigure}{.57\linewidth}
    \centering
    \includegraphics[width=\linewidth]{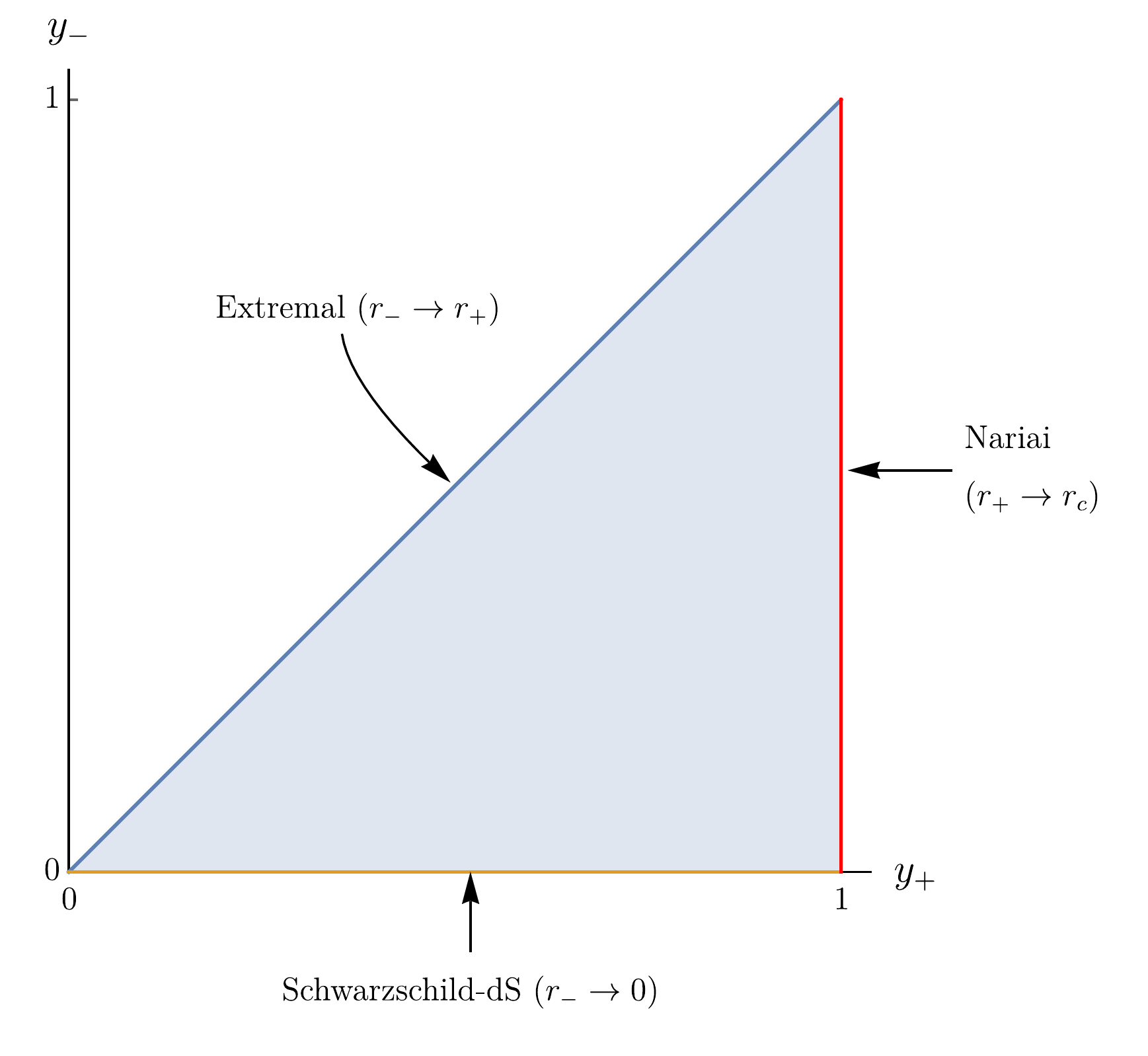}
  \end{subfigure}
  \begin{subfigure}{.37\linewidth}
    \centering
    \includegraphics[width=\linewidth]{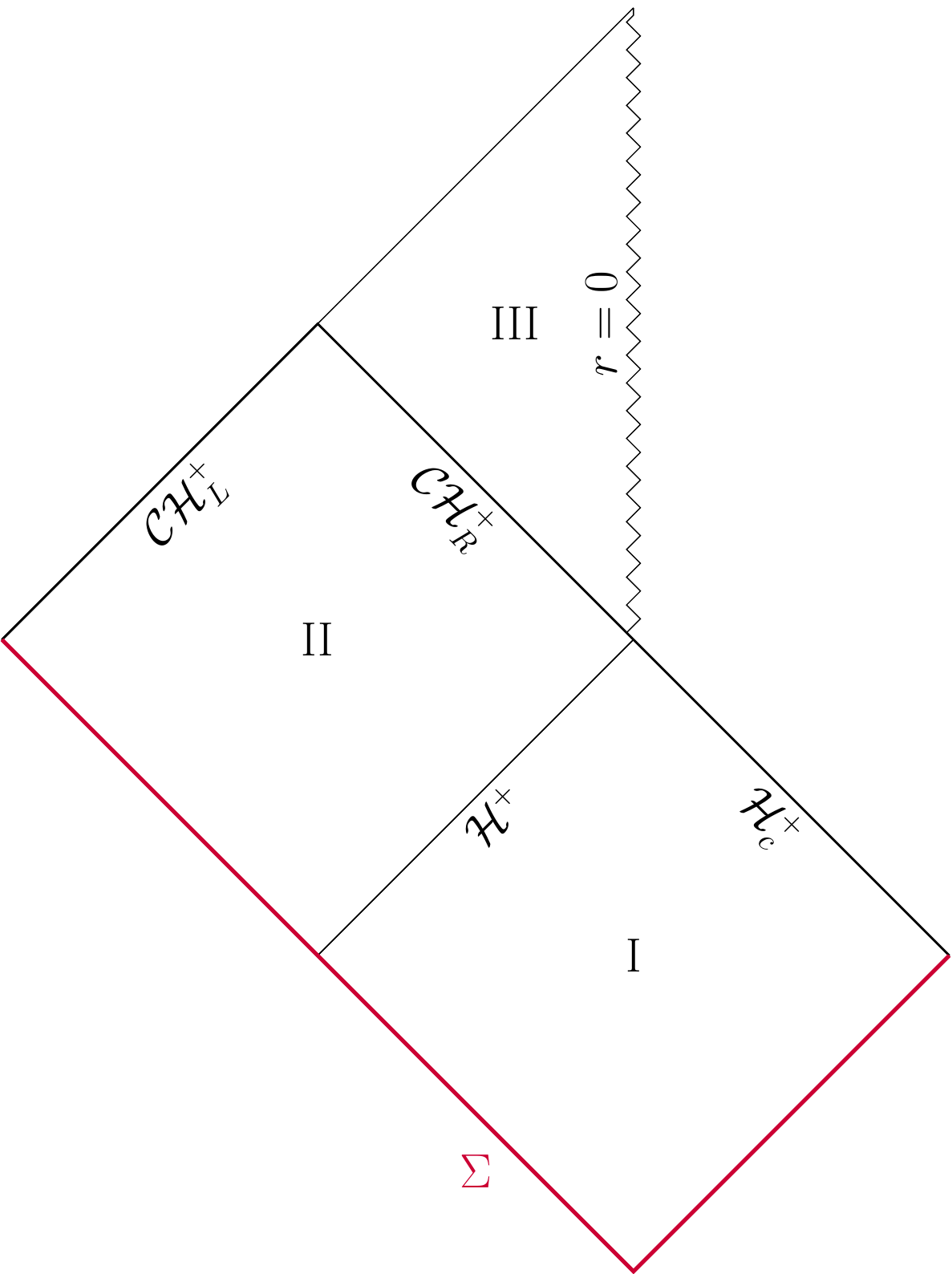}
  \end{subfigure}
  \caption{\textbf{Left panel:} Parameter space of equal angular momenta Myers-Perry$-$de Sitter in terms of the dimensionless variables (\(y_{+}, y_{-}\)), and the relevant limits. \textbf{Right panel:} Globally hyperbolic portion of non-extremal MP-dS with Cauchy surface \(\Sigma\), and its analytic extension beyond the right Cauchy horizon \(\mathcal{CH}_{R}^{+}\). The event horizon \(\mathcal{H}_{+}\) and cosmological horizon \(\mathcal{H}_{c}^{+}\) are also shown.\label{fig:parameter_space_penrose}}
\end{figure}
The surface gravity at each horizon \(r_{i} \in \{r_{-}, r_{+}, r_{c}\}\) is
\begin{equation}\label{eqn:surfGrav}
  \kappa_{i} \equiv \frac{\left| f'(r_{i}) \right|}{2 r_{i} h(r_{i})}.
\end{equation}
We can express $L$, $M$ and $a$ in terms of $r_-$, $r_+$, $r_c$ using the three conditions $f(r_i)=0$.
The Einstein equations of motion are invariant under the scaling $g\to \lambda^2 g$ and $L \to  \lambda L$, with $\lambda\in \mathbb{R}$, which we can use to construct dimensionless quantities in units of $r_c$. For example,
\begin{equation}\label{dimlessQs}
  y_{-} \equiv \frac{r_{-}}{r_{c}}, \qquad\qquad y_{+} \equiv \frac{r_{+}}{r_{c}}, \qquad\qquad \alpha \equiv \frac{a}{r_{c}}.
\end{equation}
It follows that the MP-dS black hole is a 2-parameter solution. We can choose, as we will often do, to parameterise the MP-dS solution using the dimensionless parameters $(y_{+}, y_{-})$ where \(y_{+}\in [0,1]\) and $0 \le y_{-} \le y_{+}$. This parameter space is plotted in the left panel of Fig.~\ref{fig:parameter_space_penrose}, with the relevant limits indicated. Extremality occurs when $y_- \to y_+$, at which $\kappa_+$ and $\kappa_-$ vanish, \emph{i.e.} the black hole has vanishing temperature.  

Alternatively, for $d = 5$, we will sometimes instead parameterise MP-dS using $(y_{+}, \alpha)$ where $0 \le \alpha \le \alpha_{\textrm{ext}}$ ($\alpha_{\textrm{ext}}$ is the value of $\alpha$ at extremality). We will find it useful to do so because it emphasises the relationship with Kerr-dS. In $d=5$ we can express $y_{-}$ in terms of $(y_+, \alpha)$ using\footnote{Similar relations to \eqref{eqn:ym_to_a} and \eqref{eqn:extRotd5}  hold for arbitrary odd $d$ but it is not enlightening to explicitly display them here.}
\begin{equation}
  \quad y_{-}^{2} = - \frac{1 + y_{+}^{2}}{2} + \frac{1}{2}\sqrt{(1-y_{+})^{2} - \frac{4 y_{+}^{4}}{\alpha^{2} - y_{+}^{2}(1-\alpha^{2})}}, \qquad\quad (d = 5)
  \label{eqn:ym_to_a}
\end{equation}
and the extremal value of the rotation parameter (for which the temperature of the event horizon vanishes) is given by
\begin{equation}
  \alpha_{\textrm{ext}}^{d=5} = \frac{y_{+} \sqrt{1 + 2 y_{+}^2}}{\sqrt{2} (1 + y_{+}^2)}.
  \label{eqn:extRotd5}
\end{equation}

Let us now describe the causal structure of cohomegeneity-1 MP-dS black holes in odd $d$. The relevant piece of its Penrose diagram is shown in the right panel of Fig.~\ref{fig:parameter_space_penrose}. Region I is the exterior region described by the metric~\eqref{eqn:metric_BL} with $r_{+} < r < r_{c}$. We define ingoing Eddington-Finklestein (EF) coordinates~\((v,r,\psi',x_{i})\),
\begin{equation}
  dv = dt + \frac{\sqrt{h(r)}}{f(r)} dr, \qquad\qquad d\psi' = d\psi + \Omega(r) \frac{\sqrt{h(r)}}{f(r)} dr, \label{eqn:ingoingEF}
\end{equation}
in terms of which the metric is
\begin{equation}
  ds^{2} = - \frac{f(r)}{h(r)} \, dv^{2} + \frac{2}{\sqrt{h(r)}} \,dv \,dr + r^{2} h(r) \left(d\psi' + \mathcal{A} - \Omega(r) \, dv\right)^{2} + r^{2} d\Sigma^{2}.
  \label{eqn:metric_ingoing_EF}
\end{equation} %with volume element \(\sqrt{-g} = r^{2N+1}\sqrt{\hat{g}}\). 
This is regular at the future event horizon \(\mathcal{H}^{+}\), and the metric in region I can be analytically continued to region II where $r_{-} < r < r_{+}$. Converting to outgoing Eddington-Finklestein coordinates~(\(u, r, \psi'', x_{i}\)),
\begin{equation}
  du = dt - \frac{\sqrt{h(r)}}{f(r)} dr, \qquad\qquad d\psi'' = d\psi - \Omega(r) \frac{\sqrt{h(r)}}{f(r)} dr,
  \label{eqn:outgoingEF}
\end{equation}
the metric is the same as~\eqref{eqn:metric_ingoing_EF} with the usual change of sign \(dv \,dr \to - du \, dr\), and we can use these coordinates to analytically continue beyond the Cauchy horizon into region III where $r < r_{-}$. The causal structure is the same as Kerr-de Sitter (Kerr-dS) and Reissner-Nordström-de Sitter (RN-dS). There is also a left Cauchy horizon \(\mathcal{CH}_{L}^{+}\), however in the context of gravitational collapse it is occluded by the matter region, so we do not consider it here.

\subsection{Klein-Gordon equation in MP-dS}

We want to study massive scalar field perturbations, with mass $\mu$, on a fixed MP-dS background \eqref{eqn:metric_BL}, which are governed by the Klein-Gordon equation
\begin{equation}
  \nabla_a \nabla^a \Phi - \mu^{2}\Phi= 0.
  \label{eqn:wave_eqn_abstract}
\end{equation}
In the context of Strong Cosmic Censorship, the scalar field has been found to be a good proxy for linearised gravitational perturbations in both RN-dS and Kerr-dS~\cite{diasStrongCosmicCensorship2018,cardosoQuasinormalModesStrong2018,diasStrongCosmicCensorship2018a}. To study linear mode perturbations, we make the following separation ansatz in BL coordinates~\eqref{eqn:metric_BL},
\begin{equation}
  \Phi = e^{- i \omega t + i m \psi}\mathcal{R}(r)Y(x_{i}),
  \label{eqn:separation_ansatz}
\end{equation}
which introduces the frequency $\omega$ and azimuthal quantum number $m$ of the perturbation.
Using this ansatz, the Klein-Gordon equation separates, with the angular eigenfunction $Y(x_{i})$ satisfying a charged Laplace equation on \(\mathbb{CP}^{N}\) with charge $m$ and with  eigenvalue \(\lambda\),
\begin{equation}
  (\mathcal{D}^{2} + \lambda) Y(x_{i}) = 0, \qquad\quad \mathcal{D} \equiv \hat{\nabla} - i m \mathcal{A},
  \label{eqn:CPN_equation}
\end{equation}
where \(\hat{\nabla}\) is the covariant derivative on \(\mathbb{CP}^{N}\). The eigenfunctions \(Y(x_{i})\) were studied in~\cite{hoxhaKaluzaKleinConsistencyKilling2000}, and regularity requires that the eigenvalues are quantised as \cite{hoxhaKaluzaKleinConsistencyKilling2000}
\begin{equation}
  \lambda = l(l+2N) - m^{2}, \qquad l = 2k + |m|, \qquad k = 0, 1, 2, \dots
  \label{eqn:CPN_eigenvalues}
\end{equation}
Here $l$ labels the total angular momentum of the mode. The radial equation reduces to
\begin{multline}
   \mathcal{R}''(r) +  \left( \frac{1+2N}{r} + \frac{f'}{f} \right) \mathcal{R}'(r) + \frac{1}{f}\left( \frac{h}{f}(\omega - m \Omega)^{2}- \frac{\lambda}{r^{2}} - \frac{m^{2}}{r^{2}h}  -\mu^{2} \right) \mathcal{R}(r) = 0.
  \label{eqn:waveEqnBL}
\end{multline}
We will need to solve this eigenvalue equation to find the eigenfrequencies $\omega$ of the system. For that, we require that  $\Phi$ obeys QNM boundary conditions, namely \(\Phi\) must be regular in ingoing coordinates~\eqref{eqn:ingoingEF} at the future event horizon $\mathcal{H}^+$ and regular in outgoing coordinates~\eqref{eqn:outgoingEF} at the cosmological horizon $\mathcal{H}^+_c$. In BL coordinates~\eqref{eqn:metric_BL}, this translates to the requirement that the the radial eigenfunction must behave as
\begin{equation}
  \mathcal{R}(r) \sim
  \begin{cases}
    \,\, (r-r_{+})^{-i\frac{\omega - m \Omega(r_{+})}{2\kappa_{+}}} \hat{\mathcal{R}}_{(+)} &\quad \textrm{as}\quad r \to r_{+}\,,\\
    \,\, (r_{c}-r)^{-i\frac{\omega - m \Omega(r_{c})}{2\kappa_{c}}} \hat{\mathcal{R}}_{(c)} &\quad \textrm{as}\quad r \to r_{c}\,,
  \end{cases}
  \label{eqn:QNM_BCs}
\end{equation}
for some functions $\hat{\mathcal{R}}_{(+)}$ and $\hat{\mathcal{R}}_{(c)}$ that are smooth at $r_{+}$ and $r_{c}$, respectively.

\subsection{Numerical setup}\label{sec:numerical_setup}

To prepare the radial Klein-Gordon equation~\eqref{eqn:waveEqnBL} to be solved numerically, we redefine \(\mathcal{R}(r)\) so that the only solutions which are regular at the boundaries are those which obey QNM boundary conditions \eqref{eqn:QNM_BCs}, hence the only solutions that converge numerically are QNMs. A Frobenius analysis\footnote{A similar Frobenius analysis is discussed in detail in section~\ref{sec:scc}.} at \(r=r_{+}\) yields solutions of the form
\begin{equation}
  \mathcal{R}\big|_{r=r_{+}} \, \sim \, c_{1}\, (r-r_{+})^{-i \frac{\omega - m \Omega(r_{+})}{2\kappa_{+}}} \hat{\mathcal{R}}_{(1,+)} + c_{2}\, (r-r_{+})^{i \frac{\omega - m \Omega(r_{+})}{2\kappa_{+}}} \hat{\mathcal{R}}_{(2,+)},
  \label{eqn:frobeniusRp}
\end{equation}
for some constants \(c_{i}\) and functions \(\hat{\mathcal{R}}_{(i,+)}\) which are analytic at \(r_{+}\). Similarly, at the cosmological horizon,
\begin{equation}
  \mathcal{R}\big|_{r=r_{c}}  \, \sim \, c'_{1}\,(r_{c}-r)^{i \frac{\omega - m \Omega(r_{c})}{2\kappa_{c}}} \hat{\mathcal{R}}_{(1,c)} + c'_{2}\,(r_{c}-r)^{-i \frac{\omega - m \Omega(r_{c})}{2\kappa_{c}}} \hat{\mathcal{R}}_{(2,c)}\,,
  \label{eqn:frobeniusRc}
\end{equation}
where \(\hat{\mathcal{R}}_{(i,c)}\) are analytic at \(r_{c}\). The first term in~\eqref{eqn:frobeniusRp} and the second term in~\eqref{eqn:frobeniusRc} obey the QNM boundary conditions~\eqref{eqn:QNM_BCs}, but not the other two. So we have to eliminate the latter. For that we make the redefinition
\begin{equation}
  \mathcal{R}(r) = (r-r_{+})^{-i \frac{\omega - m \Omega(r_{+})}{2\kappa_{+}}}(r_{c}-r)^{-i \frac{\omega - m \Omega(r_{c})}{\kappa_{c}}} \mathcal{Q}(r)\,,
\end{equation}
such that $\mathcal{R}(r)$ obeys QNM boundary conditions~\eqref{eqn:QNM_BCs} at both boundaries if $\mathcal{Q}(r)$ is analytic (\emph{i.e.} the singular solutions are never captured by the numerical function $\mathcal{Q}(r)$ which is necessarily analytic). We also introduce the compact dimensionless radial coordinate 
\begin{equation}
y = \frac{r-r_{+}}{r_{c}-r_{+}}
\end{equation}
such that the event horizon is at $y = 0$ and the cosmological horizon is at $y = 1$. Finally, we perform a Taylor expansion of the resulting ODE at each boundary, to get Dirichlet and Robin boundary conditions at $y = 0$ and $y = 1$, respectively, which we impose at the matrix level after discretisation. These are purely for numerical convenience, to guard against zero cancellation at the boundaries, and do not change the resulting solution, since they follow directly from the equations of motion. A detailed discussion of this type of \emph{derived boundary condition} can be found in~\cite{diasNumericalMethodsFinding2016}. The explicit expression for the final ODE fully prepared for numerical evaluation is long and unenlightening, so we do not write it here.

Working now with dimensionless quantities, as defined in \eqref{dimlessQs}, the resulting ODE is a quadratic eigenvalue problem in the frequency $\tilde{\omega}=\omega \,r_c$, \emph{i.e.} the coefficients depend quadratically on \(\tilde{\omega}\). We use two complementary approaches to numerically find these QNMs, which both use pseudospectral collocation methods. We can either solve directly for the full spectrum of eigenvalues (using e.g. \emph{Mathematica}'s built-in routine \emph{Eigensystem}), or start with a known eigenvalue/eigenvector pair and march it from one part of the parameter space to the other using a Newton-Raphson algorithm. The latter only computes a single eigenvalue family but is much more efficient, allowing us to reach parts of the parameter space that are difficult numerically. These approaches are described in detail in~\cite{Dias:2014eua,diasNumericalMethodsFinding2016}.

%=======================================================

\section{Quasinormal modes families of Myers-Perry$-$de Sitter}\label{sec:AnalyticalQNMfamilies}

%=======================================================

Ideally, the first step in classifying the QNM spectra is to identify some corner or window of the parameter space where we can find analytic expressions for the frequencies (in some approximation). This can help identify the physical nature of the modes and eventually already hint at the existence of different families of QNMs, and it can further be used to test the numerical results. In this section we derive approximations in the de Sitter ($a \to 0$, $r_{+} \to 0$), eikonal ($m = l \to \infty$) and near-extremal limit ($r_{-} \to r_{+}$).

\subsection{de Sitter modes}

When $a \to 0$, $r_{+} \to 0$, the MP-dS black hole reduces to the  pure de Sitter spacetime.
Scalar perturbations in this background,  \emph{i.e.} the pure de Sitter (dS) modes, must be regular at the origin and at the cosmological horizon.
The $d$-dimensional dS modes have been studied previously~\cite{lopez-ortegaQuasinormalModesDdimensional2006}. In $d=2N+3$ dimensions, the pure dS modes have the frequency spectrum
\begin{align}
  \omega_{dS} \, r_{c} &= -i \, ( l + 2 n), &\textrm{(for all $d$)}\label{eqn:ds_modes_odd}\\
  \omega_{dS}^{\textrm{even}} \, r_{c} &= -i \,\Big[l + 2(n+N+1)\Big], &\textrm{(even $d$ only)}\label{eqn:ds_modes_even}
\end{align}
for radial overtone \(n = 0, 1, 2, \dots\). In odd dimension $d$, \(N\) is an integer, and so the second set of modes \(\omega_{dS}^{\textrm{even}}\) is a subset of the first. However, in even dimensions $d$, $N$ is a half integer and thus the second mode set \eqref{eqn:ds_modes_even} is distinct from the first described by \eqref{eqn:ds_modes_odd}. This property of dS modes plays an important role in the mode spectrum of Schwarzschild-de Sitter and MP-dS, as we will discuss, in particular, in Appendix~\ref{sec:SdS_and_small_a}.

\subsection{Photon sphere modes}\label{sec:PS_modes}

When there is an event horizon at $r=r_+$, there is a second family of QNMs that is known as the photon sphere (PS) modes. This nomenclature derives from the fact that, in the eikonal limit  $|m| = l \to \infty$, the frequencies of these modes (which exist for any $l,m$) are related to the properties of unstable circular photon orbits of the background.

The eikonal approximation to the PS frequencies can be derived directly, via a WKB approximation of \eqref{eqn:waveEqnBL}. However, there is a well-known correspondence (established via direct comparison with the leading order WKB analysis) that relates the PS mode frequencies to the behaviour of unstable null geodesics in the equatorial plane:
\begin{equation}
  \omega_{\hbox{\tiny WKB}} = m \, \Omega_{0} - i \left(n + \frac{1}{2}\right) |\lambda_{\rm L}|
  \label{eqn:omega_WKB}
\end{equation}
where \(\Omega_{0}\) is the Kepler angular velocity of the geodesics and \(\lambda_{\rm L}\) is the principal Lyapunov exponent that describes the instability timescale of the geodesics.
In four dimensions this correspondence is well studied, both numerically and analytically~\cite{mashhoonStabilityChargedRotating1985,ferrariNewApproachQuasinormal1984,hodBlackholeQuasinormalResonances2009,dolanQuasinormalModeSpectrum2010, diasStrongCosmicCensorship2018,cardosoQuasinormalModesStrong2018,diasStrongCosmicCensorship2018a}. In higher dimensions this is not as well documented\footnote{This correspondence has been derived explicitly for static spacetimes in arbitrary dimensions~\cite{cardosoGeodesicStabilityLyapunov2009}, but this was not generalised to rotating spacetimes.}. We will verify that this correspondence also holds for cohomogeneity-1 MP-dS black holes.

We use the Lagrangian formalism, solving directly for null geodesics which are independent of the $\mathbb{CP}^{N}$ coordinates, but only rotate in the fiber direction $\psi$. For example, in the $N = 1$ ($d=5$) case, this corresponds to geodesics with $\dot{\theta} = \dot{\phi} = 0$ in the $(\theta, \phi)$ coordinates of~\eqref{eqn:CP1_coords}. For all $N$, the Lagrangian describing such geodesics is
\begin{equation}
  \mathcal{L} = - \frac{f(r)}{2 h(r)} \, \dot{t}^2+\frac{1}{2 f(r)} \, \dot{r}^2+\frac{r^{2} h(r)}{2} \left(\dot{\psi} - \Omega(r) \, \dot{t}\right)^2
\end{equation}
where the dot indicates a derivative with respect to the affine parameter \(\tau\) of the null geodesic and the background functions $f,h,\Omega$ are defined in \eqref{eqn:metric_BL2}. We can associate a conserved energy $E$ and angular momentum $L_{\psi}$ to translations in the $t$ and $\psi$ Killing directions, respectively:
\begin{equation}
  E \equiv \frac{f}{h}\, \dot{t} + \Omega \, L_{\psi}, \qquad\qquad L_{\psi} \equiv r^{2} h(\dot{\psi} -\Omega \, \dot{t}).
  \label{eqn:conserved_quantities}
\end{equation}
Substituting these into the Lagrangian and noting that null geodesics have $\mathcal{L} = 0$, we arrive at the radial Schr\"oedinger equation with potential $V_{\textrm{eff}}$:
\begin{equation}
  \dot{r}^{2} + V_{\textrm{eff}}(r) = 0, \qquad V_{\textrm{eff}}(r) \equiv {L_{\psi}}^{2} f \left( \frac{1}{r^{2} h} - \frac{h}{f}\left(b^{-1}-\Omega\right)^{2} \right)
  \label{eqn:effective_potential}
\end{equation}
where we have defined the impact parameter \(b \equiv \frac{L_{\psi}}{E}\). Unstable circular orbits have \(V_{\textrm{eff}}(r_{0}) = V_{\textrm{eff}}'(r_{0}) = 0\). First solving \(V_{\textrm{eff}}'(r_{0}) = 0\) gives two possible values for the impact parameter
\begin{equation}
  b^{\pm} = \frac{a \sqrt{2 (N+1) M}}{\sqrt{2 (N+1) M} \pm r_{0}^{N}}.
  \label{eqn:impact_parameter_soln}
\end{equation}
Substituting this back into the effective potential and now requiring \(V_{\textrm{eff}}(r_{0}) = 0\), we find that the orbit radii \(r_{0}\) are the (relevant) roots of a polynomial of order $2 (N+1)$:
\begin{equation}
  L^{2} r_{0}^{2} \Big[2 (N+1) M + r_{0}^{N}(r_{0}^{N} \pm 2 \sqrt{2 (N+1) M})\Big] = 2 a^{2} M \Big[N L^{2} - (N + 1) r_{0}^{2}\Big]
  \label{eqn:orbit_polynomial}
\end{equation}
with the $\pm$ sign corresponding to the signs of \(b^{\pm}\). For $N = 1$ this is a quartic polynomial, which we can solve explicitly, but for $N > 1$ we can only find the orbit radii $r_{0}$ numerically.\footnote{Note that using \eqref{eqn:impact_parameter_soln} and \eqref{eqn:orbit_polynomial} we can express the mass and AdS radius in terms of the orbit parameters as $M = \frac{(b^\pm)^2 r_0^{2 N}}{2 (N+1)
   (a-b^\pm)^2}$ and  $L = r_0 \sqrt{\frac{N+1}{
   N-(b^\pm)^{-2}(N+1) r_0^2}}$, which are useful to simplify our final expressions.} We find that only two solutions have \(r_{+}<r_{0}<r_{c}\), with the physical solution corresponding to the minus sign in \eqref{eqn:orbit_polynomial} and thus to $b^{-}$.
The imaginary part of the photon sphere modes~\eqref{eqn:omega_WKB} is proportional to the principal Lyapunov exponent, which characterises the instability time scale of the geodesic. It can be computed from the second derivative of the effective potential~\cite{cardosoGeodesicStabilityLyapunov2009}
\begin{equation}
  \lambda_{\rm L} = \sqrt{- \frac{V_{\textrm{eff}}''(r)}{2 \dot{t}^{2}}} \biggr|_{r=r_{0}}
\end{equation}
and the real part is proportional to the angular velocity of the null orbit, \(\Omega_{0} = \frac{\mathrm{d}\psi}{\mathrm{d}t} = \frac{\dot{\psi}}{\dot{t}}\), which can be computed from~\eqref{eqn:conserved_quantities}. Explicitly, these two quantities are given by
\begin{equation}
  \Omega_{0} = \frac{1}{b^-}, \qquad\quad
  \lambda_{\rm L} = \frac{\sqrt{2N}}{|b^-|} \left| 1 + \frac{a \, (b^-)^2}{r_0^2 (N+1) (a-b^-) }\right| .
   \label{eqn:Omega_Lyapunov}
\end{equation}

Note that expressions involving the black hole parameters $(M, a, L)$ can be written in terms of $(y_{+}, y_{-})$ using~\eqref{eqn:explicit_horizon_radii} in Appendix~\ref{sec:explicit_N}. The solution to~\eqref{eqn:orbit_polynomial} with the larger angular velocity $\Omega_{0}$ (\emph{i.e.} larger real part) corresponds to the \emph{corotating} PS modes, while the other with smaller $\Omega_{0}$ describes the \emph{counter-rotating} PS modes. The corotating modes have smaller $r_{0}$, \emph{i.e.} are closer to the event horizon, and so intuitively they are less stable, with smaller $\lambda_L$. Indeed, the corotating modes always dominate, \emph{i.e.} they have smaller $|\operatorname{Im}(\omega)|$ than their counter-rotating partner.

In Section~\ref{sec:scc}, we will verify that  \eqref{eqn:omega_WKB} and \eqref{eqn:Omega_Lyapunov}, although strictly valid in the limit $|m| = l \to \infty$, also give a very good approximation for the PS modes even when $|m| = l$ is of $\mathcal{O}(1)$.

%============================================
\subsection{Near-horizon geometry and near-horizon modes}\label{sec:near-horizon}
%============================================

There is another limit where we can compute QNMs using analytical methods. Indeed, near-extremal black holes typically have a set of modes known as near-horizon (NH) modes. This nomenclature follows from the fact that these NH modes characteristically have a wavefunction that is highly localised around the event horizon (when near extremality), and they have frequencies approaching $\operatorname{Im}(\omega) \to 0$ and $\operatorname{Re}(\omega) \to m \Omega(r_{+})$ in the strict extremal limit. In subsection \ref{sec:NH_modes}, we will use a matched asymptotic expansion method to analytically capture these NH modes, whereby we match the solution of the Klein-Gordon equation in the near-horizon region of near-extremal MP-dS with a trivial solution in the far-region.

Before deriving the near-horizon modes, we do a small digression in subsections \ref{sec:near-horizon_geometry}$-$\ref{sec:BFbound}, and we find the near-horizon geometry of extremal MP-dS and the associated effective AdS$_2$ Breitenl\"ohner-Freedman (BF) bound. The motivation to do so is twofold. Firstly, the BF bound naturally appears in the expressions for the NH frequencies and it will ultimately provide a criterion to find QNMs that preserve Strong Cosmic Censorship. Moreover, for completeness, we take the opportunity to discuss and test a theorem about instabilities arising from perturbations of near-horizon geometries~\cite{durkeePerturbationsNearhorizonGeometries2011,hollandsInstabilitiesExtremalRotating2015} that is relevant in the context of our study, as we now explain. 

In the near-horizon limit, the geometry of extremal black holes can be expressed locally as a product of AdS$_2$ times a compact space~\cite{durkeePerturbationsNearhorizonGeometries2011}. This is true even if the original spacetime is asymptotically de Sitter. In this limit, the Klein-Gordon equation in the near-horizon geometry reduces to an effective scalar field equation on pure AdS$_2$ space with a certain effective mass $\mu_{\rm eff}$ and charge  $q_{\hbox{\tiny AdS}}$. It is well known that in  AdS$_2$ (with radius $L_{\hbox{\tiny AdS}}$), a scalar field perturbation is normalisable even if its squared mass $\mu_{\hbox{\tiny AdS}}^{2}$ is negative, provided that it obeys the 2-dimensional BF bound \(\mu_{\hbox{\tiny AdS}}^{2} L_{\hbox{\tiny AdS}}^{2} \ge -\frac{1}{4}\)~\cite{breitenlohnerStabilityGaugedExtended,mezincescuStabilityLocalMaximum1985}. On the other hand, the scalar field on AdS$_2$ is not stable if its mass is below the 2-dimensional BF bound.
However, a violation of the effective AdS$_2$ BF bound of the near-horizon geometry of extreme MP-dS does not \emph{necessarily} imply an instability of the scalar field on the full $d$-dimensional MP-dS black hole geometry.

For asymptotically flat or AdS black holes, a conjecture by Durkee and Reall~\cite{durkeePerturbationsNearhorizonGeometries2011} (proven by Hollands and Ishibashi in~\cite{hollandsInstabilitiesExtremalRotating2015}) states that a sufficient (but not necessary) condition for this near-horizon AdS$_2$ BF bound violation to develop into a instability of the extremal black hole is that the unstable mode preserves a certain symmetry already present in the background geometry. In the case of rotating black holes, this is that the perturbation is axisymmetric, \(m= 0\). Assuming there is such an instability, one expects that it also extends away from extremality, by continuity. For Myers-Perry-AdS, near-horizon instabilities triggered by a violation of the near-horizon AdS$_2$ BF bound have been studied in detail in~\cite{diasScalarFieldCondensation2010,durkeePerturbationsNearhorizonGeometries2011}. Strictly speaking, the proof established in \cite{hollandsInstabilitiesExtremalRotating2015} only applies to asymptotically flat or AdS black holes, but is somehow trivial to extend the proof in the asymptotically flat context to black holes living in the static patch of dS. Indeed, in \cite{hollandsInstabilitiesExtremalRotating2015}, slices that extend from $\mathcal{H}^+$ to $\mathcal{I}^+$ were considered and appropriate boundary conditions were given so that the \emph{canonical energy} of \cite{Hollands:2012sf} obeys a certain balance equation that plays a crucial role in the proof. Such a balance equation can still be obtained in the context of black holes living in the static patch of dS by imposing boundary conditions on $\mathcal{H}_c^+$ that are similar to those imposed on  $\mathcal{H}^+$. If one further restricts to perturbations that preserve axisymmetry, the desired result follows.\footnote{We thank S.~Hollands for discussions on this point.} Recent numerical results seem to corroborate the previous extension and show that an effective AdS$_2$ BF bound violation explains the instability of \(d \ge 6\) Reissner-Nordstr\"om$-$de Sitter black holes~\cite{diasOriginReissnerNordstrOm2020} so it is worthwhile to check whether $m=0$ modes in MP-dS can lead to a violation of the 2-dimensional BF bound, and eventually to an instability in MP-dS (in subsection \ref{sec:BFbound} we will find that this is not the case). 

\subsubsection{Near-horizon geometry of MP-dS}\label{sec:near-horizon_geometry}

To find the near-horizon geometry of the extremal MP-dS black hole, we start with~\eqref{eqn:metric_ingoing_EF} at extremality $r_{-} = r_{+}$, where $f(r)$ has a double root $r = r_{-} = r_{+}$, and zoom into the horizon by the coordinate transformations
\begin{equation}
  r \to r_{+} + \epsilon \, R, \qquad t \to \frac{\tilde{T}}{\epsilon}, \qquad \psi \to \Psi + \Omega(r_{+}) \frac{\tilde{T}}{\epsilon}.
  \label{eqn:near_horizon_transformation}
\end{equation}
Taking the limit \(\epsilon \to 0\), i.e. keeping only the leading contribution of an expansion in small $\epsilon$, we obtain the near-horizon geometry
\begin{equation}
  ds^{2} = -\frac{f^{''}(r_{+})}{2 h(r_{+})} R^{2}\, {d\tilde{T}\,}^{2} + \frac{2}{f^{''}(r_{+})} \, \frac{dR^{2}}{R^{2}} + r_{+}^{2} h(r_{+}) \left( d\Psi + \mathcal{A} - R \Omega'(r_{+}) \, d\tilde{T} \right)^{2} + r_{+}^{2} \, d\Sigma^{2}.
\end{equation}
We can make the AdS$_2$ structure more explicit by a further change of coordinates, similarly to~\cite{durkeePerturbationsNearhorizonGeometries2011}. We define the constants
\begin{equation}
  L_{\hbox{\tiny AdS}}^{2} \equiv \frac{2}{f''(r_{+})}\biggr|_{\rm ext}, \qquad\quad \tilde{\Omega} \equiv \Omega'(r_{+}) \sqrt{h(r_{+})}\biggr|_{\rm ext},
  \label{eqn:ads2_L_eff}
\end{equation}
and rescale the time coordinate $\tilde{T} \to L_{\hbox{\tiny AdS}}^{2} \sqrt{h(r_{+})} T$, to get the near-horizon geometry in \((T,R,\Psi,x_{i})\) coordinates
\begin{equation}
  ds^{2} = L_{\hbox{\tiny AdS}}^{2} \left( -R^{2} \, dT^{2} + \frac{dR^{2}}{R^{2}} \right) + r_{+}^{2} h(r_{+}) \left( d\Psi + \mathcal{A} - R \, \tilde{\Omega} L_{\hbox{\tiny AdS}}^{2} \, dT \right)^{2} + r_{+}^{2} \, d\Sigma^{2}\,.
  \label{eqn:near-horizon_geometry}
\end{equation}
This near-horizon geometry is still a solution of the $d$-dimensional Einstein-dS equations. On the other hand, the AdS$_2$ part parameterised by \((T,R)\) satisfies the 2-dimensional Einstein-AdS equations with \(R = - 2 L_{\hbox{\tiny AdS}}^{-2}\).

\subsubsection{Perturbations of the near-horizon geometry and the AdS$_2$ BF bound}\label{sec:BFbound}

Linear mode solutions of the Klein-Gordon equation on the near-horizon geometry~\eqref{eqn:near-horizon_geometry}, with the Fourier decomposition $\Phi = e^{- i \omega T + i m \Psi}\chi(R)$, must satisfy the ODE
\begin{equation}
  R^2 \, \chi'' + 2 R \, \chi' - \left[ \frac{(\omega^{2} - m \tilde{\Omega} L_{\hbox{\tiny AdS}}^{2} R)^{2}}{R^{2}} - L_{\hbox{\tiny AdS}}^{2}\left( \mu^{2} + \frac{\lambda}{r_{+}^{2}} + \frac{m^{2}}{r_{+}^{2}h(r_{+})} \right) \right] \chi = 0.
  \label{eqn:near-horizon_wave_eqn}
\end{equation}
We can write this as a massive charged Klein-Gordon equation on pure AdS$_2$,
\begin{equation}
  \left( \tilde{\nabla} - i q_{\hbox{\tiny AdS}}A(R) \right)^{2} \Phi = L_{\hbox{\tiny AdS}}^{2} \left(\mu^{2} + \frac{\lambda}{r_{+}^{2}} + \frac{m^{2}}{r_{+}^{2}h(r_{+})}\right) \Phi, \qquad A(R) = - R \, dT\,,
  \label{eqn:generic_wave_eqn}
\end{equation}
if we make the identification $q_{\hbox{\tiny AdS}}= - m \, L_{\hbox{\tiny AdS}}^{2} \tilde{\Omega}$. Here \(\tilde{\nabla}\) is the covariant derivative on pure AdS$_2$ associated to the metric
\begin{equation}
  {ds}_{\textrm{AdS}_2}^{2} = L_{\hbox{\tiny AdS}}^{2} \left( -R^{2} \, dT^{2} + \frac{dR^{2}}{R^{2}} \right),
\end{equation}
and $ \tilde{\nabla} - i q_{\hbox{\tiny AdS}}A(R)$ is the gauge covariant derivative of a scalar field with effective charge $q_{\hbox{\tiny AdS}}$ in the AdS$_2$ background with a  homogeneous electric field \(A(R)\). The latter descends from the  \(dT \, d\Psi\) component of the metric in the original near-horizon solution \eqref{eqn:near-horizon_geometry}.

Asymptotically, as \(R \to \infty\), the solutions of \eqref{eqn:near-horizon_wave_eqn} behave as \(\chi\sim R^{-\Delta_{\pm}}\) where the 2-dimensional conformal dimensions $\Delta_\pm$ are
\begin{equation}
  \Delta_{\pm} = \frac{1}{2} \pm \frac{1}{2} \sqrt{1 + 4 {\mu_{\rm eff}}^{2} L_{\hbox{\tiny AdS}}^{2}}, \qquad {\mu_{\rm eff}}^{2} \equiv \mu^{2} + \frac{\lambda}{r_{+}^{2}} + \frac{m^{2}}{r_{+}^{2}h(r_{+})} - L_{\hbox{\tiny AdS}}^{2} m^{2} \tilde{\Omega}^{2}.
  \label{eqn:asymptotic_soln_and_effective_mass}
\end{equation}
In order for such solutions to not oscillate at infinity (\emph{i.e.} to be normalisable; with finite energy), we require that $\Delta_{\pm}$ is real. This requirement defines the  AdS$_2$ BF bound of the near-horizon geometry:
\begin{equation}
  {\mu_{\rm eff}}^{2} L_{\hbox{\tiny AdS}}^{2} \ge - \frac{1}{4}.
  \label{eqn:BF_bound}
\end{equation}
In summary, by taking the near-horizon limit of extreme MP-dS, we have found the effective near-horizon AdS$_2$ radius $L_{\hbox{\tiny AdS}}$, charge $q_{\hbox{\tiny AdS}}$ and mass $\mu_{\rm eff}$, which are explicitly given in terms of $(N, y_+)$ by:
\begin{align}
  L_{\hbox{\tiny AdS}}^{2} &= \frac{r_{c}^{2} y_{+}^{2}}{2(N+1)} \frac{1 - y_{+}^{2N+2}(2-y_{+}^{2}+N(1-y_{+}^{2}))}{N(1-y_{+}^{2})(1+y_{+}^{2N +2}) - 2y_{+}^{2} (1-y_{+}^{2N})}, \label{eqn:L_AdS_explicit}\\
  q_{\hbox{\tiny AdS}} &= \frac{L_{\hbox{\tiny AdS}}^{2}}{r_{c}^{2}} \frac{2m}{y_{+}^{2}} \sqrt{\frac{N(1-y_{+}^2)-y_{+}^{2}(1-y_{+}^{2N})}{(1-y_{+}^{2})(1-y_{+}^{2N+2})}}, \label{eqn:q_AdS_explicit}\\
  {\mu_{\rm eff}}^{2} &= \mu^{2} + \frac{1}{r_{c}^{2}} \left(\frac{\lambda}{y_{+}^{2}} - \frac{m^{2}}{y_{+}^{2}}\frac{N}{N+1} \frac{1-y_{+}^{2N+2}(2-y_{+}^{2}+N(1-y_{+}^{2}))}{N(1-y_{+}^{2})(1+y_{+}^{2N+2}) - 2 y_{+}^{2}(1-y_{+}^{2N})}\right). \label{eqn:mu_eff_explicit}
\end{align}
In the UV region of the full geometry, excitations in asymptotically $d$-dimensional dS spacetimes have finite energy (\emph{i.e.} are stable) if and only if $\mu \ge 0$. However, in the IR region, these can correspond to an effective mass $\mu_{\rm eff}$ in the near-horizon region, as defined in~\eqref{eqn:asymptotic_soln_and_effective_mass} and \eqref{eqn:mu_eff_explicit}, that violates the AdS$_2$ BF bound~\eqref{eqn:BF_bound} of the near-horizon geometry. Since \({\mu_{\rm eff}}^{2}\) is minimised when \(\mu = 0\), we will restrict considerations to the massless scalar field $\mu = 0$ on MP-dS from now on. 

For the axisymmetric modes \(m = 0\) the effective mass~\eqref{eqn:asymptotic_soln_and_effective_mass} is always non-negative, and hence there is no AdS$_2$ BF bound violation (that could be relevant to the previously discussed theorem about near-horizon instabilities~\cite{durkeePerturbationsNearhorizonGeometries2011,hollandsInstabilitiesExtremalRotating2015}). Since that theorem provides a sufficient but not necessary condition for instability, we cannot make any conclusions about the stability of MP-dS, but we indeed do not find any instabilities when $m=0$.

We will also be interested in modes with non-zero \(m\), which \emph{can} violate the BF bound~\eqref{eqn:BF_bound}. Of particular interest will be the behaviour in the eikonal limit where $m = l$ is large. Recall that, analogously to the spherical harmonics, the \(\mathbb{CP}^{N}\) angular eigenvalues $\lambda$ of~\eqref{eqn:CPN_equation} can be labelled by $m$ and $l$, with $|m| \le l$. For a fixed $m$, one can show that ${\mu_{\rm eff}}^{2}$ is minimised when $l = |m|$, \emph{i.e.} a BF bound violation will first occur for the maximally corotating modes. Furthermore, the BF bound can always be violated for sufficiently large $m = l$. Recall that in this $m\neq 0$ case the background symmetry is not preserved, and it follows from the analysis of Durkee-Reall and Hollands-Ishibashi~\cite{durkeePerturbationsNearhorizonGeometries2011,hollandsInstabilitiesExtremalRotating2015} that a violation of the AdS$_2$ BF bound says nothing about the existence of eventual instabilities in the full MP-dS geometry. Yet, one might expect that a BF bound violation can signal some transition boundary of the physical properties of the system. This will be indeed the case, as we will see in the discussion of the results of Fig.~\ref{fig:beta_d11_varying_m} and Table~\ref{tab:BF_comparison} of section~\ref{sec:scc}.

\subsubsection{Near-horizon modes}\label{sec:NH_modes}

To find the near-horizon (NH) modes in an off-extremality expansion, we use a matched asymptotic expansion, which is motivated by the following considerations. From our numerical results we find that close to extremality, where $r_{-} \to r_{+}$, NH eigenfunctions are very much localized near the event horizon and very quickly decay as we move away from it towards the cosmological horizon. To obtain a good analytical approximation that well describes the NH mode solutions of the Klein-Gordon equation we can then split the spacetime into a near-region, localized around the horizon, and a far-region, that extends all the way up to the cosmological horizon. In the near-region, a double series expansion of the Klein-Gordon equation around the extremal black hole and, simultaneously, about the event horizon yields an hypergeometric equation which we can solve analytically to find the near-region eigenfunction. We then match this solution with the far-region eigenfunction which, from the above observations and in a ``poor-man'' approximation, can be taken to be approximately the trivial vanishing solution to leading order in the expansion. The matching and boundary conditions fix the amplitudes of the eigenfunctions and quantise the frequency of the NH modes. To validate our matched asymptotic expansion and to simultaneously identify the NH modes, we compare this analytical expression for the frequency with the numerical data. 
In the literature there are systems where a similar strategy proved to be very useful and successful~\cite{diasStrongCosmicCensorship2018a,diasHuntingFermionicInstabilities2020, diasStrongCosmicCensorship2019,yangQuasinormalModesNearly2013}. 

The explicit derivation presented here is for $N = 1$ (for clarity of the presentation), but the approach generalises, and the main result that we present in the end is valid for all $N$. We start by defining
\begin{equation}
  \sigma = 1 - \frac{r_{-}}{r_{+}}, \qquad x = 1 - \frac{r}{r_{+}}, \qquad z = x \, \sigma\,.
  \label{eqn:MAE_variables}
\end{equation}
Small $\sigma$ corresponds to taking the near-extremal limit, while small $x=z/\sigma$ corresponds to a zoom into the horizon (note that $x\leq 0$ and $z\leq 0$). We will take the $\sigma \to 0$ limit while holding $z$ fixed to zoom in the near-extremal solution around the horizon. To zeroth order in the $\sigma$-expansion, we look for modes (the NH modes) whose frequency at extremality is purely real and satisfies the superradiant bound, \(\omega = m \, \Omega(r_{+})|_{\rm ext}\). This suggests that \eqref{eqn:MAE_variables} should be accompanied by the $\sigma$-expansion in the frequency,
\begin{equation}
  \omega = m \, \Omega(r_{+})\big|_{\rm ext} + \sigma \, \delta \omega\,,
  \label{eqn:NH_ansatz}
\end{equation}
where we will have to determine the next-to-leading order frequency correction $\delta\omega$.
Inserting~\eqref{eqn:MAE_variables}$-$\eqref{eqn:NH_ansatz} into  the (massless) Klein-Gordon equation~\eqref{eqn:waveEqnBL} and taking the limit \(\sigma \to 0\) while holding \(z\) fixed, we can show that the leading order contribution of the expansion is a hypergeometric equation for $\chi(z)$ if we perform the field redefinition
\begin{equation}
  R(z) = z^{A}(1-z)^{B} \chi(z)
\end{equation}
where $A$ and $B$ are given by
\begin{align}
  A &= - i\left( \frac{ m \sqrt{1+y_{+}^{2}}(1 + 3 y_{+}^{2} + 4 y_{+}^{4})}{8(1 + 2 y_{+}^{2})(1 - y_{+}^{4})} + \frac{y_{+} \sqrt{(1+y_{+}^{2})(1+2 y_{+}^{2})}}{2 \sqrt{2} (1-y_{+}^{2})} \, \delta \tilde{\omega} \right), \\
  B &=  i\left( \frac{m (1 + 3 y_{+}^{2})(3 + 4 y_{+}^{2})}{8 \sqrt{1 + y_{+}^{2}}(1 + 2 y_{+}^{2})(1 - y_{+}^{2})} - \frac{y_{+} \sqrt{(1 + y_{+}^{2})(1 + 2 y_{+}^{2})}}{2 \sqrt{2} (1 - y_{+}^{2})} \, \delta \tilde{\omega}  \right),
\end{align}
where we have introduced the dimensionless frequency correction $\delta \tilde{\omega}\equiv r_{c} \, \delta \omega$.
In these conditions, the general solution of the system is a sum of hypergeometric functions $\, _2F_1$~\cite{NIST:DLMF}:
\begin{equation}\label{eqn:solnChi}
  \chi(z) = C_{(1)} \,{}_{2}\operatorname{F}_{1}(a_{+},a_{-},c;z) + C_{(2)} z^{1-c} {}_{2}\operatorname{F}_{1}(a_{+} - c + 1, a_{-} - c + 1, 2-c; z),
\end{equation}
for some constants $C_{(1)} $ and $C_{(2)} $, and the coefficients \(a_{\pm}\) and \(c\) are expressed in terms of the effective mass $\mu_{\rm eff}$ and AdS$_2$ radius $L_{\hbox{\tiny AdS}}$ given in~\eqref{eqn:L_AdS_explicit}-\eqref{eqn:mu_eff_explicit}, as well as $y_+$ and $m$ as:
\begin{align}
  a_{\pm} &= \frac{1}{2} \pm \frac{1}{2} \sqrt{1 + 4 {\mu_{\rm eff}^{2} {L_{\hbox{\tiny AdS}}}^2}} + \frac{i m \sqrt{1 + y_{+}^{2}} (1 + 4 y_{+}^{2})}{4(1-y_{+}^{2})(1+2 y_{+}^{2})} - \frac{i y_{+} \sqrt{(1 + y_{+}^{2})(1+2 y_{+}^{2})}}{\sqrt{2}(1 - y_{+}^{2})} \, \delta \tilde{\omega} ,\nonumber \\
  c &= 1 - \frac{i m \sqrt{1 + y_{+}^{2}}(1 + 3 y_{+}^{2} + 4 y_{+}^{4})}{4\sqrt{1 + y_{+}^{2}}(1 - y_{+}^{2})(1 + 2 y_{+}^{2})} - \frac{i y_{+} \sqrt{1 + 3 y_{+}^{2} + 2 y_{+}^{4}}}{\sqrt{2} (1 - y_{+}^{2})} \, \delta \tilde{\omega} .
  \label{eqn:near-horizon_coefficients_ac}
\end{align}
Using ${}_{2}\operatorname{F}_{1}(\alpha, \beta, \gamma, 0) = 1$, the leading order behaviour of $R(z)$ near the event horizon $z = 0$ is
\begin{equation}
  R\big|_{z\to 0^-} \, \simeq \, C_{(1)} \, z^{A} + C_{(2)} \, z^{-A}.
  \label{eqn:NH_general_soln}
\end{equation}
The first (second) term describes an ingoing (outgoing) wave at the event horizon $z=0$. We want the solution that is regular in ingoing Eddington-Finklestein coordinates \eqref{eqn:ingoingEF} so we 
set  $C_{(2)} = 0$ in~\eqref{eqn:solnChi}. 

Formally, we should now find the \emph{far-region} wavefunction in some approximation (tailored to an analytical treatment) that is valid far from the event horizon all the way up to the cosmological horizon, and match it with the near-horizon solution to find the QNMs. In our case it is difficult to solve the far-region equations analytically, so we will take the simpler heuristic approach of matching the near-region eigenfunction with a vanishing far-region wavefunction, motivated by our observation that the near-horizon modes are highly peaked near the horizon. In spite of being a ``poor-man'' matched asymptotic expansion, we will find {\it \`a posteriori} that this simple analysis yields an approximation that agrees extremely well with our numerics. The edge of the near-horizon region where the matching is done is at $z \to - \infty$. Using the following relationship between the coefficients,
\begin{equation}
  A + B - a_{\pm} = - \frac{1}{2} \mp \frac{1}{2} \sqrt{1 + 4 {\mu_{\rm eff}}^{2} L_{\hbox{\tiny AdS}}^{2}},
\end{equation}
we can expand the near-region hypergeometric function for large negative \(z\), to get
\begin{align}
\begin{split}
R\big|_{|z|\gg 1} \simeq & \:(-z)^{-\frac{1}{2} - \frac{1}{2} \sqrt{1 + 4 {\mu_{\rm eff}}^{2} L_{\hbox{\tiny AdS}}^{2}}} \frac{\Gamma(a_{-}-a_{+})\Gamma(c)}{\Gamma(c-a_{+})\Gamma(a_{-})} \\
& + (-z)^{-\frac{1}{2} + \frac{1}{2} \sqrt{1 + 4 {\mu_{\rm eff}}^{2} L_{\hbox{\tiny AdS}}^{2}}} \frac{\Gamma(a_{+}-a_{-}) \Gamma(c)}{\Gamma(c-a_{-})\Gamma(a_{+})}.\qquad\qquad
  \label{eqn:near-horizon_asymptotic}
  \end{split}
\end{align}
Thus, the behaviour of the two contributions depends on the real part of each exponent. The expression \(\sqrt{1 + 4 {\mu_{\rm eff}}^{2} L_{\hbox{\tiny AdS}}^{2}}\) is always either positive or imaginary, and so the first term in~\eqref{eqn:near-horizon_asymptotic} always vanishes far away from the event horizon. When \({\mu_{\rm eff}}^{2} L_{\hbox{\tiny AdS}}^{2} \ge 0\) the second term diverges as $|z|$ grows large. Since we want to match the large radius expansion \eqref{eqn:near-horizon_asymptotic} of the near-region with a vanishing far-region wavefunction, we must require that the coefficient of the second term vanishes identically. This happens if one of the arguments of the gamma functions in the denominator is a non-positive integer since $\Gamma(-n)=\infty$, $n\in \mathbb{N}_0$. That is to say, we require \(a_{+} = -n\), for $n = 0, 1, 2, \dots$ which gives a quantisation condition for $\delta\tilde{\omega}$.
Namely, for $N=1$,  the frequency of the NH modes should be well approximated by 
\begin{align}
\begin{split}
  \omega_{\hbox{\tiny NH}}^{(N=1)} r_{c} \simeq &\: \frac{m}{y_{+} \sqrt{2(1+2 y_{+}^{2})}} - i\,\frac{1 - y_{+}^{2}}{y_{+} \sqrt{2(1+y_{+}^{2})(1+2y_{+}^{2})}}   \Biggr[  i\,\frac{ m (1 + 4 y_{+}^{2}) \sqrt{1 + y_{+}^{2}}}{2(1-y_{+}^{2})(1+2y_{+}^{2})}  \\
& +1 + 2 n + \sqrt{1 - \frac{m^{2} (1+2y_{+})^{2}}{2(1-y_{+}^{2})(1+2y_{+}^{2})} + \frac{\lambda (1+2 y_{+}^{2})}{1 - y_{+}^{2}}}  \Biggr] \sigma +\mathcal{O}\left(\sigma^2 \right)\,. 
  \label{eqn:NH_modes_MP-dS_d5}
  \end{split}
\end{align}
Note that $\sigma = 1 - y_{-}/y_{+}$ can be expressed in terms of $(y_{+}, \alpha)$ using~\eqref{eqn:ym_to_a}. The calculation so far is strictly valid for $N=1$ but it generalizes {\it mutatis mutandis} for all $N$. At the end of the day, for any $d=2N+3$, the frequency of the NH modes  can be written as
\begin{equation}
  \omega_{\hbox{\tiny NH}} \simeq m \, \Omega(r_{+})|_{\rm ext}+\! \left[m \Omega_{(1)} - \frac{i}{2} \left( 1 + 2n + 2 i q_{\hbox{\tiny AdS}} + \sqrt{1 + 4 {\mu_{\rm eff}}^{2} {L_{\hbox{\tiny AdS}}}^2} \right) \kappa_{(1)} \right] \sigma +\mathcal{O}\left(\sigma^2 \right)\,
  \label{eqn:NH_modes_MP-dS_abstract}
\end{equation}
where we have defined the first-order coefficients of the Taylor expansion of $\Omega(r_{+})$ and $\kappa_{+}$:
\begin{equation}\label{eqn:NHparameters}
  \Omega_{(1)} \equiv \frac{d\Omega(r_{+})}{d\sigma} \bigg|_{\sigma = 0}, \qquad \kappa_{(1)} \equiv \frac{d\kappa_{+}}{d\sigma} \bigg|_{\sigma = 0}.
\end{equation}
Explicit expressions for $\Omega_{(1)}$ and $\kappa_{(1)}$ as a function of $N$ and $y_+$ are given in \eqref{eqn:kappaOmegaNH} of Appendix~\ref{sec:explicit_N}. Apart from the angular velocity and surface gravity, $\omega_{\hbox{\tiny NH}}$ only depends on the black hole parameters via the effective mass $\mu_{\rm eff}$, charge $q_{\hbox{\tiny AdS}}$ and AdS$_2$ radius $L_{\hbox{\tiny AdS}}$ which characterize  the near-horizon geometry and its perturbations. In general, $\omega_{\hbox{\tiny NH}}$ is complex, but in the axisymmetric case $m = 0$, $q_{\hbox{\tiny AdS}}$ vanishes and the BF bound~\eqref{eqn:BF_bound} is never violated, so the resulting modes are purely imaginary:
\begin{equation}
  \omega_{\hbox{\tiny NH}}^{(m = 0)} r_{c} \simeq - \frac{i}{2} \left( 1 + 2n + \sqrt{1 + 4 {\mu_{\rm eff}}^{2} {L_{\hbox{\tiny AdS}}}^2} \right) \kappa_{(1)}\, \sigma  +\mathcal{O}\left(\sigma^2 \right).
  \label{eqn:NH_modes_ml0}
\end{equation}
This expression is again valid for all $N$.

In the next section, we will use  \eqref{eqn:NH_modes_MP-dS_abstract} to help identify the NH family of QNMs, and simultaneously  \eqref{eqn:NH_modes_MP-dS_abstract} will verify some of our numerical results.

%===========================================================
%===========================================================
\newpage
\section{Quasinormal mode spectra of equal angular momenta MP-dS}\label{sec:QNMspectra}
%===========================================================
%===========================================================

In the previous Section \ref{sec:AnalyticalQNMfamilies} we used analytical methods, strictly valid in certain windows of the black hole or wave parameters, to identify three possible families of quasinormal modes: the de Sitter (dS), photon sphere (PS) and near-horizon (NH) QNM families. 
Strictly speaking we do not know if these 3 families are distinct or whether e.g. two of them describe the same family that happens to be captured by two distinct analytical analysis in different ``corners'' of the parameter space.   
In this section (and in section \ref{sec:scc}) we do a numerical search of the quasinormal modes of MP-dS black holes. This numerical scan of the QNMs is done completely independently of the analysis of the previous section. However, after collecting the data we compare our numerical results with the analytical approximations of Section \ref{sec:AnalyticalQNMfamilies}, in the regime of parameters where the analytical approximations are valid, to identify the origin of each family of QNMs that we find.

Recently, it was shown that in (asymptotically-flat) Kerr-Newman black holes, where the PS and NH families of QNMs exist, there is only a sharp distinction between the PS and NH modes in certain regions of the parameter space. In other regions, the distinction between the two families is much less clear because a phenomenon known as {\it eigenvalue repulsion} is present~\cite{diasEigenvalueRepulsionsQuasinormal2021}. In particular, this means that we can have e.g. PS surfaces that, when approaching the NH surface, `break' into two branches and each one of the two branches then merges smoothly with what was (in other regions) a NH branch (the NH surface itself also breaks into two pieces)~\cite{diasEigenvalueRepulsionsQuasinormal2021}.  Instead of ending with one PS and one NH family of modes we have what we can call two `combined PS-NH' families (describing different overtones) with a frequency gap between them.
This phenomenon of eigenvalue repulsion is commonly observed in solid state physics, for example in the form of an energy gap between different energy bands of simple lattice models (see e.g. section~7 of~\cite{kittelIntroductionSolidState2004}). In our study of the QNM spectra of MP-dS, we also observe eigenvalue repulsions, similar to those in Kerr-Newman (although not just between the PS and NH families).

It is not our aim to do a detailed study of QNMs of MP-dS, since we just need to identify the modes that enforce SCC in MP-dS, and this only requires finding a dominant QNM family that does the job. Instead, we present a selection of results that illustrate the key features of MP-dS QNMs. Recall, from the discussions in Section \ref{sec:myers-perry}, that the MP-dS black hole is a 2-parameter solution and we can take these two parameters to \mbox{be $(y_{+}, \alpha)\equiv (r_+,a)/r_c$} (where $0 \le \alpha \le \alpha_{\textrm{ext}}$, with $\alpha_{\textrm{ext}}$ being the value of $\alpha$ at extremality) or $(y_+,y_-)$. Typically we will display 2-dimensional plots where we plot the frequency as a function of one of the parameters while keeping the second black hole parameter fixed. Altogether, our selection of plots allows us to infer how the complete 3-dimensional plot $(y_+,\alpha,\omega \, r_c)$ looks like.

\begin{figure}
  \includegraphics[width=\textwidth]{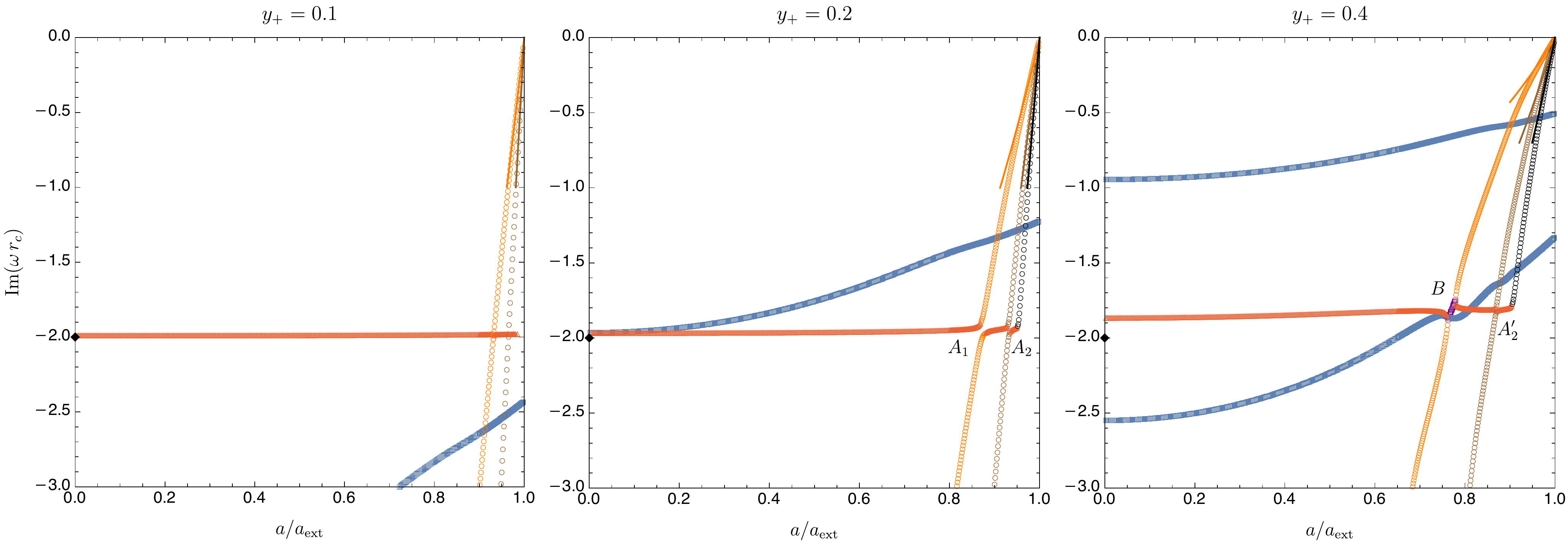}
  \caption{QNM spectrum for $d = 5$ MP-dS with $m = l = 0$ and $y_{+} = 0.1$ (left), $0.2$ (middle) and $0.4$ (right). The red triangle curve is the dS mode (with $n=0$). In the limit $(y_{+} \to 0, a \to 0)$ these reduce to the QNMs of pure dS space~\eqref{eqn:ds_modes_odd}, indicated by a black diamond. The blue square curves are the complex PS modes (with $n=0$; we also show the PS curve with $n=1$ in the right panel). The orange/brown/black circle curves are the $n = 0,1,2$ harmonics of the NH modes, respectively, with analytic approximations~\eqref{eqn:NH_modes_MP-dS_d5} given by solid lines. Eigenvalue repulsions occur at $A_{1}, A_{2}, A_{2}'$ and further subdominant modes (not shown). The complex purple mode at $B$ does not fit into any of the three families: it simply provides a ``bridge'' between two points where, at each one, 3 curves bifurcate from.\label{fig:eigRepulsion}}
\end{figure}

Let us start with the axisymmetric $m = l = 0$ modes in $d = 5$ MP-dS, displayed in Fig.~\ref{fig:eigRepulsion}. For small $y_{+} = 0.1$ (left panel) we identify 3 distinct QNM families: the dS (red triangle curve), PS (blue square curve) and NH (orange and brown circles) families. The PS family has complex frequencies. In the NH case, we display not only the curve with radial overtone $n=0$ (orange disks) but also the family with $n=1$ (brown circles), and they all have purely imaginary frequencies (for $m=0$). The dS red triangle curve (with purely imaginary frequencies) approaches the pure ($y_+=0$) dS normal mode \eqref{eqn:ds_modes_even} when $\alpha\to 0$ (black diamond). On the other hand the  $n=0$ (orange) and $n=1$ (brown) circle NH curves match very well with the NH analytic approximation~\eqref{eqn:NH_modes_MP-dS_d5}  (described by the solid lines) which are valid near extremality $\alpha/\alpha_{\rm ext} \lesssim 1$.

Still in Fig.~\ref{fig:eigRepulsion}, as we increase $y_{+}$ to $0.2$ (middle panel), the dS modes clearly merge with the NH modes as $\alpha$ approaches extremality. This first occurs near $A_{1}$, between the $n = 0$ dS and $n = 0$ NH modes, leaving a gap region in the eigenvalue spectrum of the `old' $n = 0$ NH curve (by `old' we mean w.r.t. the left panel). To the right of $A_{1}$ we see that a small branch of the `old' $n = 0$ dS curve now provides a bridge that connects the bottom section of the `old' $n = 0$ NH curve (on the left/bottom) with the top/right half of the `old' $n = 1$ NH curve. Then, for even larger $\alpha$ we see that around $A_{2}$ there is a branch of the `old' $n = 0$ dS curve that now provides a bridge between a branch of the `old' $n = 1$ NH mode (on the left) and a branch of the `old' $n = 2$ NH mode (on the right). In fact, similar bridges continue to exist (although not displayed) as we approach $\alpha \to \alpha_{\rm ext}$ between the `old' $n$ and $n+1$ NH overtones, not just the $n = 0\to n = 1$ and $n = 1\to n = 2$ overtones that are displayed. Altogether, these features are characteristic of the phenomenon of eigenvalue repulsion~\cite{diasEigenvalueRepulsionsQuasinormal2021}, and Fig.~\ref{fig:eigRepulsion} illustrates how intricate this phenomenon can be.

The spectra become even more intricate when $y_{+}$ increases further, e.g. at $y_{+} = 0.4$ (right panel of Fig.~\ref{fig:eigRepulsion}). Indeed, we find that in region $B$ the $n = 0$ dS mode (red triangles to the left of $B$) merges with the `old' $n =0$ NH mode (orange circles below $B$) and, at the very same point, a small purple bridge bifurcates and extends to the right and up till a new point where 3 lines merge again: this time it is the purple bridge, and the other `halves' of the `old' $n = 0$ NH (orange circles) and the `old' $n = 0$ dS curve  (red triangles). For larger $\alpha$, the latter then merges with the `old' $n = 1$ NH family (much like the middle panel). Note that all of this occurs in a region where one also finds the $n=1$ PS curve (blue square curve on the bottom) that seems to go through crossovers without significant interaction. Again, we see how eigenvalue repulsions can make the spectra very elaborate. 

\begin{figure}
  \centering
  \includegraphics[width=0.75\textwidth]{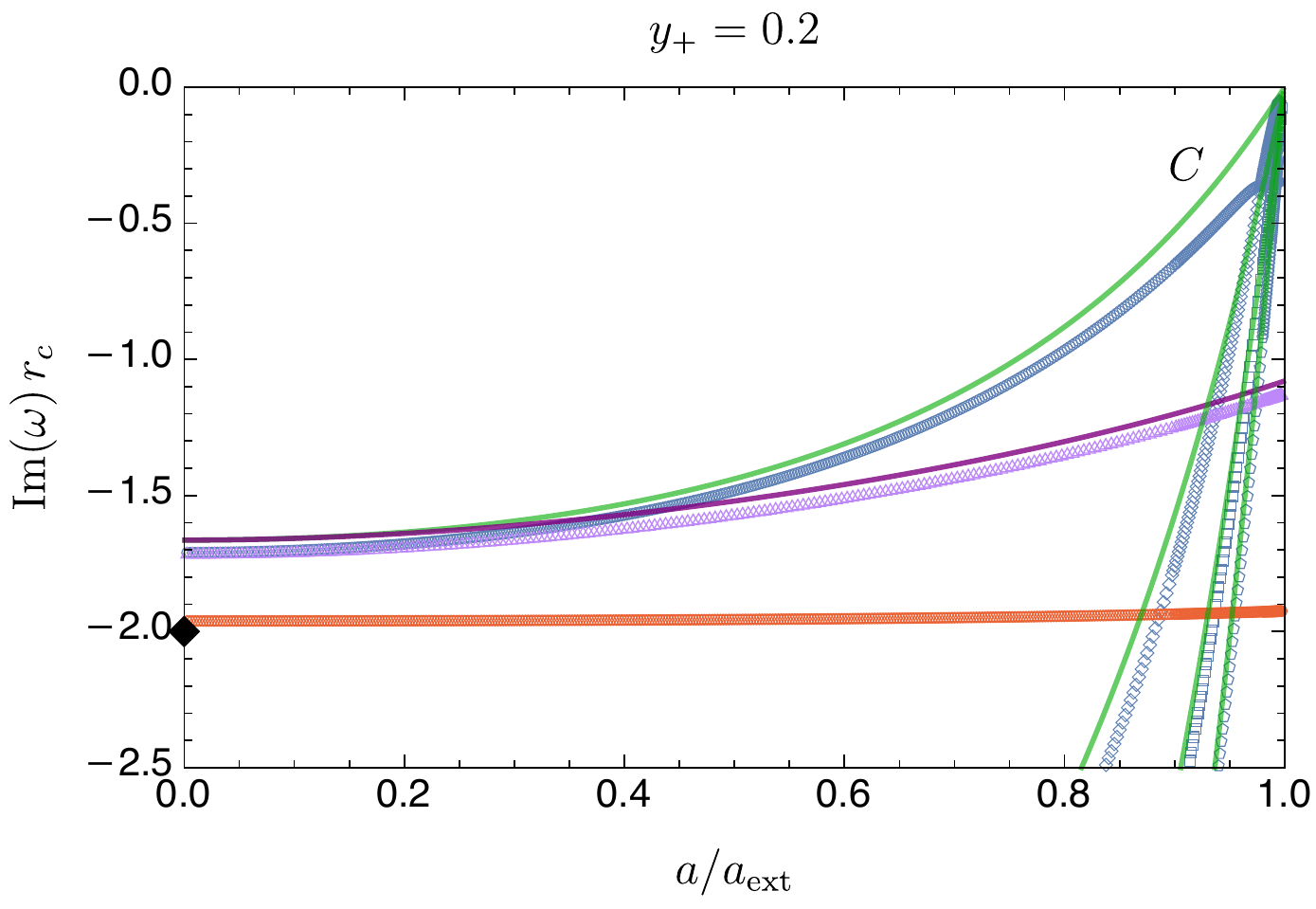}
  \vspace{0.2cm}
  \includegraphics[width=\textwidth]{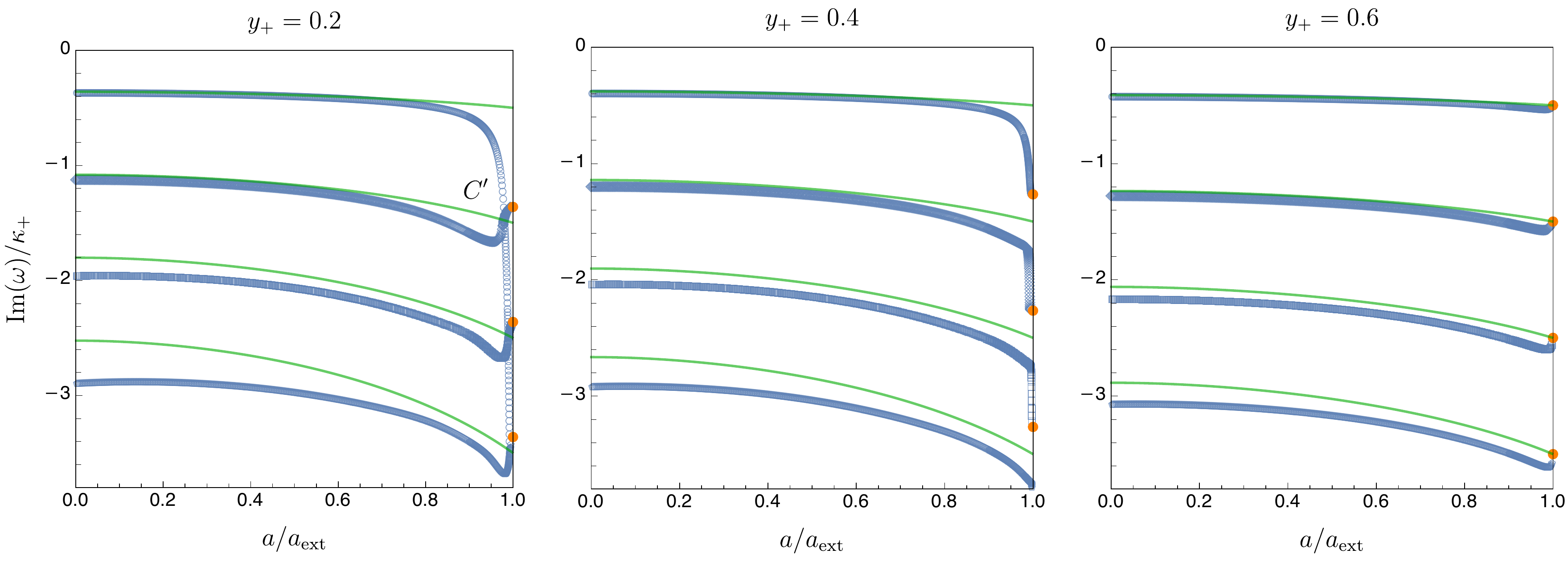}
  \caption{QNM spectrum for $d = 5$ MP-dS with $m = l = 2$. Typically, the corotating PS and NH modes (blue circles, diamonds, squares and pentagons) have merged to become a single PS-NH family: in all plots we display the first 4 overtones of this unified family. The only exception is the blue circle curve in the top of the left panel which describes a $n=0$ corotating PS mode, not a PS-NH mode since it is not captured by a NH analysis.
  The top panel is for $y_{+} = 0.2$ and the frequency is normalized in units of $r_c$.
In this top panel we also display the counter-rotating PS QNMs (magenta triangle curve), and the dS modes (red circles). The corotating and counter-rotating eikonal approximations~\eqref{eqn:omega_WKB} are described by the the solid green and purple lines, respectively. For reference, the pure dS frequency~\eqref{eqn:ds_modes_odd} with $y_+=a = 0$ is represented by the black diamond. The bottom panels are normalised in units of $\kappa_{+}$ to differentiate the near-extremal behaviour of the different overtones $n=0,1,2,3$ of the PS-NH family with $y_{+} = 0.2$ (left), $y_{+} = 0.4$ (middle) and $y_{+} = 0.6$ (right). The NH approximation~\eqref{eqn:NH_modes_MP-dS_d5} at $a = a_{\rm ext}$ (for $n=0,1,2,3$ from top to bottom) are  represented by the orange disks in the bottom panels. (In the bottom panel the dS modes and counter-rotating PS modes are not shown).
  \label{fig:PS_NH_connection}}
\end{figure}

To start discussing the modes with $m>0$ it is important to first recall that when $m = l = 0$, the NH frequencies are purely imaginary while the PS frequencies are complex, so they clearly form two distinct families of QNM, as was seen in Fig.~\ref{fig:eigRepulsion}. However, for $m \ne 0$, the PS modes with $m=l$ split into corotating and counter-rotating modes, as first discussed in the eikonal limit in section \ref{sec:PS_modes}. As in the eikonal limit, the counter-rotating mode always has a frequency with smaller imaginary part than the corotating PS mode for a given overtone $n$.\footnote{Note that the $t-\psi$ symmetry of MP-dS means that we need only consider modes with $m\geq 0$, as long as we study both signs of Re$(\omega)$. When $a=0$ this enhances to a $t \to -t $ symmetry and the QNM frequencies form pairs of $\{\omega,-\omega^* \}$.} 
Remarkably, we find that when $m\neq 0$, {\it typically} ({\it with an exception} to this rule discussed below) the corotating PS modes turn out to {\it merge} with the NH modes and they form a single unified family (plus its overtones) that we can denote as the `PS-NH' family. This is a property that was already found in the asymptotically flat Kerr-Newman black hole (but not in their Reissner-Nordstr\"om and Kerr limits where the PS and NH QNM families remain distinct in the whole parameter  space)~\cite{diasEigenvalueRepulsionsQuasinormal2021}. To illustrate this unification of the corotating PS and NH modes, we plot the frequency spectra for $d=5$ MP-dS with $m = l = 2$ and $y_{+} = 0.2$, in the top panel of Fig.~\ref{fig:PS_NH_connection}. The four blue curves with circles, diamonds, squares and pentagons are the corotating PS modes with radial overtone $n=0,1,2,3$ (from left/top to right). As expected, this classification is consistent with the eikonal analytical result  $\omega_{\hbox{\tiny WKB}}$ obtained in \eqref{eqn:omega_WKB} which is represented by the solid green curves for $n=0,1,2,3$.  The magenta triangle curve is the $n=0$ counter-rotating PS curve and it is also  well approximated by the counter-rotating eikonal frequency  $\omega_{\hbox{\tiny WKB}}$ of \eqref{eqn:omega_WKB} (solid purple curve). Finally, the red square curve in the top panel of Fig.~\ref{fig:PS_NH_connection} is the $n=0$ dS QNM family, clearly identified by the fact that it approaches the pure dS frequency~\eqref{eqn:ds_modes_odd} when $a\to 0$ (black diamond). 

An important property of the top panel of Fig.~\ref{fig:PS_NH_connection} is the fact that there are no extra curves which we could associate to a third independent family\footnote{We emphasise that we did an exhaustive direct numerical search for eigenvalues using {\it Mathematica's built-in routine {\it Eigensystem}} (as described in the end of Section \ref{sec:numerical_setup}) but we found no other frequencies besides the ones that are displayed in Fig.~\ref{fig:PS_NH_connection} (in the range shown and excluding even higher overtones $n\geq 4$). That is, we found no third family of QNMs besides the two main families (dS and unified PS-NH) displayed in the top panel of  Fig.~\ref{fig:PS_NH_connection}.}. We only have the dS family and the unified `PS-NH' family (and their overtones); the PS and NH modes do not exist separately. This is because, unlike the $m = l = 0$ case in Fig.~\ref{fig:eigRepulsion}, the NH modes at extremality (\emph{i.e.} at $a = a_{\rm ext}$) or nearby $-$ as unambiguously identified by the analytical approximation    $\omega_{\hbox{\tiny NH}}$ in \eqref{eqn:NH_modes_MP-dS_abstract} $-$ can always be traced back to a corotating PS mode when we move away from extremality. Indeed, the blue PS-NH curves in Fig.~\ref{fig:PS_NH_connection} (top and bottom panels) are {\it simultaneously} well approximated by $\omega_{\hbox{\tiny WKB}}$ in \eqref{eqn:omega_WKB} and, near-extremality, by $\omega_{\hbox{\tiny NH}}$ of \eqref{eqn:NH_modes_MP-dS_abstract}. This is better seen in the bottom-left panel of Fig.~\ref{fig:PS_NH_connection} for $y_+=0.2$. Here we choose a different normalization for the frequency: we plot the dimensionless frequency in units of the surface gravity $\kappa_+$. We make this choice because at extremality both $\omega_{\hbox{\tiny NH}}$ and $\kappa_+$ go to zero but their ratio is finite (and changes with $n$). Therefore, these properties help identifying the NH approximation $\omega_{\hbox{\tiny NH}}$ of \eqref{eqn:NH_modes_MP-dS_abstract} (orange disks at $a/a_{\rm ext}=1$ for $n=0,1,2,3$)\footnote{Note that, since we are plotting $\omega/\kappa_{+}$, the first-order accurate approximation~\eqref{eqn:NH_modes_MP-dS_d5} only gives us the value of $\omega_{\hbox{\tiny NH}}/\kappa_{+}$ at extremality, and not away from it.}. We indeed see that, typically, the unified PS-NH blue  curves terminate at extremality at the NH orange disks and, away from extremality, are also well approximated by the solid green corotating PS line described by 
$\omega_{\hbox{\tiny WKB}}$ in \eqref{eqn:omega_WKB}. Still in the bottom panel of Fig.~\ref{fig:PS_NH_connection} we see that these conclusions also hold for $y_+=0.4$ (middle panel)  and $y_+=0.6$ (right panel) and actually for all other values of $y_+$ (not shown).  

A second important property that is observed in the plots of Fig.~\ref{fig:PS_NH_connection} is that there is an ``exception to the rule'' described in the previous two paragraphs.
Namely, for small $y_+ \lesssim 0.3$, e.g. $y_+=0.2$ in the top and bottom-left panels, we see that the $n=0$ corotating PS mode is the only solution that is {\it not} also captured by the NH description. Indeed, as seen on the top panel, the $n=0$ corotating PS curve (unlike for $n\geq 1$) does {\it not} have ${\rm Im}(\omega r_c)\to 0$ (neither does it have $\operatorname{Re}(\omega) \to m \Omega|_{\rm ext}$, although this is not shown) as extremality is approached. Instead ${\rm Im}(\omega r_c)$ goes to a finite value as $a \to a_{\rm ext}$. This is better seen in the bottom-left panel,  since the $n=0$ corotating PS curve plunges into ${\rm Im}(\omega)/\kappa_+ \to -\infty$ as $a \to a_{\rm ext}$ because $\kappa_+ \to 0$ in this limit but ${\rm Im}(\omega)$ is finite. In particular, this means that this particular mode, and only this  one (and only for small $y_+$), is not described by $\omega_{\hbox{\tiny NH}}$ with $n=0$ in \eqref{eqn:NH_modes_MP-dS_abstract}.\footnote{Again, a similar behaviour can be found in asymptotically flat Kerr-Newman black holes~\cite{diasEigenvalueRepulsionsQuasinormal2021}. Indeed, very far away from extremality the PS and NH families are distinct families but they become unified PS-NH families as we approach extremality: see Fig.~1 of~\cite{diasEigenvalueRepulsionsQuasinormal2021}.} This also means that near extremality (see regions $C$ in top panel or $C'$ in bottom-left panel) the $n=0$ corotating PS mode trades dominance with the $n=1$ corotating PS mode. Indeed, from $\alpha=0$ all the way up to a critical $\alpha$ near-extremality, the $n=0$ corotating PS QNM is the one with the smallest $|{\rm Im}(\omega r_c)|$, but above this critical $\alpha$ and all the way to extremality, it is instead the $n=1$ corotating PS QNM that has the smallest $|{\rm Im}(\omega r_c)|$.
An interesting property that follows from the previous one is that the $n=0$ NH   
approximation $\omega_{\hbox{\tiny NH}}$ of \eqref{eqn:NH_modes_MP-dS_abstract} actually describes the extremal limit of the $n=1$ (not $n=0$) PS curve: see orange disk nearby point $C'$ in bottom-left panel. Similarly, the $n=1,2$ NH  approximation $\omega_{\hbox{\tiny NH}}$ describes the extremal limit of the $n=2,3$ (not $n=1,2$) unified PS-NH curves, respectively. 
Interestingly this ``exception to the rule'' ceases to hold for larger values of $y_+$ namely for $y_+\gtrsim 0.3$: see e.g. the cases $y_+=0.4$ (middle panel) and $y_+=0.6$ (right panel) of Fig.~\ref{fig:PS_NH_connection}. That is, in these cases, we have unified PS-NH curves, with $\omega_{\hbox{\tiny NH}}$ (with overtone $n$) describing the extremal limit of the blue curves (with the same overtone $n$), and the $n=0$ PS-NH QNM is the one
that dominates the spectra for all values of $a/a_{\rm ext}$. 

Note that, as  Fig.~\ref{fig:PS_NH_connection} illustrates, the eikonal approximation $\omega_{\hbox{\tiny WKB}}$ of \eqref{eqn:omega_WKB} (solid green curves), although strictly valid only for $|m|=l \to \infty$, is nevertheless already a good approximation for $m=l=2$ as long as we are away from extremality.  However the PS-NH family is no longer well approximated by the eikonal approximation in the near-extremal limit, although the approximation gets better even in this region as $y_{+}$ increases. Close to extremality, the PS-NH frequencies are better approximated by $\omega_{\hbox{\tiny NH}}$ in \eqref{eqn:NH_modes_MP-dS_abstract} (orange disks in Fig.~\ref{fig:PS_NH_connection}). In fact, we can find the difference between the analytical NH prediction \eqref{eqn:NH_modes_MP-dS_abstract} and the eikonal prediction \eqref{eqn:omega_WKB} exactly at extremality. Interestingly, we find that the value of $y_{+}$ at which this difference vanishes turns out to be given by the value of $y_{+}$ that saturates the AdS$_2$ BF bound of the near-horizon geometry, $1 + 4{\mu_{\rm eff}}^2 L_{\rm AdS}^2 = 0$, where $\mu_{\rm eff}$ is given by~\eqref{eqn:mu_eff_explicit}. For $(N = 1, m = l = 2)$, the AdS$_2$ BF bound is saturated at $y_{+} \sim 0.54$. We postpone a detailed discussion of this observation to section~\ref{sec:scc}.

\begin{figure}
  \centering
  \includegraphics[width=\textwidth]{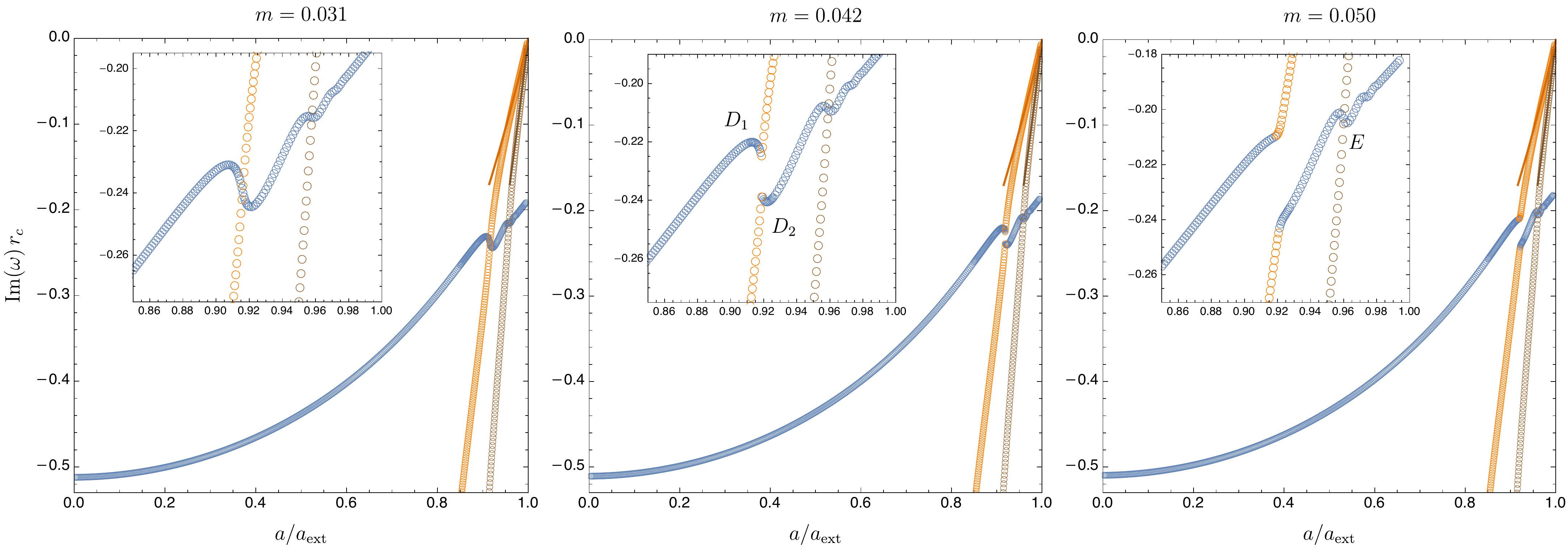}
  \caption{PS modes (blue circles) merging with the $n = 0$ NH mode (orange circles) as $m = l$ increases, in $d = 5$ MP-dS with $y_{+} = 0.6$. The $n = 1$ NH mode (brown circles) merges later, by $m \simeq 0.125$. The inset figures are enlargements of the merging region. The NH  approximations~\eqref{eqn:NH_modes_MP-dS_d5} for the $n = 0$ and $n = 1$ overtones are indicated by solid orange and brown lines, respectively. The dS modes and counter-rotating PS modes are not shown in these plots.\label{fig:PS_NH_merge}}
\end{figure}

The fact that the PS and NH modes typically merge in a single PS-NH family for $m>0$ in MP-dS might be considered puzzling. How can it be that for $m=0$ the system has three distinct QNM families (dS, PS and NH) and for $m>0$ there are typically only two (dS and PS-NH)?
We can address and settle this question with the following strategy. Regularity of the \(\mathbb{CP}^{N}\) harmonics requires that the angular eigenvalues $\lambda$ are given by~\eqref{eqn:CPN_equation} with {\it integer} $m$. But nothing prevents us from doing an exercise where we search numerically for the eigenfrequencies of the problem when the value of $m$ is not an integer. This would simply describe the eigenfrequencies of singular modes which are physically irrelevant. But we can learn important lessons from this academic exercise: we start with the PS and NH modes of $m=0$ (displayed e.g. in Fig.~\ref{fig:eigRepulsion}) and see how these curves evolve as we increase $m$ incrementally. We do this in Fig.~\ref{fig:PS_NH_merge} for $d = 5$ MP-dS with $y_{+} = 0.6$. In the left panel we display the spectra for $m=0.031$. In this case, the $n=0$ corotating PS family  (blue circles), the $n=0$ NH family (orange circles) and the $n=1$ NH family (brown circles) are still very much distinct families as in the $m=0$ case of Fig.~\ref{fig:eigRepulsion} (we do not show the dS modes in Fig.~\ref{fig:PS_NH_merge}). However, as we increase $m$, the $n=0$ PS curve breaks into two branches, and the same happens to the $n=0$ NH curve. This is clearly seen e.g. in the middle panel of Fig.~\ref{fig:PS_NH_merge} for $m=0.042$. And the left-branch of the `old' (w.r.t. the left panel) $n=0$ PS curve merges smoothly with the upper-branch of the `old'  $n=0$ NH curve at  point $D_1$, while the lower-branch of the `old' $n=0$ NH curve merges smoothly with the right-branch of the `old' $n=0$ PS curve at the point $D_2$. Effectively, the PS and NH families loose their individual identities and the two `old' PS and NH curves become two PS-NH curves with a eigenfrequency gap $D_1 D_2$ appearing between the two new PS-NH curves. This is nothing but another manifestation of the eigenvalue repulsion phenomenon observed in Fig.~\ref{fig:eigRepulsion}.
As $m$ keeps increasing, the gap $D_1D_2$ keeps increasing  and similar breakups, merges and gaps will keep happening between the $n=0$ PS curve and `old' NH curves with overtone $n\geq 1$ as suggested in the right panel of  Fig.~\ref{fig:PS_NH_merge}  for $m=0.05$. For example, although not shown, the breakout/merger between the $n=0$ PS curve and the $n=1$ NH curve occurs for $m\simeq 0.125$, as the region $E$ in the right panel already suggests will happen. After this exercise, we finally understand why for $m=0$ we have three families of QNMs but only two families for $m>0$.
Once we reach $m = 1$ all of the sub-dominant PS and NH modes  (at least those with $n=0,1,2,3$) will have merged in the same fashion and we get the homogeneous picture previously presented for $m = 2$ in the bottom-right panel of Fig.~\ref{fig:PS_NH_connection}. There is a striking similarity to the eigenvalue repulsions observed in Kerr-Newman (compare for example with Fig. 1 of~\cite{diasEigenvalueRepulsionsQuasinormal2021}). It seems likely that the underlying mechanism is the same.

Although we have focused our discussion on $d=5$ black holes in this section, similar properties occur for higher spacetime dimensions $d$. However, there are also differences, some of which can be traced back to the fact that the QNM spectrum of Schwarzschild-dS black holes (\emph{i.e.} the limit $a=0$) changes when $d$ increases, especially when $d$ changes from odd to even. The dominant QNMs of $d> 4$  Schwarzschild-dS are very similar to the $d=4$ case, with two distinct mode families (dS and PS) which do not interact (\emph{i.e.} there is no eigenvalue repulsion), in all dimensions. However, if we also consider the subdominant modes we find an intricate network of mode interactions, particularly in $d = 5$, and even modes which break the standard mode classification in $d \ge 9$. For completeness, we discuss this in more detail in Appendix~\ref{sec:SdS_and_small_a}. However, these effects are only present for subdominant modes, and with $m = l = 0$. Hence these effects are not relevant to SCC, which is ultimately the main focus of this manuscript, and that we discuss in the next section.

%===========================================================
%===========================================================
\section{Strong Cosmic Censorship  in MP-dS. Discussion of the results}\label{sec:scc}
%===========================================================
%===========================================================

We are finally ready to discuss Strong Cosmic Censorship in cohomogeneity-1 Myers-Perry$-$de Sitter black holes. The Christodoulou formulation of Strong Cosmic Censorship states that the maximal Cauchy development cannot be extended beyond the Cauchy horizon as a weak solution of the Einstein equations or matter fields~\cite{christodoulouFormationBlackHoles2008}. For the scalar field, this translates to the requirement that the scalar field is not in the Sobolev space $H_{\rm loc}^{1}$ near the Cauchy horizon. In four dimensional de Sitter black holes, it has been shown that the decay rate of generic linear perturbations is governed by the \emph{spectral gap} $\beta$, \emph{i.e.} the imaginary part of the {\it slowest}-decaying quasinormal mode, relative to the surface gravity $\kappa_{-}$ at the Cauchy horizon~\cite{hintzAnalysisLinearWaves2017}. Specifically, defining the spectral gap as
\begin{equation}
  \beta \equiv - \frac{\operatorname{Im}(\omega)}{\kappa_{-}},
  \label{eqn:beta_definition}
\end{equation}
it was shown in~\cite{hintzAnalysisLinearWaves2017} that the scalar field is in $H_{\rm loc}^{1}$ if $\beta > \frac{1}{2}$. We will now argue that the requirement is the same for MP-dS. Consider a quasinormal mode defined in region I (see the left panel of Fig.~\ref{fig:parameter_space_penrose}),
\begin{equation}
  \Phi = e^{-i \omega t}e^{i m \psi} Y(x_{i}) \tilde{R}(r).
\end{equation}
Changing to ingoing EF coordinates \eqref{eqn:ingoingEF}, the metric is regular at the event horizon $\mathcal{H}^+$ and \(\Phi\) can be analytically continued into region II of Fig.~\ref{fig:parameter_space_penrose}. Then, using \eqref{eqn:outgoingEF} to change to outgoing EF coordinates $(v, r, \psi'', x_{i})$, which are regular at the Cauchy horizon $\mathcal{CH}^+$, we get
\begin{equation}
  \Phi = e^{-i \omega v} e^{i m \psi''} Y(x_{i})R(r),
\end{equation}
where \(R(r)\) includes the original contribution \(\tilde{R}(r)\) from region I but has additional factors from the coordinate transformations. In outgoing EF coordinates, the massless radial equation \eqref{eqn:waveEqnBL} reads
\begin{multline}
  R''(r) + \left( \frac{2 N+1}{r} + \frac{2 i \sqrt{h}}{f} (\omega -m \Omega) + \frac{f'}{f} \right)R'(r) \\
  - \frac{1}{f}\left[ \frac{m^2}{r^{2} h} + \frac{\lambda}{r^2} - \frac{i}{r^{2 N+1}} \partial_{r} \left(r^{2 N+1} \sqrt{h}  \, (\omega -m \Omega)\right) \right] R(r) = 0.
  \label{eqn:wave_eqn_outgoing}
\end{multline}
% \begin{multline}
%   R''(r) + \left( \frac{2 N+1}{r}+\frac{2 i g^{2} h}{r} (\omega -m \Omega) - \frac{{g^{2}}'}{g^{2}} \right)R'(r) \\
%   - g^{2}  \left( \frac{m^2}{h^2} + \frac{\lambda}{r^2} - \frac{i}{r^{2 N+1}} \partial_{r} \left(h \, r^{2 N} (\omega -m \Omega)\right) \right) R(r) = 0
% \end{multline}
This equation has regular singular points at the roots of \(f(r)\), \emph{i.e.} at the horizon radii. In particular, we know from section~\ref{sec:myers-perry} that \(f(r)\) has a single zero at \(r = r_{-}\), so we can factor \(f(r)\) as \(f(r)=(r-r_{-}) \Delta(r) \). The remaining terms in~\eqref{eqn:wave_eqn_outgoing}, including \(\Delta \), are all analytic and non-zero at $r_{-}$. Hence, we can perform a Frobenius expansion around $r=r_{-}$. Fuch's theorem~\cite{NIST:DLMF} asserts that there exists a solution with a non-zero radius of convergence
\begin{equation}
  R(r) = A \, \hat{R}_{(1)}(r) + B \, (r-r_{-})^s \hat{R}_{(2)}(r),
\end{equation}
for some constants \(A\) and \(B\), where \(\hat{R}_{(1,2)}(r)\) are analytic and non-zero at \(r = r_{-}\), and \(s \equiv i(\omega - m \Omega_{-})/\kappa_{-}\) is the non-trivial solution of the indicial equation. \(\Phi\) is in the Sobolev space \(H_{\textrm{loc}}^{1}\) if \(R(r)\) and its first derivative are locally square integrable. Since \(\hat{R}_{(i)}\) are analytic, the only relevant factor is \({(r-r_{-})}^{s}\), which is locally square integrable if and only if \(\operatorname{Re}(2s)>-1\). In terms of \(\beta\), we can write this condition as $\beta > \frac{1}{2}$. In other words, to prove that Christodoulou's formulation of Strong Cosmic Censorship is respected, we `just' need to show that, for the whole parameter space, at least one family of QNMs is not in \(H_{\textrm{loc}}^{1}\), \emph{i.e.} there exists a QNM family with \(\beta \le \frac{1}{2}\).

\begin{figure}
  \centering
  \begin{subfigure}{.6\linewidth}
    \includegraphics[width=\textwidth]{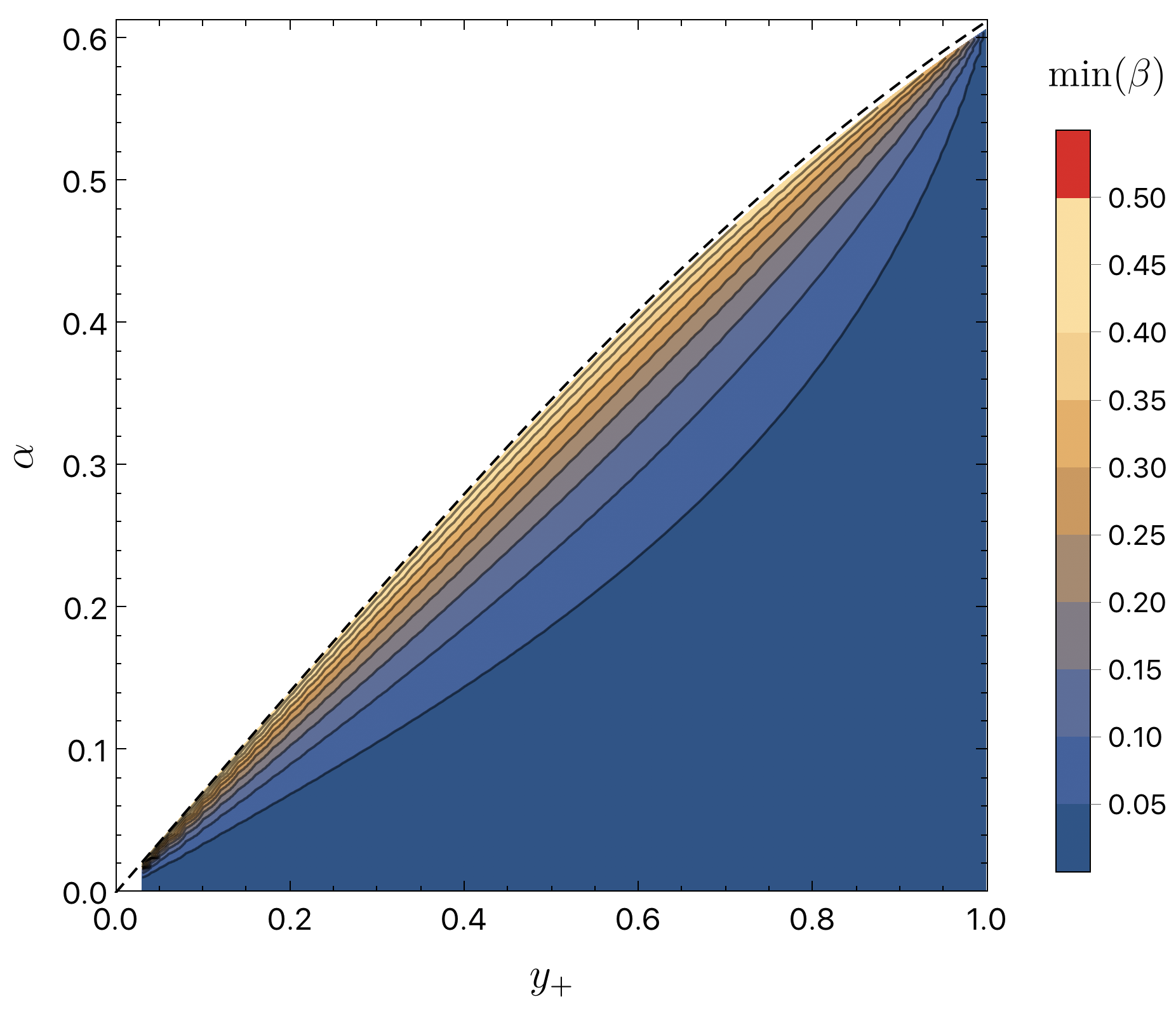}
  \end{subfigure}
    \caption{Plot of \(\operatorname{min}(\beta)\) for the full spectrum of quasinormal modes of $d=5$ MP-dS with \(m = l = 10\) as computed by numerically solving the eigenvalue problem. There are two families of modes, the PS-NH modes and the dS modes. We find that the PS-NH modes are dominant for the entire parameter space, since the dS modes are suppressed for large $m = l$. The dashed black line represents extremality $r_{-} \to r_{+}$. One finds that ${\rm min}(\beta)<1/2$ for the whole parameter space.}
\label{fig:beta_plot}
\end{figure}

In the eikonal limit $|m| = l \to \infty$, the quasinormal mode spectrum is dominated by the PS family of modes (actually, the PS-NH modes, following the findings of the previous section), which are well approximated by $\omega_{\rm WKB}$, as defined in~\eqref{eqn:omega_WKB}. We first check that the corresponding $\beta_{\rm WKB}$ computed using~\eqref{eqn:beta_definition} satisfies $\beta_{\rm WKB} < \frac{1}{2}$. This result {\it per se} should establish that  Christodoulou's SCC is preserved in equal angular momenta MP-dS since we would have found a QNM family with $\beta\leq 1/2$ in the whole range of the parameter space. However, to make such a strong claim we must ensure that the eikonal approximation is really valid, \emph{i.e.} we have to compare it with the exact numerical frequencies, a task that can be completed only at finite $m$. From the previous section we already know that the eikonal approximation is reasonably good when extrapolated to finite $m$ but here it is fundamental that we find a family that has exactly $\beta\leq 1/2$ everywhere.
So, in practice, we need to numerically compute the dominant QNMs of MP-dS at finite $m$ for the whole parameter space and check that there is indeed at least one $m$ for which $\beta \leq 1/2$ everywhere. In this process, we will have the opportunity to further quantify how good the eikonal approximation~\eqref{eqn:omega_WKB}  is when extrapolated to finite $m$. Moreover, we want to complete this exercise for several dimensions $d$ to find whether there is a critical dimension where the validity of SCC could change.

Recall that we have corotating and counter-rotating PS modes, including in the eikonal limit, but the norm of the imaginary part of the frequency of the corotating modes is always smaller than the counter-rotating ones (for a given overtone). So we just need to consider the corotating PS modes (\emph{i.e.}, typically, the corotating PS-NH modes).
Inserting $\omega_{\hbox{\tiny WKB}}$, as read from~\eqref{eqn:omega_WKB} and \eqref{eqn:Omega_Lyapunov}, and the surface gravity $\kappa_-$, as computed from~\eqref{eqn:surfGrav}, into \eqref{eqn:beta_definition} we can compute $\beta_{\rm WKB}$ for any odd spacetime dimension $d$. We find that, just as in Kerr-dS~\cite{diasStrongCosmicCensorship2018}, $\beta_{\rm WKB}$ is bounded less than \(\frac{1}{2}\) away from extremality, only approaching \(\beta_{\rm WKB} = \frac{1}{2}\) at extremality. It turns out that, across the range of dimensions we tested ($d\leq 15$), $\beta_{\rm WKB}$ is a non-increasing function of the dimension, \emph{i.e.} for every point $(y_{+},y_{-})$ in the parameter space, $\beta_{\rm WKB}(y_{+},y_{-};d) \le \beta_{\rm WKB}(y_{+},y_{-};d+2)$. Hence we expect that $\beta_{\rm WKB} \leq \frac{1}{2}$ also holds true for $d>15$. However, this conclusion does not necessarily extend to the exact PS or PS-NH modes at finite $m$ near-extremality, because the eikonal result fails to be a good approximation in the near-extremal regime for small $m$ (more below).

After this simple but enlightening and encouraging eikonal exercise, we should now confirm that the exact numerical solutions of the eigenvalue problem indeed yield $\beta\leq 1/2$, {\it at least} for a sufficiently high $m=l$ family of QNMs. We start by doing this for  $l = m = 10$ and for the whole parameter space $(y_{+}, \alpha)$ of $d = 5$ MP-dS (with the parameter space discretised into about 2700 points). As discussed in section~\ref{sec:QNMspectra}, when $m > 0$ the individual PS and NH families that exist for $m=0$ typically lose their identity (except for small values of $y_+$ if $m=l$ is small) and become a single PS-NH family for each radial overtone $n$. Here, since we are working in the eikonal limit $m=l\to\infty$, we are only interested in the PS-NH family with the lowest overtone (since it has smaller $\beta$), and this family of modes dominates for the entire parameter space over the second QNM family of the system (the dS family). The smallest value of $\beta$ at each point of the phase space is plotted in Fig.~\ref{fig:beta_plot}. The closest we approach extremality in this plot  is $\alpha/\alpha_{\rm ext} = 0.99$ or $r_{-}/r_{+} = 0.98$. All the points tested have $\beta < \frac{1}{2}$, with a maximum of $\beta \simeq 0.488$. Hence, we conclude that for $l = m = 10$ one has $\beta<1/2$, as predicted by the eikonal approximation. To quantify the accuracy of the WKB approximation when extrapolated to such a finite $m$, we compute $\Delta_{\rm WKB} \equiv \frac{\beta_{\rm WKB} - \beta}{\frac{1}{2} - \beta}$, where $\beta_{\rm WKB}$ is the eikonal approximation~\eqref{eqn:omega_WKB} and $\beta$ is the numerical value. We find that $-0.07 < \Delta_{\rm WKB} < 10^{-11}$ for $d = 5$, \emph{i.e.} up to numerical accuracy the true value of $\beta$ is never larger than that of the eikonal approximation $\beta_{\rm WKB}$.

\begin{figure}
  \centering
  \begin{subfigure}{.43\textwidth}
      \includegraphics[width=\textwidth]{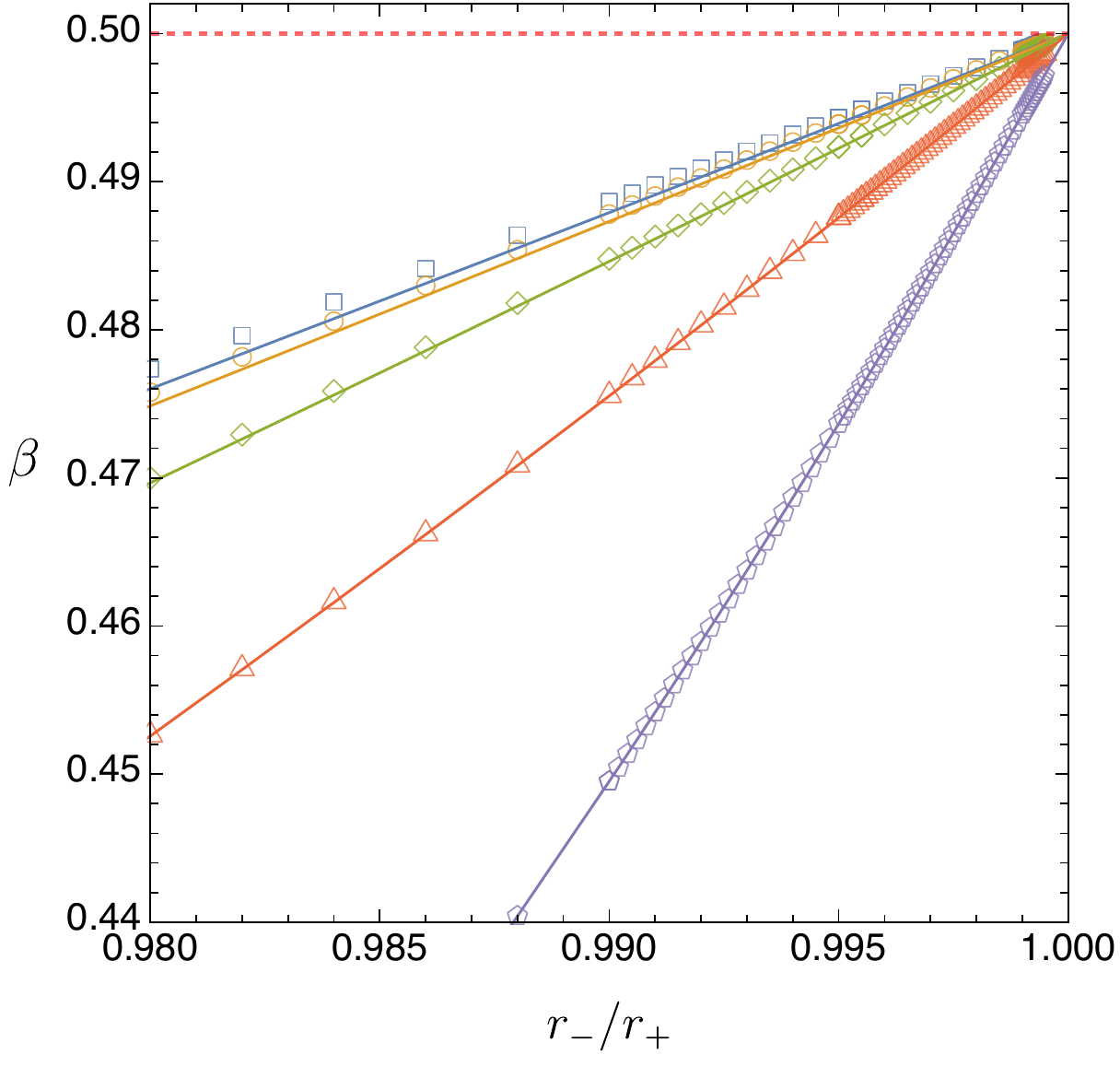}
  \end{subfigure}
  \begin{subfigure}{.56\textwidth}
      \includegraphics[width=\textwidth]{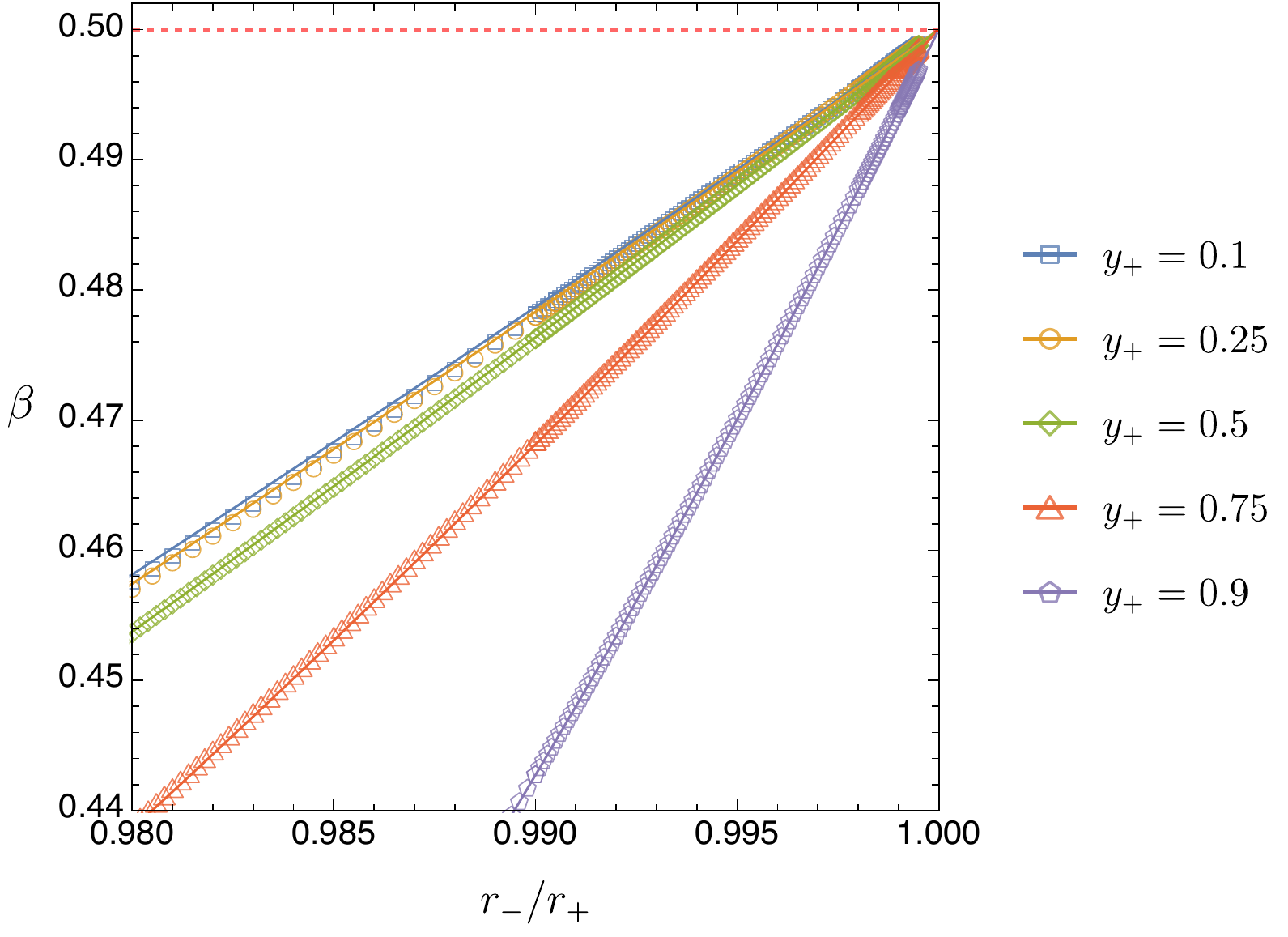}
  \end{subfigure}
  \caption{$\beta$ for the dominant QNMs near extremality in $d = 5$ MP-dS with $m = l = 10$ (left panel) and $d = 11$ with $m = l = 30$ (right panel), compared to the eikonal approximation~\eqref{eqn:omega_WKB} of the PS modes (solid lines). The horizontal red dashed line indicates $\beta = \frac{1}{2}$. In both plots, the closest we approach extremality is $r_-/r_+ = 0.9995$.  \label{fig:beta_near_extremality_all_N}}
\end{figure}

However, to claim that $\beta\leq 1/2$ for the whole parameter space we should still stretch our numerical analysis even closer to extremality, \emph{i.e.} even closer than what we do in Fig.~\ref{fig:beta_plot} where we have reached `only' $\alpha/\alpha_{\rm ext} = 0.99$ or $r_{-}/r_{+} = 0.98$. For that we can focus our attention on lines of constant  $y_+=r_{+}/r_{c}$ and push the numerical collection of data as close as possible to extremality where $r_-=r_+$. We do this for $d=5$ and $m = l = 10$ in the left panel of Fig.~\ref{fig:beta_near_extremality_all_N} for several lines of constant $0<y_+<1$ identified in the legend\footnote{A convergence test is given in Appendix~\ref{sec:convergence_tests} for the $d = 5$ case at $r_{-} = 0.9995\, r_{+}$.}. We also display, as solid lines, the eikonal approximation $\beta_{\rm WKB} $ as read from ~\eqref{eqn:omega_WKB}. We conclude that all solutions approach $\beta=1/2$ from below as extremality is approached. Moreover, we also find that the eikonal approximation very well describes this approach even for a relatively small value such as $m = 10$ (note that the approximation is better for large $y_+$). 

We find similar results when we repeat the analysis but this time for $d=7,9,11$, although we need to pick families with higher $m=l$ as $d$ increases to still have $\beta\leq 1/2$ everywhere (for reasons explained below). As an example, in the right panel of Fig.~\ref{fig:beta_near_extremality_all_N} we present the results for $m=l=30$ in $d=11$.

Altogether, we conclude that there is at least one family of $m=l$ QNMs for which the spectral gap satisfies the condition $\beta\leq 1/2$  in the whole parameter space of MP-dS for odd $d\leq 11$ (and most probably also above). It follows that  Christodoulou's formulation of Strong Cosmic Censorship holds for equal angular momenta $d>4$ MP-dS black holes, very much alike in the $d=4$ Kerr-dS case~\cite{diasStrongCosmicCensorship2018}. This is the main result of our study.  

\begin{figure}
    \centering
    \includegraphics[width=0.6\textwidth]{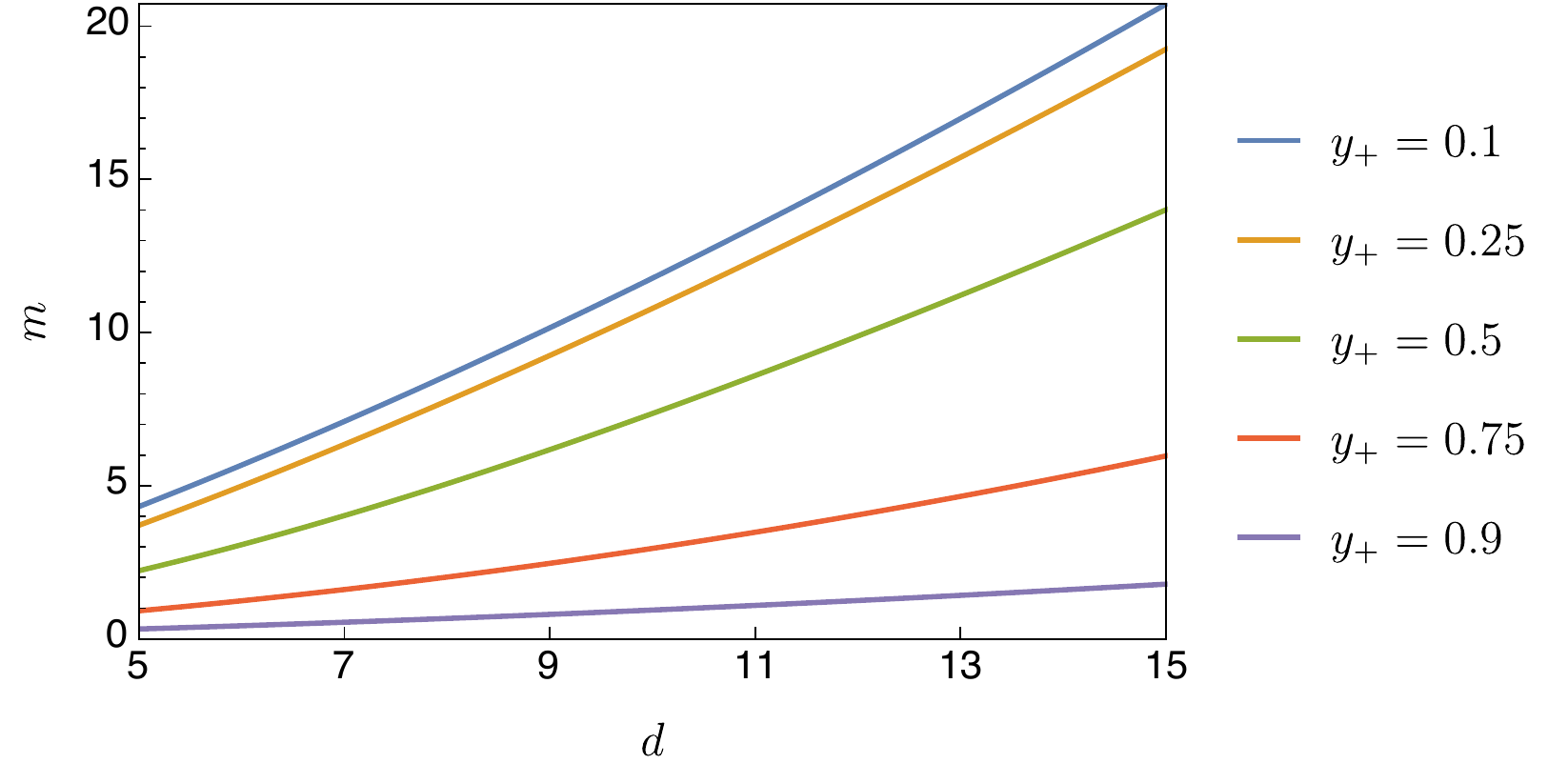}
    \caption{Critical value $m_{\rm crit}$ given by~\eqref{eqn:m_crit} above which we have a BF bound violation. $m_{\rm crit}$ is plotted as a function of dimension $d$ for a range of fixed $y_{+}$.\label{fig:critical_m}}
\end{figure}

As the above discussions indicate, it is very easy to find that $\beta<1/2$ away from extremality; however it is much more difficult to stretch the numerical code to prove that we have $\beta\leq 1/2$ all the way up to extremality. However, even without resorting to a numerical analysis, we can establish analytically that there are modes that have $\beta \leq 1/2$ in the whole parameter space if $m=l$ is sufficiently large, for any $d$. While doing so, we can also find a criterion that tells us how large $m=l$ needs to be (for a given $d$) to have a family of QNMs that approach $\beta = 1/2$ at extremality. We discuss how this can be done in the rest of this section. 
As emphasized previously, for $m\neq 0$ and sufficiently large $y_+$, the PS and NH modes do not exist as separate families; instead they combine to form what we call the PS-NH family. This means that the PS-NH QNMs are simultaneously well approximated by the eikonal approximation $\omega_{\hbox{\tiny WKB}}$ in \eqref{eqn:omega_WKB}  and by the NH approximation $\omega_{\hbox{\tiny NH}}$ of \eqref{eqn:NH_modes_MP-dS_abstract}. The eikonal approximation~\eqref{eqn:omega_WKB}  is a good approximation as long as we are far away from extremality but it deviates from the exact result as we approach extremality and this deviation gets higher for small $y_+$ and higher $d$. On the other hand $-$ and this is a key observation for our purposes $-$ the NH approximation~\eqref{eqn:NH_modes_MP-dS_abstract} becomes more and more accurate as we approach extremality and this is precisely the region where we want to have a solid proof that $\beta$ does not exceed $1/2$ for at least a family of modes. Thus, using the fact that $q_{\hbox{\tiny AdS}}$ defined in~\eqref{eqn:q_AdS_explicit} is real, it follows from~\eqref{eqn:NH_modes_MP-dS_abstract} and \eqref{eqn:beta_definition} that, for all $d=2N+3$, $\beta_{\hbox{\tiny NH}}$ is given by
\begin{equation}
  \beta_{\hbox{\tiny NH}} \simeq \frac{1}{2} + \frac{1}{2} \operatorname{Re}\left( \sqrt{1 + 4 {\mu_{\rm eff}}^{2} L_{\hbox{\tiny AdS}}^{2}} \right).
  \label{eqn:beta_NH}
\end{equation}
Based on this near-horizon approximation, the PS-NH modes will have $\beta > \frac{1}{2}$ at extremality unless the near-horizon AdS$_2$ BF bound is violated. Conversely, to have $\beta \le 1/2$ (at and away from extremality) and thus a family of modes that enforce SCC, one must violate the AdS$_2$ BF bound. For a MP-dS BH of fixed $y_+$ and dimension $d = 2N + 3$, the violation of the AdS$_2$ BF bound can occur if $m$ is above a  critical value $m_{\rm crit}$. More concretely, choosing $l = m$ and using $L_{\hbox{\tiny AdS}}$ and $\mu_{\rm eff}$  as given in~\eqref{eqn:mu_eff_explicit}, we find that in order to have an AdS$_2$ BF bound violation, we must have $m > m_{\rm crit}$ where
\begin{equation}
m_{\rm crit} = \frac{(N+1) \left(\sqrt{2} +\sqrt{N+2}\right)}{\sqrt{2} N} \frac{2 y_{+}^2 \left(y_{+}^{2 N}-1\right)-N \left(y_{+}^2-1\right)
   (y_{+}^{2 N+2}+1)}{1 -(N+2) y_{+}^{2 N+2}+(N+1) y_{+}^{2 N+4}}\,,\label{eqn:m_crit}
\end{equation}
which, for a fixed dimension, has a finite maximum given by
\begin{equation}
  \max(m_{\rm crit}) = \frac{N+1}{\sqrt{2}}\left(\sqrt{2} + \sqrt{N + 2}\right).\label{eqn:max_m}
\end{equation}
We plot $m_{\rm crit}$ as a function of the dimension $d$ for several values of $y_{+}$ in Fig.~\ref{fig:critical_m}. Note that $m_{\rm crit}$ increases for higher dimensions, and smaller $y_{+}$. For example, for $d=5$ one has $m_{\rm crit} \sim 5$ but for $d=11$ one has $m_{\rm crit} \sim 13$ (or even higher if $y_+\to 0$). Coming back to Fig.~\ref{fig:beta_near_extremality_all_N} this explains why for $d=5$ it was sufficient to look at $m=l=10$ modes to attain $\beta\leq 1/2$ everywhere, but for $d=11$ we had to use a higher $m=l=30$ to obtain modes with $\beta\leq 1/2$ everywhere.

To illustrate how $m_{\rm crit}$ and the associated AdS$_2$ BF bound violation is closely (but not sharply) related to modes with $\beta \leq \frac{1}{2}$ near extremality, we give $\beta$ for the dominant QNM at $r_{-}/r_{+} = 0.99$ in Table~\ref{tab:BF_comparison}, while varying both $m = l$ and the dimension $d$, for $y_{+} = 0.25$ (left table) and $y_{+} = 0.75$ (right table). We  conclude that  AdS$_2$ BF bound violation is a necessary (but not sufficient) condition for $\beta \leq \frac{1}{2}$ near  extremality. Indeed, modes that violate the AdS$_2$ BF bound are those below the zigzag line $m = m_{\rm crit}$ in Table.~\ref{tab:BF_comparison}, and we see that these modes with $m > m_{\rm crit}$ typically  have $\beta \leq \frac{1}{2}$. However, this is not always the case: some modes with $m$ just above $m_{\rm crit}$ can still have $\beta > \frac{1}{2}$. Once $\beta$ crosses below $\frac{1}{2}$, there is a sharp change in behaviour (in the sense that increasing $m$ produces very small changes in the value of $\beta$) and all of the modes with larger $m$ are approaching $\frac{1}{2}$ from below.

\begin{table}
\footnotesize
\hspace{-0.4cm}
\begin{subtable}{.5\linewidth}
\begin{tabular}{ccccccc}
\toprule
    & \multicolumn{6}{c}{dimension $d$} \\ \cmidrule{2-7}
$m$ & 5 & 7 & 9 & 11 & 13 & 15 \\ \midrule
0 & \textbf{0.99} & \textbf{0.99} & \textbf{0.99} & \textbf{0.99} & \textbf{0.99} & \textbf{0.99} \\
2 & \textbf{1.34} & \textbf{1.18} & \textbf{1.11} & \textbf{1.07} & \textbf{1.05} & \textbf{1.02} \\ \cline{2-2}
4 & \multicolumn{1}{c|}{\textbf{0.54}} & \textbf{1.13} & \textbf{1.13} & \textbf{1.10} & \textbf{1.07} & \textbf{1.06} \\
6 & \multicolumn{1}{c|}{0.49} & \textbf{0.63} & \textbf{1.05} & \textbf{1.09} & \textbf{1.08} & \textbf{1.07} \\ \cline{3-3}
8 & 0.49 & \multicolumn{1}{c|}{0.49} & \textbf{0.73} & \textbf{1.02} & \textbf{1.06} & \textbf{1.06} \\ \cline{4-4}
10 & 0.49 & 0.49 & \multicolumn{1}{c|}{\textbf{0.54}} & \textbf{0.81} & \textbf{1.01} & \textbf{1.05} \\
12 & 0.49 & 0.49 & \multicolumn{1}{c|}{0.49} & \textbf{0.61} & \textbf{0.87} & \textbf{1.00} \\ \cline{5-5}
14 & 0.49 & 0.48 & 0.48 & \multicolumn{1}{c|}{\textbf{0.52}} & \textbf{0.68} & \textbf{0.91} \\ \cline{6-6}
16 & 0.49 & 0.48 & 0.48 & 0.48 & \multicolumn{1}{c|}{\textbf{0.57}} & \textbf{0.76} \\
18 & 0.49 & 0.48 & 0.48 & 0.48 & \multicolumn{1}{c|}{\textbf{0.51}} & \textbf{0.63} \\ \cline{7-7}
20 & 0.49 & 0.48 & 0.48 & 0.48 & 0.49 & \textbf{0.56} \\ \bottomrule
\end{tabular}
\end{subtable}%
\hspace{0.8cm}
\begin{subtable}{.5\linewidth}
\begin{tabular}{ccccccc}
\toprule
    & \multicolumn{6}{c}{dimension $d$} \\ \cmidrule{2-7}
$m$ & 5 & 7 & 9 & 11 & 13 & 15 \\ \midrule
0 & \multicolumn{1}{l}{\textbf{0.96}} & \multicolumn{1}{l}{\textbf{0.96}} & \multicolumn{1}{l}{\textbf{0.96}} & \multicolumn{1}{l}{\textbf{0.97}} & \multicolumn{1}{l}{\textbf{0.97}} & \multicolumn{1}{l}{\textbf{0.97}} \\ \cline{2-3}
2 & 0.48 & \multicolumn{1}{c|}{0.50} & \textbf{0.72} & \textbf{1.00} & \textbf{1.05} & \textbf{1.05} \\ \cline{4-5}
4 & 0.48 & 0.47 & 0.47 & \multicolumn{1}{c|}{\textbf{0.52}} & \textbf{0.67} & \textbf{0.88} \\ \cline{6-6}
6 & 0.48 & 0.47 & 0.47 & 0.47 & \multicolumn{1}{c|}{0.49} & \textbf{0.57} \\ \cline{7-7}
8 & 0.48 & 0.47 & 0.47 & 0.47 & 0.47 & 0.48 \\
10 & 0.48 & 0.47 & 0.47 & 0.47 & 0.47 & 0.46 \\
12 & 0.48 & 0.47 & 0.47 & 0.47 & 0.47 & 0.46 \\
14 & 0.48 & 0.47 & 0.47 & 0.47 & 0.47 & 0.46 \\
16 & 0.48 & 0.47 & 0.47 & 0.47 & 0.47 & 0.46 \\
18 & 0.48 & 0.47 & 0.47 & 0.47 & 0.47 & 0.46 \\
20 & 0.48 & 0.47 & 0.47 & 0.47 & 0.47 & 0.46 \\ \bottomrule
\end{tabular}
\end{subtable}
\caption{$\beta$ for the dominant QNM of MP-dS at $r_-/r_+ = 0.99$ (\emph{i.e.} at 99\% of extremality) and $y_{+} = 0.25$ (left) and $y_{+} = 0.75$ (right), for a range of $m = l$ and dimensions $d$. The zigzag line describes the boundary $m = m_{\rm crit}$ as given by~\eqref{eqn:m_crit}. Modes above this line (\emph{i.e.} those with smaller $m$) respect the AdS$_2$ BF bound; below it the BF bound is violated. Bold values have $\beta > \frac{1}{2}$ and the others have $\beta\leq  \frac{1}{2}$.\label{tab:BF_comparison}}
\end{table}

\begin{figure}
  \begin{subfigure}{.57\textwidth}
    \includegraphics[width=\textwidth]{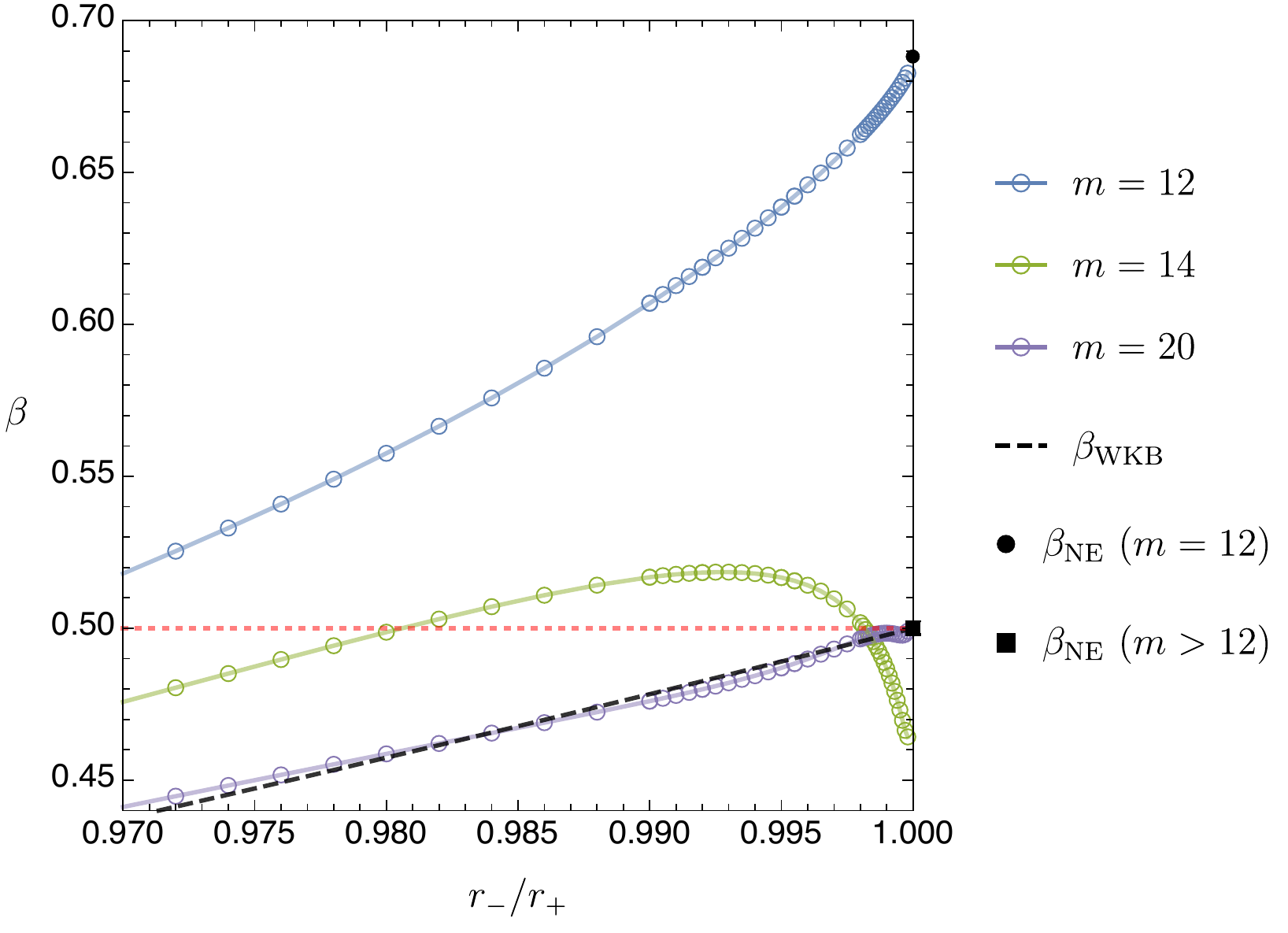}
  \end{subfigure}
  \begin{subfigure}{.42\textwidth}
    \includegraphics[width=\textwidth]{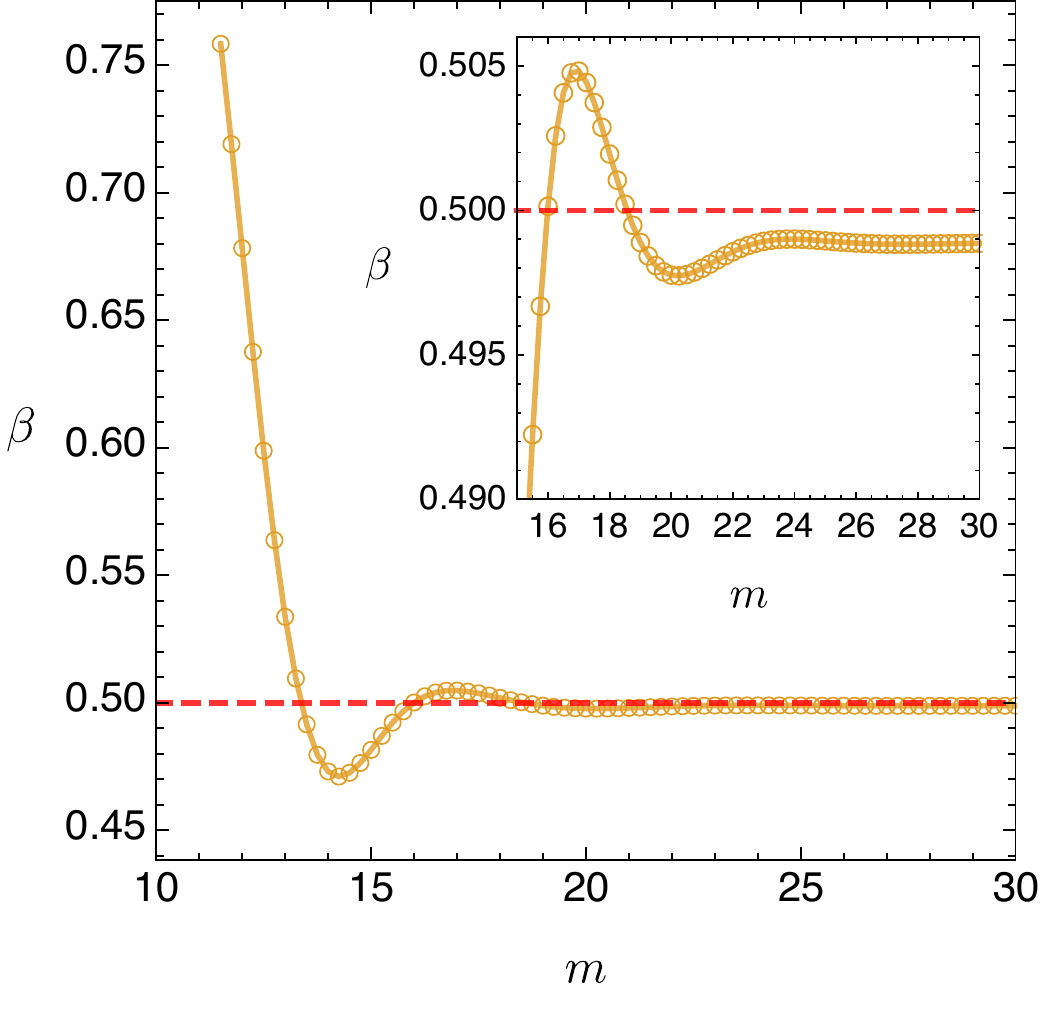}
  \end{subfigure}
  \caption{\textbf{Left panel:} Dominant QNMs (PS-NH modes) near extremality for $d = 11$ MP-dS with $y_{+} = 0.25$, for varying $m = l$. The black dashed line is the eikonal approximation $\beta_{\rm WKB}$. The black disk and black square at $r_{-}/r_{+} = 1$ are $\beta_{\hbox{\tiny NH}}$ for $m = 12$ and $m > 12$, respectively, as given by~\eqref{eqn:beta_NH}. The BF bound is saturated at $m \simeq 12.4$ so $\beta_{\hbox{\tiny NH}} = \frac{1}{2}$ for $m \geq 13$. \textbf{Right panel:}  $\beta$ as a function of $m$ for $d=11$ MP-dS with $y_+ = 0.25$ and $r_{-} = 0.9995 \, r_{+}$. For large $m$, $\beta$ converges to a value with $\beta < \frac{1}{2}$ (see the inset plot, which is an enlargement), however $\beta$ oscillates around this value for smaller $m$.}
  \label{fig:beta_d11_varying_m}
\end{figure}

We now analyse what happens when we approach extremality even closer, namely as close as $r_-/r_+ = 0.9998$. As an illustrative example, we do this in Fig.~\ref{fig:beta_d11_varying_m} for a $d = 11$ ($N =4$) MP-dS black hole with $y_+ = 0.25$, for which $m_{\rm crit} \simeq 12.4$. For modes with $m < m_{\rm crit}$, e.g. the blue curve $m = 12$ in Fig.~\ref{fig:beta_d11_varying_m}, there is no BF violation and accordingly we have $\beta > \frac{1}{2}$ at extremality (note that the near-horizon approximation~\eqref{eqn:beta_NH} is highly accurate in this case: see black disk $\beta_{\hbox{\tiny NE}}\: (m=12)$ in the plot).
For $m > m_{\rm crit}$ the situation is more complicated. For $m$ slightly above $m_{\rm crit}$, e.g. for the $m=14$ green curve  in the left panel of Fig.~\ref{fig:beta_d11_varying_m},  we find that the behaviour of $\beta(r_{-}/r_{+})$ is not monotonic; $\beta$ can first reach a maximum above $\frac{1}{2}$ before decreasing below $\frac{1}{2}$ at extremality. However, when $m$ is well above $m_{\rm crit}$ e.g. for the $m=20$ red curve  in the left panel of Fig.~\ref{fig:beta_d11_varying_m},  the modes have $\beta \leq \frac{1}{2}$ everywhere as they approach extremality. 
To understand this non-monotonicity better, in the right panel of Fig.~\ref{fig:beta_d11_varying_m} we fix the distance to extremality to be close to the minimum we reached, namely $r_{-}/ r_{+} = 0.9995$ and we plot $\beta$ as a function of $m$ for $d=11$ MP-dS with $y_+ = 0.25$. We see that $\beta$ oscillates around $\frac{1}{2}$ for moderate values of $m$ but ultimately converges to a value below  $\frac{1}{2}$ as $m$ grows large. Interestingly, but not perhaps not surprisingly (since the underlying physics is similar), a similar behaviour is observed in the discussion of SCC for a charged scalar field in Reissner-Nordstr\"om-dS when we plot $\beta$ as a function of the scalar field charge (which is the analogue of $m$ in charged system); see Fig.~9 of \cite{diasStrongCosmicCensorship2019}). Just like in \cite{diasStrongCosmicCensorship2019}, we could probably try to capture the behaviour of $\beta(m)$ using a WKB expansion at large $m$ with the oscillations around $\beta=1/2$ only captured after including non-perturbative contributions to the analysis via a Borel resummation (\emph{i.e.} a resurgence analysis) of the WKB expansion.

To conclude, we summarize our main SCC results.
MP-dS black holes with $m>0$ typically have two families of QNMs: the dS and PS-NH families. For sufficiently large $m$, the latter always has smaller $|{\rm Im}(\omega)|$ than the dS family so the PS-NH modes are the ones relevant for Strong Cosmic Censorship.
We found that there is at least one family of $m=l$ QNMs for which the spectral gap satisfies the condition $\beta\leq 1/2$ in the whole parameter space of MP-dS for odd $d\leq 11$ (and most probably even higher dimensions). It follows that  Christodoulou's formulation of Strong Cosmic Censorship holds for equal angular momenta $d>4$ MP-dS black holes, very much like in the $d=4$ Kerr-dS case~\cite{diasStrongCosmicCensorship2018}. This is the main result of our study. 
For each dimension $d=2N+3$ we found a (necessary but not sufficient) criterion, based on the violation of the AdS$_2$ BF bound associated to the near-horizon geometry of the extremal MP-dS, to find how large $m=l$ needs to be to ensure that we have at least one family of PS-NH modes with $\beta\leq1/2$ everywhere. 
Strictly speaking, our numerical analysis covered only the range $0.1 \leq y_{+} \leq 0.9$ and $0\leq a/a_{\rm ext}\lesssim 1$ of the 2-dimensional parameter space of MP-dS. So it seems that we missed cases near the endpoints of $y_+\in [0,1]$. However, we have complemented our numerical analysis with an (approximate) analytic analysis that covers the corners of the phase space which are not easy to explore numerically. Namely, we used the eikonal approximation~\eqref{eqn:omega_WKB} and the NH approximation~\eqref{eqn:beta_NH}-\eqref{eqn:m_crit}. 
We have only discussed SCC in equal angular momenta MP-dS black holes in odd spacetime dimensions $d$. However, the generic considerations of \cite{rahmanFateStrongCosmic2019} further indicate that this result extends to other, perhaps all, MP-dS solutions. Together with \cite{liuStrongCosmicCensorship2019} we thus have strong evidence that for arbitrary spacetime dimensions in de Sitter and for scalar induced perturbations, Christodoulou's formulation of SCC holds in dynamically stable, vacuum, rotating black hole solutions of the Einstein equations, but can be violated in the presence of charged matter. We stress, however, that the leading eikonal behaviour is spin independent, and thus the result quoted above could indeed also be true for gravitational perturbations.

%===========================================================
\acknowledgments

The authors acknowledge the use of the IRIDIS High Performance Computing Facility and associated support services at the University of Southampton in the completion of this work.
O.~C.~D. acknowledges financial support from the STFC ``Particle Physics Grants Panel (PPGP) 2018'' Grant No.~ST/T000775/1. J.~E.~S. has been partially supported by STFC consolidated grants ST/P000681/1, ST/T000694/1. The research leading to these results has received funding from the European Research Council under the European Community's Seventh Framework Programme (FP7/2007-2013) / ERC grant agreement no. [247252].

%===========================================================
%===========================================================
%===========================================================
%\clearpage
\appendix
%{\centering\textbf{\huge Appendices}
%\vspace{0.25in}}
%===========================================================
%===========================================================
%===========================================================

\section{Quasinormal modes of higher-dimensional Schwarzschild-de Sitter}\label{sec:SdS_and_small_a}

To discuss Strong Cosmic Censorship in equal angular momenta MP-dS black holes, it was necessary to carefully study the quasinormal spectra of these black holes, and we identified and highlighted the main features in Section~\ref{sec:QNMspectra}. However, even before studying the QNMs of the MP-dS black hole, we have to start by identifying the QNM spectra of its non-rotating limit, namely of the Schwarzschild-dS black hole. Since there are no detailed studies of the QNMs of higher-dimensional Schwarzschild-dS in the literature, for completeness we highlight some key properties of this spectra in this appendix. Although in the main text we restrict our analysis to odd spacetime dimensions $d=2N+3$, in this appendix we also consider even dimensions $d$ because the Schwarzschild-dS QNM spectra is substantially different for $d=4$ and $d\geq 5$, and so it is important to emphasize these differences. We do not aim to present the full QNM spectra of Schwarzschild-dS for all dimensions, but just the main features that we have identified and that seem worth highlighting. In particular, those that help us further understand features of the MP-dS spectra.  

 Taking the limit \(a \to 0\) of the MP-dS metric~\eqref{eqn:metric_BL}, we recover the metric for \(d=5\) Schwarzschild-dS, with a \(\mathbb{CP}^{1}\) angular part that is isomorphic to a 2-sphere $S^2$:
\begin{align}
  ds^{2} &= - f(r) dt^{2} + \frac{1}{f(r)} \, dr^{2} + r^{2}(d\psi + \frac{1}{2} \cos \theta \, d\phi)^{2} + \frac{r^{2}}{4}(d\theta^{2} + \sin^{2}\theta \, d\phi^{2}), \\
  f(r) &=  1 - \frac{2 M}{r^{2}} - \frac{r^{2}}{L^{2}}.
\end{align}
Note that due to spherical symmetry we can label perturbations by the total angular momentum $l$ alone, using \(\lambda = l(l+2N) - m^{2}\) from equation~\eqref{eqn:CPN_eigenvalues} to eliminate the explicit dependence on $m$. In general, there are two distinct mode families in Schwarzschild-dS. These are the de Sitter (dS) modes, which reduce to~\eqref{eqn:ds_modes_odd}-\eqref{eqn:ds_modes_even} in the limit $y_{+} \to 0$, and the photon sphere (PS) modes, which are well approximated by~\eqref{eqn:omega_WKB} in the eikonal limit $|m| = l \to \infty$. Schwarzschild-dS has no extremal limit and thus  there are no near-horizon (NH) modes in its QNM spectra. 

Taking the limit $a \to 0$ of the effective potential for the wave equation in MP-dS~\eqref{eqn:effective_potential} and setting $N = 1$, the solutions to $V_{\rm eff}(r_{0}) = V_{\rm eff}'(r_{0}) = 0$ are $r_{0} = 2 \sqrt{M}$ and $b^{\pm} = \pm 2 L \big(\frac{L^{2}}{2M} - 4\big)^{-1/2}$. The corresponding eikonal approximation~\eqref{eqn:omega_WKB}  to the PS modes is
\begin{align}
  \omega_{\textrm{PS}} \, r_{c} &\simeq \frac{y_{+}^{2}-1}{2 y_{+} \sqrt{1+y_{+}^{2}}}\left[ \pm l + i\sqrt{2} \left(n + \frac{1}{2}\right) \right]. \label{eqn:SdS_PS_modes}
\end{align}
One can also derive an approximation for the modes in the Nariai limit $y_{+} \to 1$ (see e.g. the supplementary material in~\cite{cardosoQuasinormalModesStrong2018}), but we will not do this here, since there is no extra family of QNMs associated to it: the Nariai analysis simply captures the PS family. 

\begin{figure}
  \centering
  \includegraphics[width=0.48\linewidth]{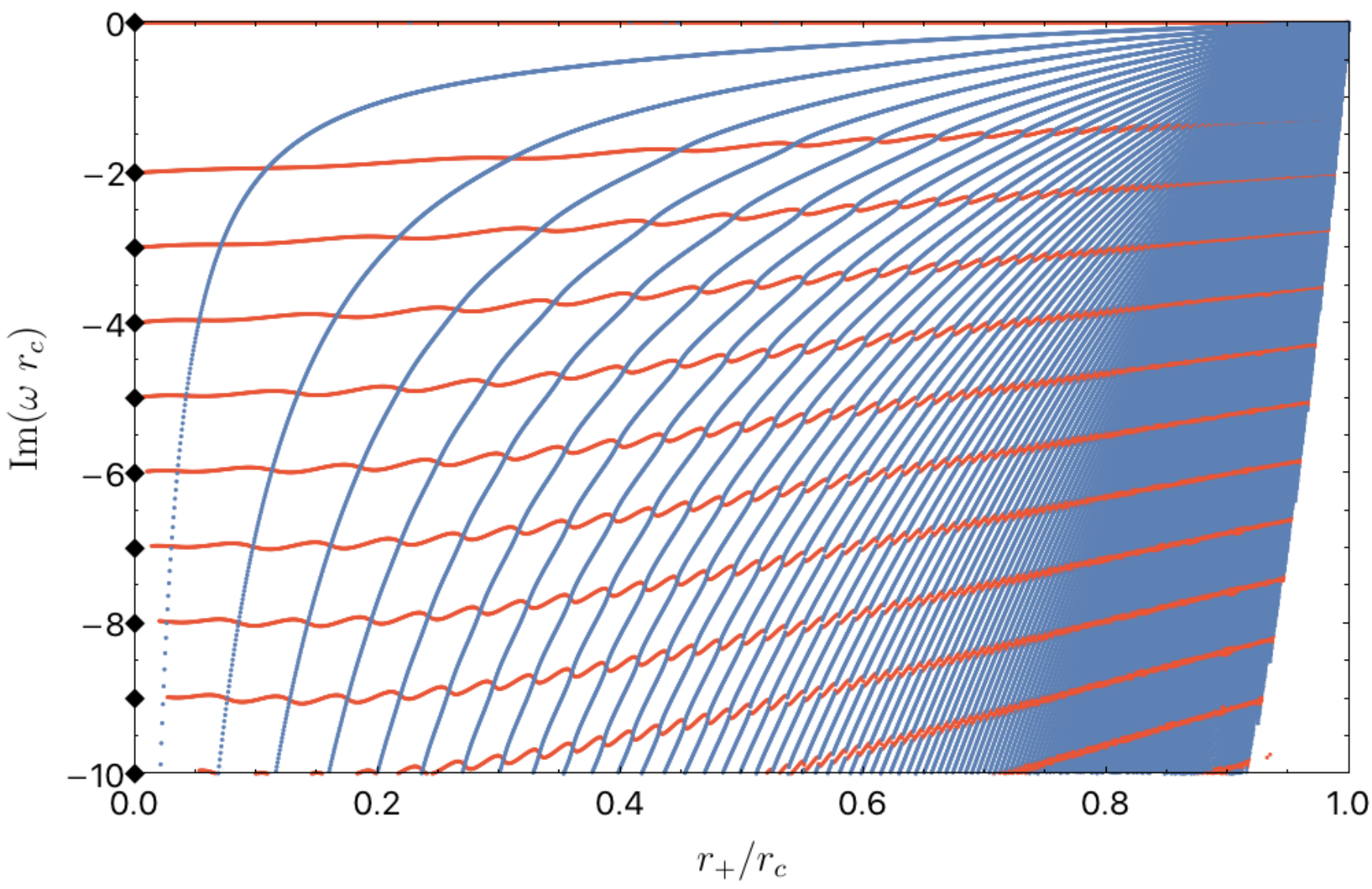}
  \includegraphics[width=0.48\linewidth]{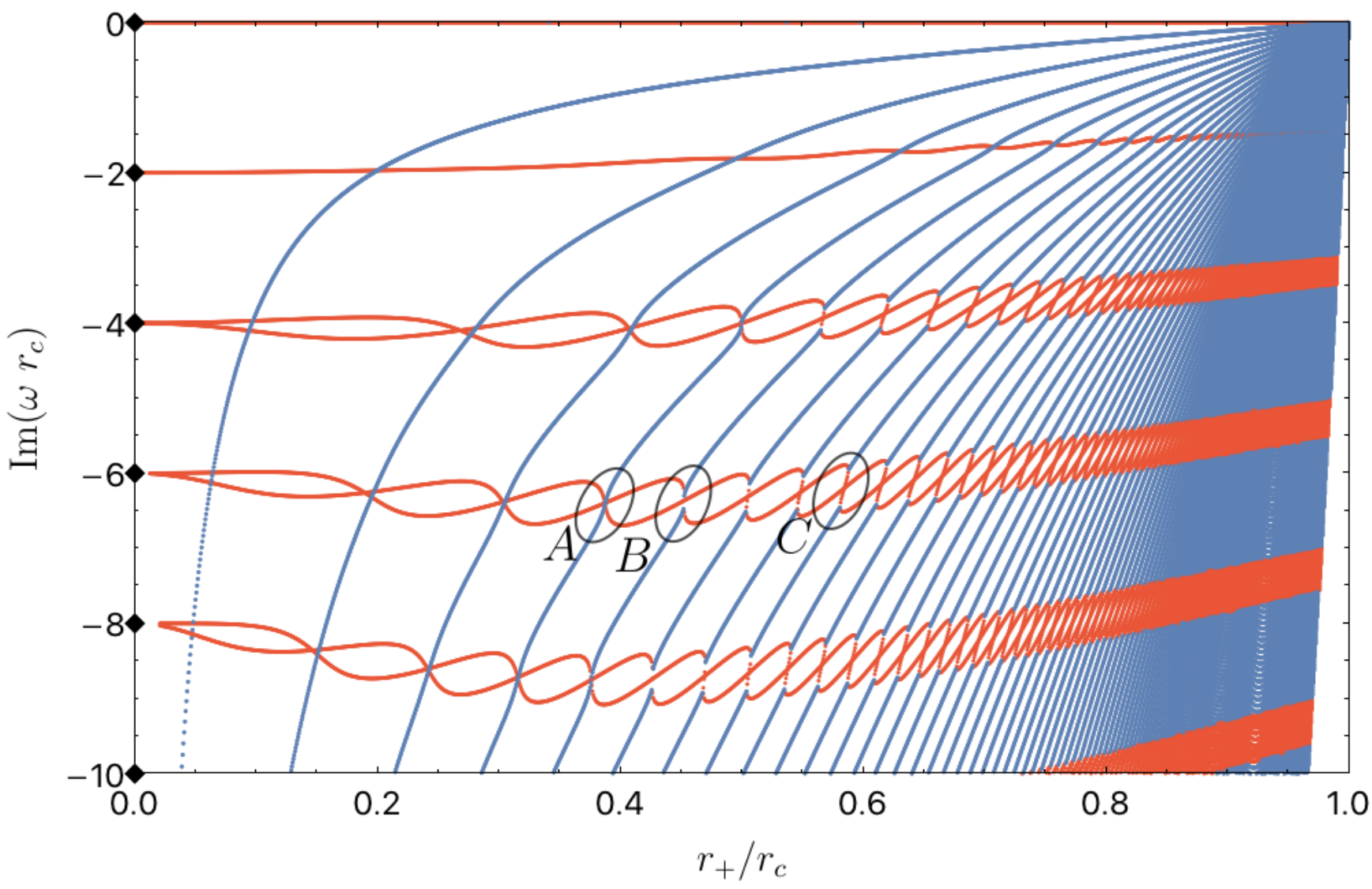}
  \caption{QNM spectrum for \(d=4\) (left) and \(d = 5\) (right) Schwarzschild-dS with \(m = l=0\) (several radial overtones $n=0,1,2,3,4,5,\dots$ are displayed). There are purely imaginary dS modes (red) and complex PS modes (blue). The modes of pure dS space~\eqref{eqn:ds_modes_odd}-\eqref{eqn:ds_modes_even} are indicated by black diamonds at $r_{+} = 0$. In \(d=5\) (right panel) the pure dS frequencies are degenerate, but the dS curves split as $r_+/r_c$ increases above zero, while they are always non-degenerate in \(d=4\)  (left panel). The points \(A, B, C\) indicate three mode crossings/mergers plotted in detail in Fig.~\ref{fig:QNM_schwarzschild_5d_merge}.\label{fig:QNM_schwarzschild_5d}}
\end{figure}

The $d = 5$ QNM spectra are shown in Fig.~\ref{fig:QNM_schwarzschild_5d} for $d = 4$ (left panel) and $d = 5$ (right panel) Schwarzschild-dS with $m = l = 0$.\footnote{It becomes very difficult to find the full quasinormal mode spectrum as we approach the two limits of the parameter space: de Sitter ($r_+ = 0$) and Nariai ($r_+ = r_c$), so some modes have not been resolved in those limits, although they certainly do exist.} There are  dS modes with purely imaginary frequencies (red curves with increasing radial overtone $n=0,1,2,3,4,5,\dots$ from top to bottom), whose eigenvalues are smoothly connected to the frequencies of pure de Sitter space~\eqref{eqn:ds_modes_even}-\eqref{eqn:ds_modes_odd} (indicated by black diamonds) in the limit $r_{+} \to 0$. There also exist PS modes with complex frequencies (blue curves with increasing radial overtone $n=0,1,2,3,4,5,\dots$ from top-left to bottom-right), which we have identified by marching each QNM to $m = l = 20$, where they are in excellent agreement with the eikonal approximation~\eqref{eqn:SdS_PS_modes}. 

Comparing the left ($d=4$) and right ($d=5$) panels of Fig.~\ref{fig:QNM_schwarzschild_5d}, the main  difference going from \(d = 4\) to \(d = 5\) can be found in the dS curves. To start with, in $d=4$ we have roughly two times more dS curves than in $d=5$, in agreement with the discussion of~\eqref{eqn:ds_modes_even}-\eqref{eqn:ds_modes_odd}. Indeed, in the $r_+\to 0$ limit, all negative imaginary integers (except $-i$) are frequencies of pure dS spacetime in $d=4$ but, in $d=5$, the frequencies of pure dS are given only by the even negative integers. The next difference occurs when we let $r_+/r_c$ increase. For $d=4$, there is a single curve departing from each overtone of pure dS but, in $d=5$, the pure dS mode is degenerate and two intertwining curves emerge from it as $r_+/r_c$ increases.   
This can be partially explained by the degeneracy of the pure de Sitter modes~(\ref{eqn:ds_modes_odd}-\ref{eqn:ds_modes_even}): in odd dimensions $d = 2 N + 3$, the $(N+1)$-th pure de Sitter mode (counting the zero mode) and higher overtones are degenerate and  these degenerate modes split as we move away from pure de Sitter. However, this intertwining behaviour of the dS modes appears to be unique to \(d=5\) (and not shared by the $d=7, 9, 11, \dots$ case) as we will see later.

\begin{figure}[!t]
  \vspace{1cm}
  \centering
  \begin{subfigure}{.3415\linewidth}
    \includegraphics[width=\linewidth]{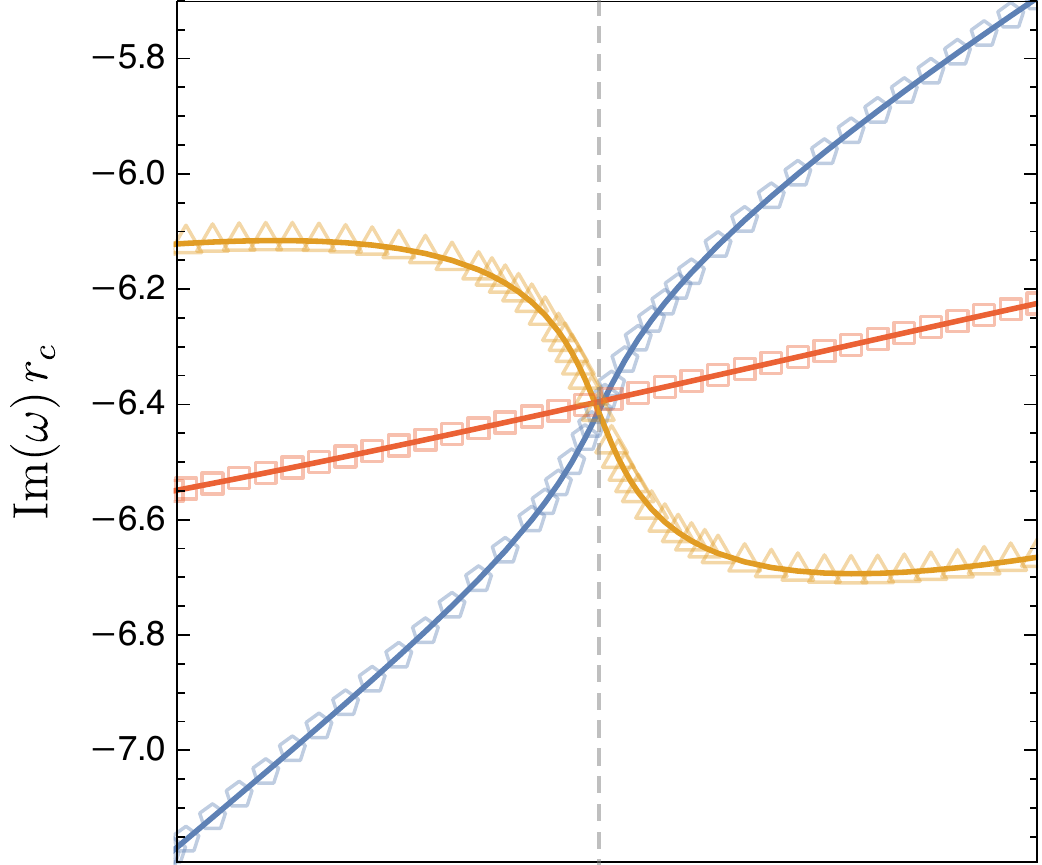}
    \raggedleft
    \includegraphics[width=0.97\linewidth]{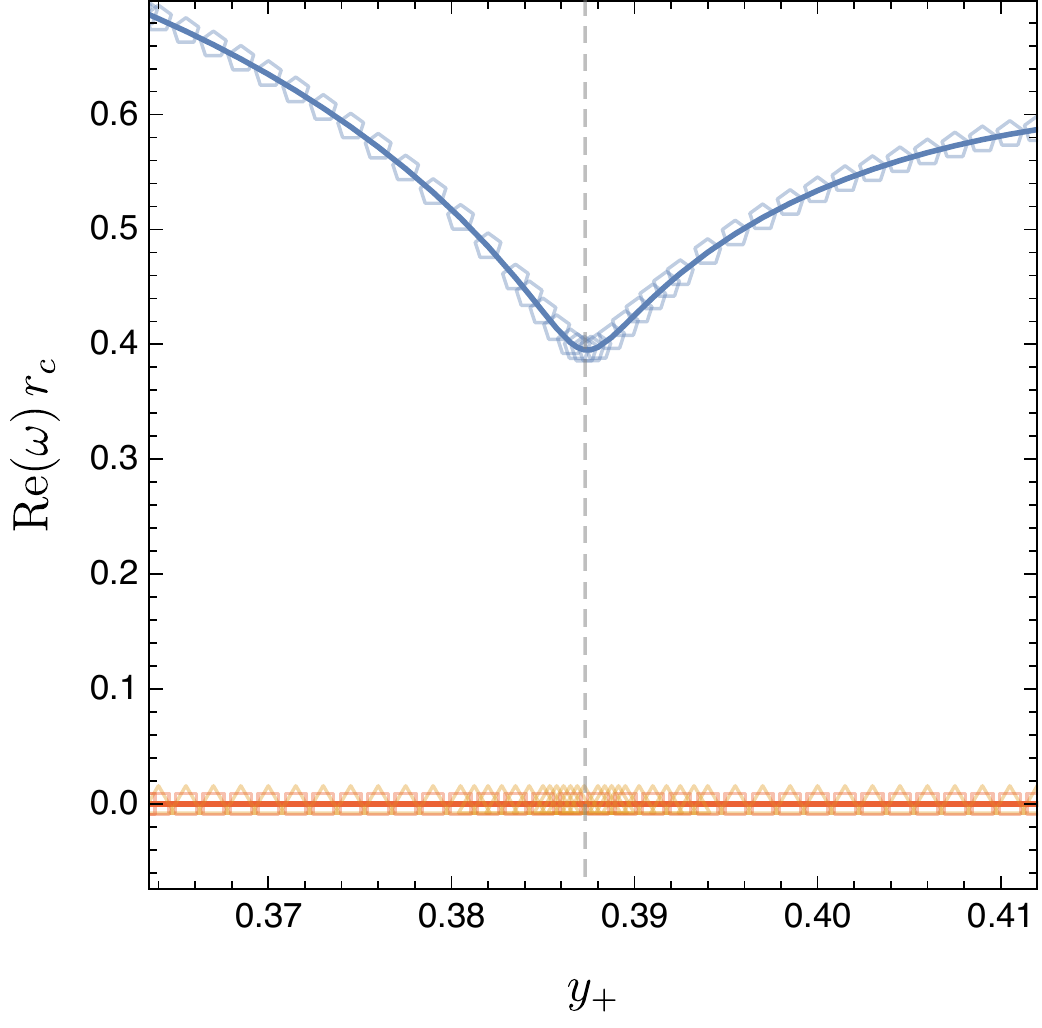}
    \caption{Mode crossing at A\label{fig:QNM_schwarzschild_5d_merge_A}}
  \end{subfigure}
  \begin{subfigure}{.31\linewidth}
    \includegraphics[width=\linewidth]{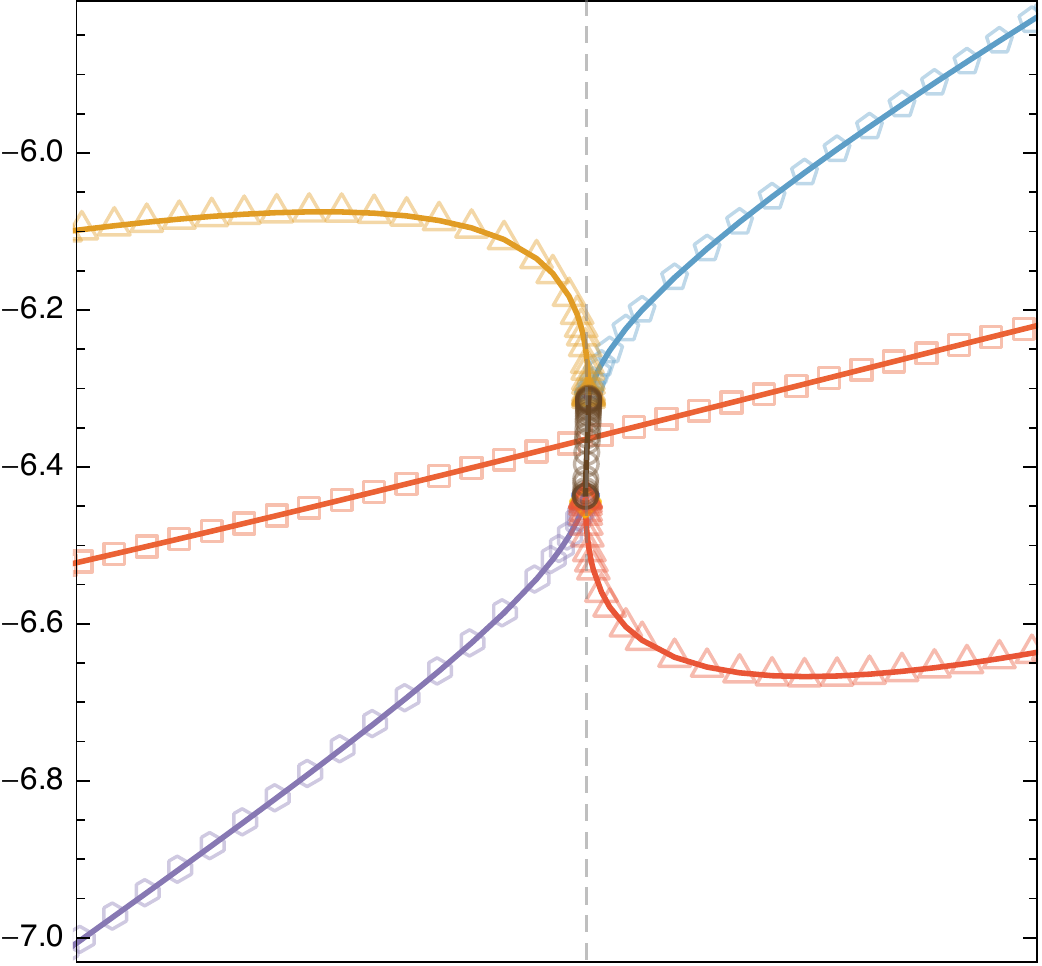}
    \raggedleft
    \includegraphics[width=0.98\linewidth]{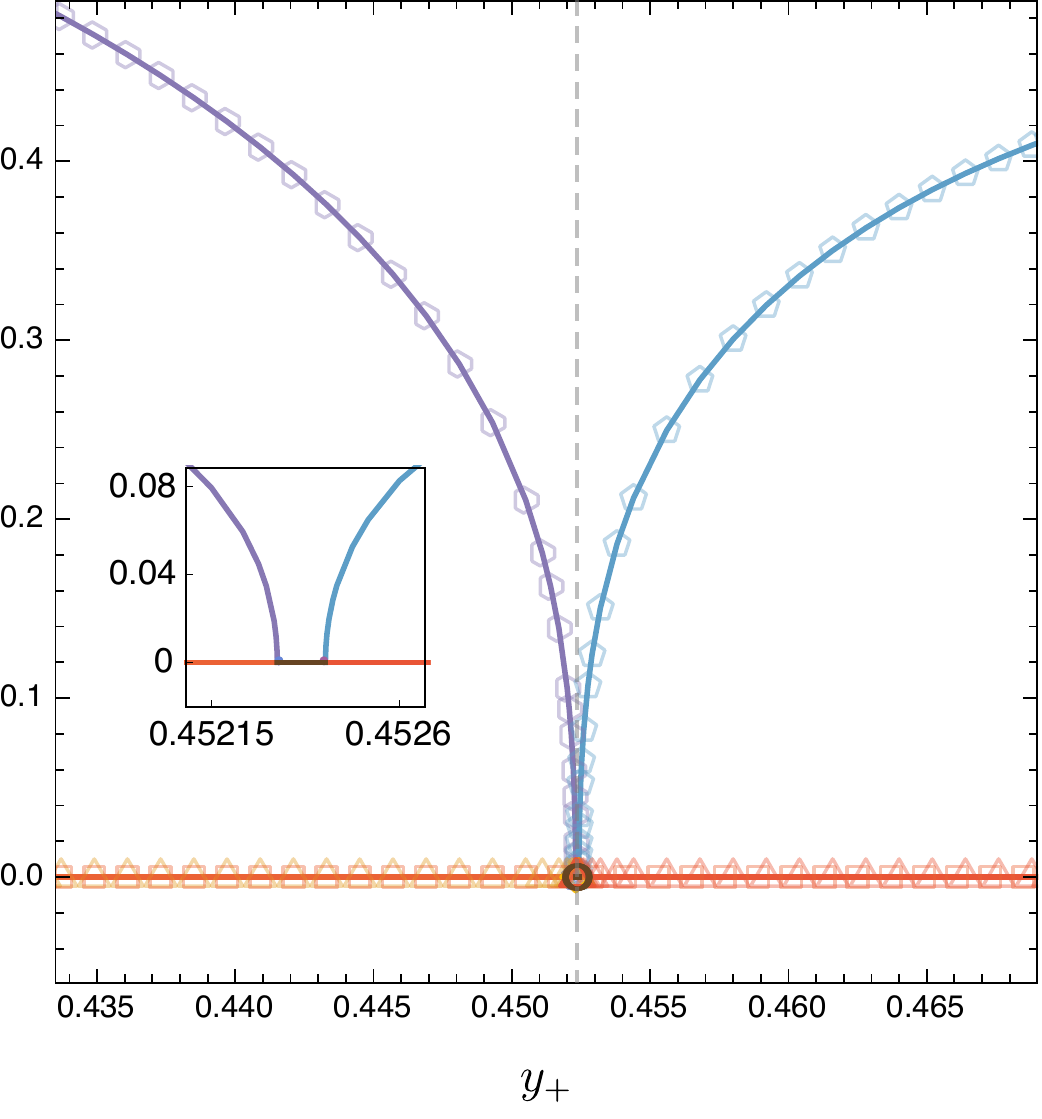}
    \caption{Close merger at B\label{fig:QNM_schwarzschild_5d_merge_B}}
  \end{subfigure}
  \begin{subfigure}{.31\linewidth}
    \includegraphics[width=\linewidth]{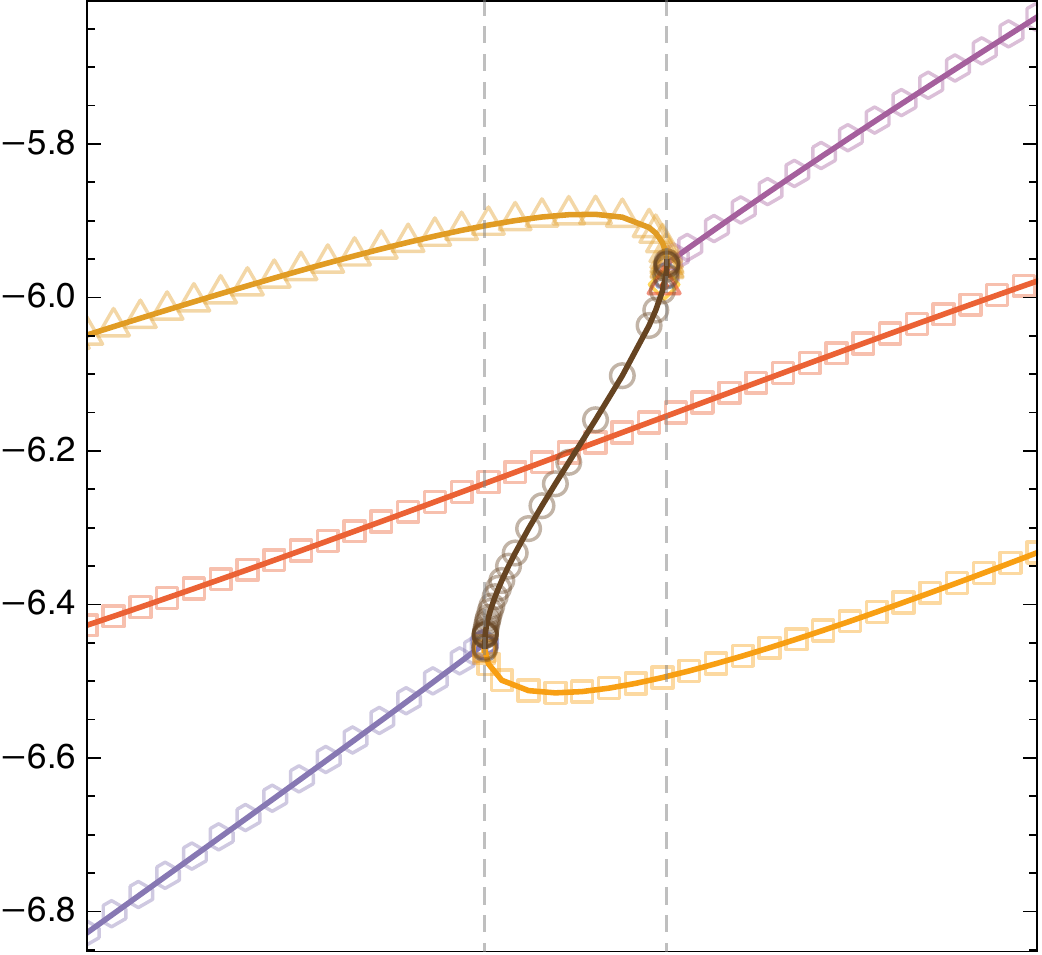}
    \raggedleft
    \includegraphics[width=0.99\linewidth]{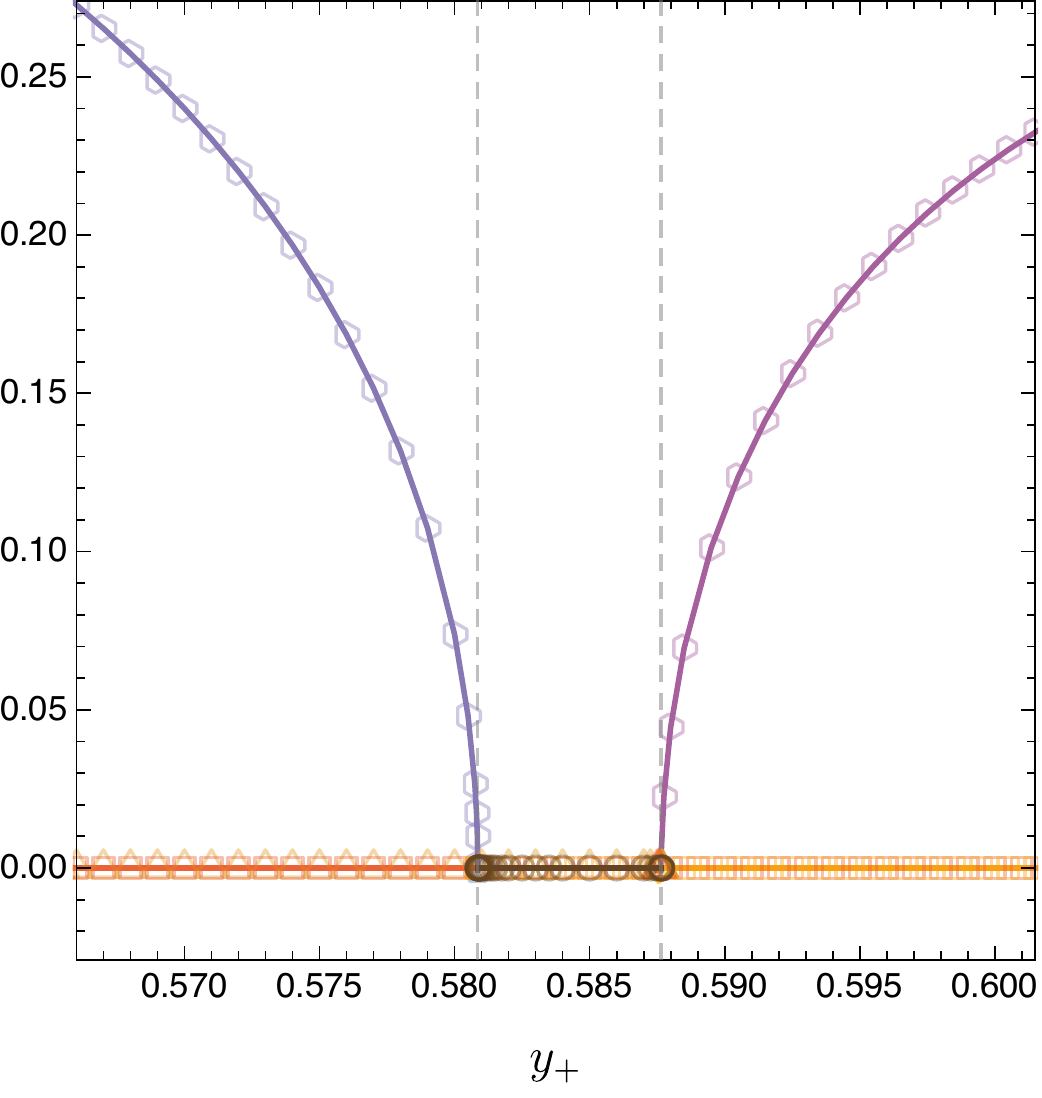}
    \caption{Mode merger at C\label{fig:QNM_schwarzschild_5d_merge_C}}
  \end{subfigure}
  \caption{Imaginary (top panel) and real (bottom panel) parts of the $l=m=0$ QNM frequency spectra of the $d=5$ Schwarzschild-dS black hole of the right panel of Fig.~\ref{fig:QNM_schwarzschild_5d}, but this time zoomed-in around the regions $A$, $B$ and $C$ identified in Fig.~\ref{fig:QNM_schwarzschild_5d}. The red/yellow/brown modes are purely imaginary frequency dS modes and the blue/purple pentagon curves describe  complex frequency PS modes. The inset plot in the middle-bottom figure (b) is an enlarged plot at the merge region to show that it is a very short-lived merge rather than a crossing. \label{fig:QNM_schwarzschild_5d_merge}}
\end{figure}

To better understand the intertwining structure of \(d=5\) Schwarzschild-dS, the three regions $A$, $B$ and $C$ in the right panel of Fig.~\ref{fig:QNM_schwarzschild_5d} are enlarged in Fig.~\ref{fig:QNM_schwarzschild_5d_merge}. The dS modes are represented by  orange triangles and/or red squares while the PS modes are described by the blue or purple pentagons. At region $A$ (top-left panel) both dS modes merely cross each other and the PS mode. The PS modes have a real part that dips to a finite value at the crossing (bottom-left panel), but don't become purely imaginary (the dS modes always have frequencies with zero real part). 
Increasing \(y_{+}\) to region $B$  (middle panels), the real part of the PS mode drops to zero (bottom-middle panel), and the orange triangle dS curve merges with the PS curve (middle panels). The red square dS curve  passes through the other two without interaction (top-middle panel)
Further increasing \(y_{+}\) till region $C$  (right panels), we see two bifurcation points (with 3 curves departing from each one). Looking at the imaginary part (top-right panel) of Fig~\ref{fig:QNM_schwarzschild_5d_merge_C}, the PS mode in the bottom-left splits (at the first bifurcation point) into the orange triangle dS curve and a new `bridge' mode (brown circles bridging the two bifurcation points in the right panels) with purely imaginary frequencies. This bridge mode then extends up and to the right till the second bifurcation point where the other branches of the orange triangle dS and PS curves also meet. Again, the red square dS curve  passes through the other two without interaction. Similar bridge modes were observed in the QNM spectra of Reissner-Nordst\"om$-$dS black holes in higher dimensions \cite{diasOriginReissnerNordstrOm2020}.

\begin{figure}
  \centering
  \begin{subfigure}{.48\linewidth}
    \includegraphics[width=\linewidth]{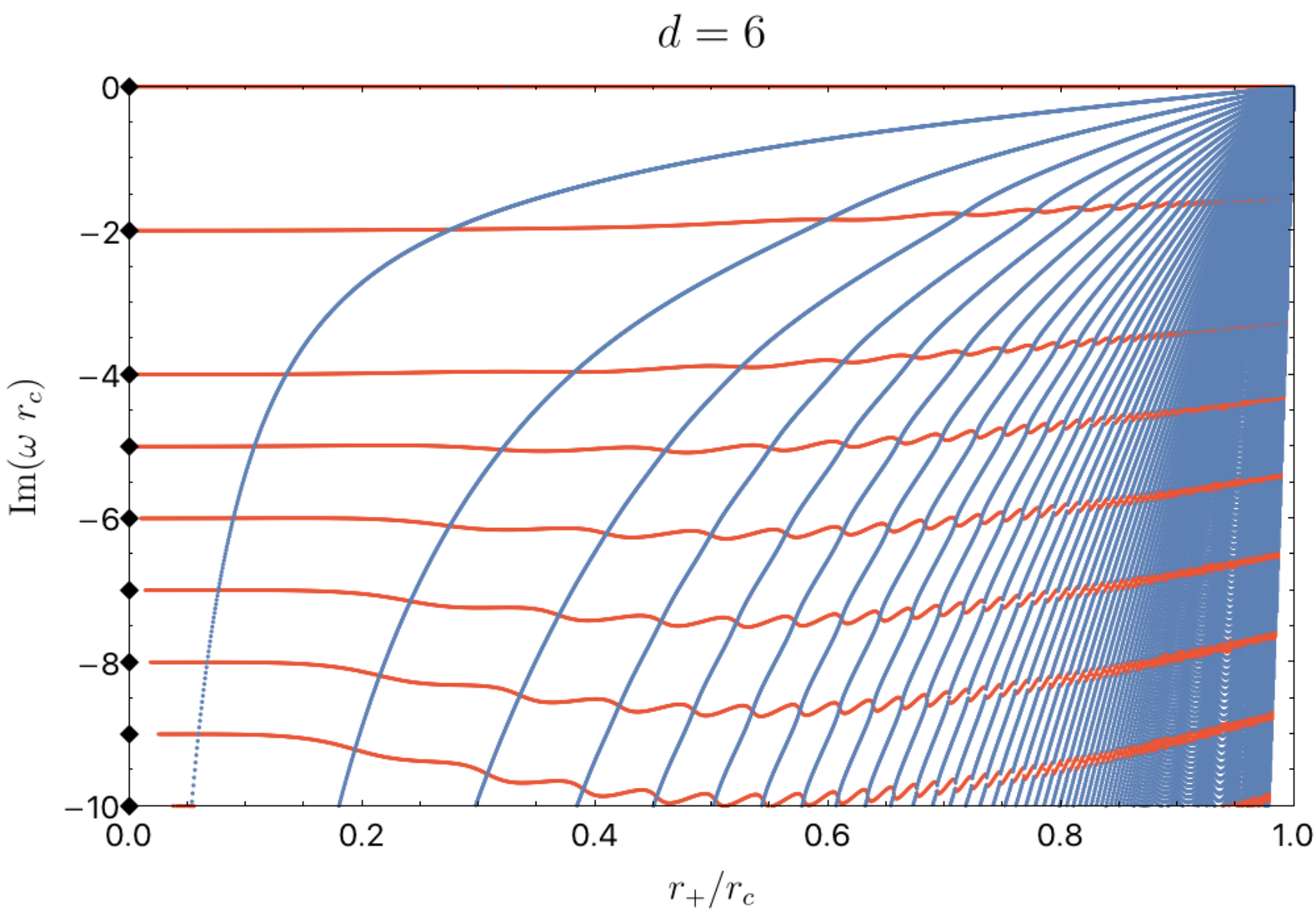}

    \vspace{0.5cm}

    \includegraphics[width=\linewidth]{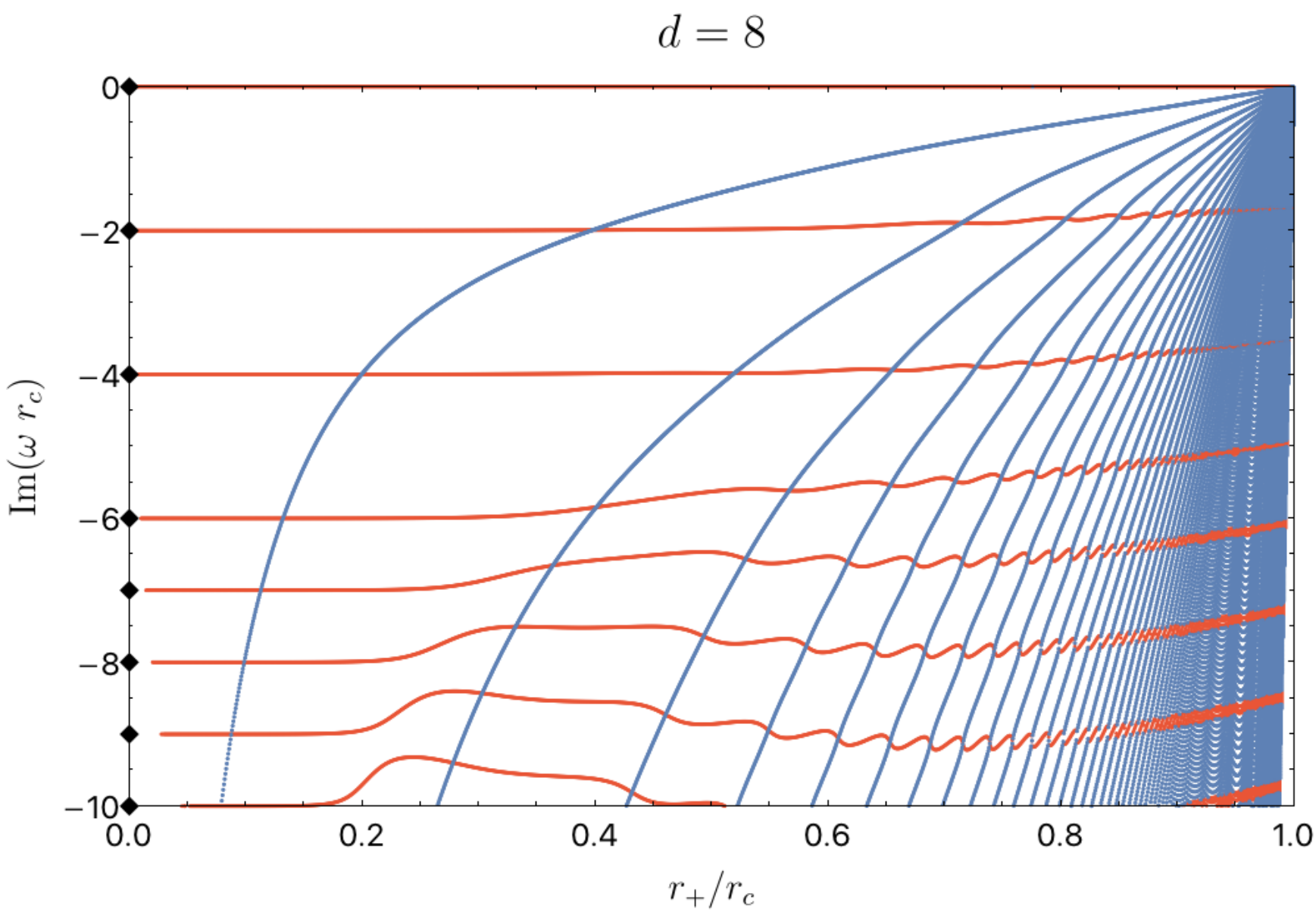}
  \end{subfigure}
  \begin{subfigure}{.48\linewidth}
    \hspace{0.5cm}
    \includegraphics[width=\linewidth]{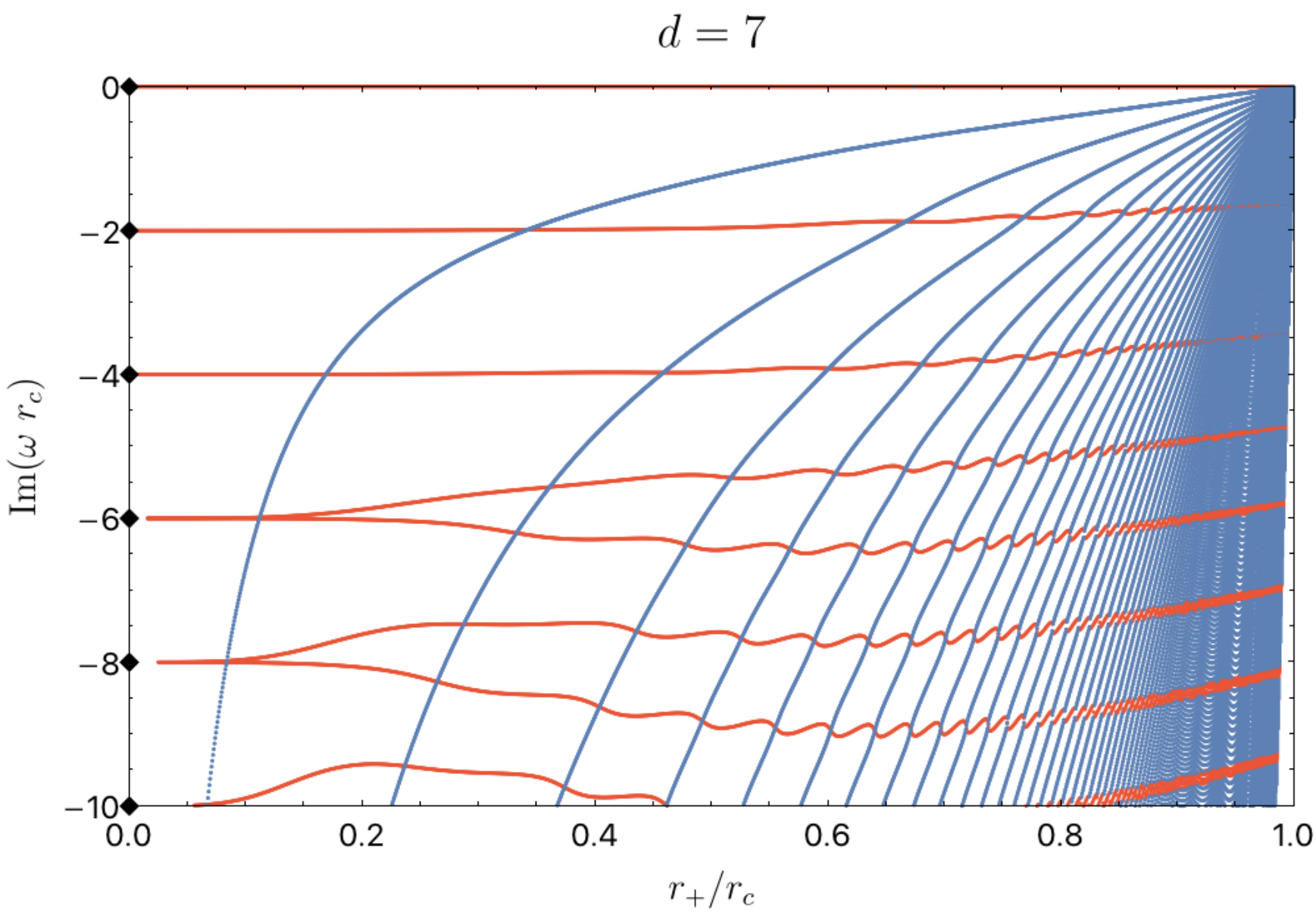}

    \vspace{0.5cm}

    \hspace{0.5cm}
    \includegraphics[width=\linewidth]{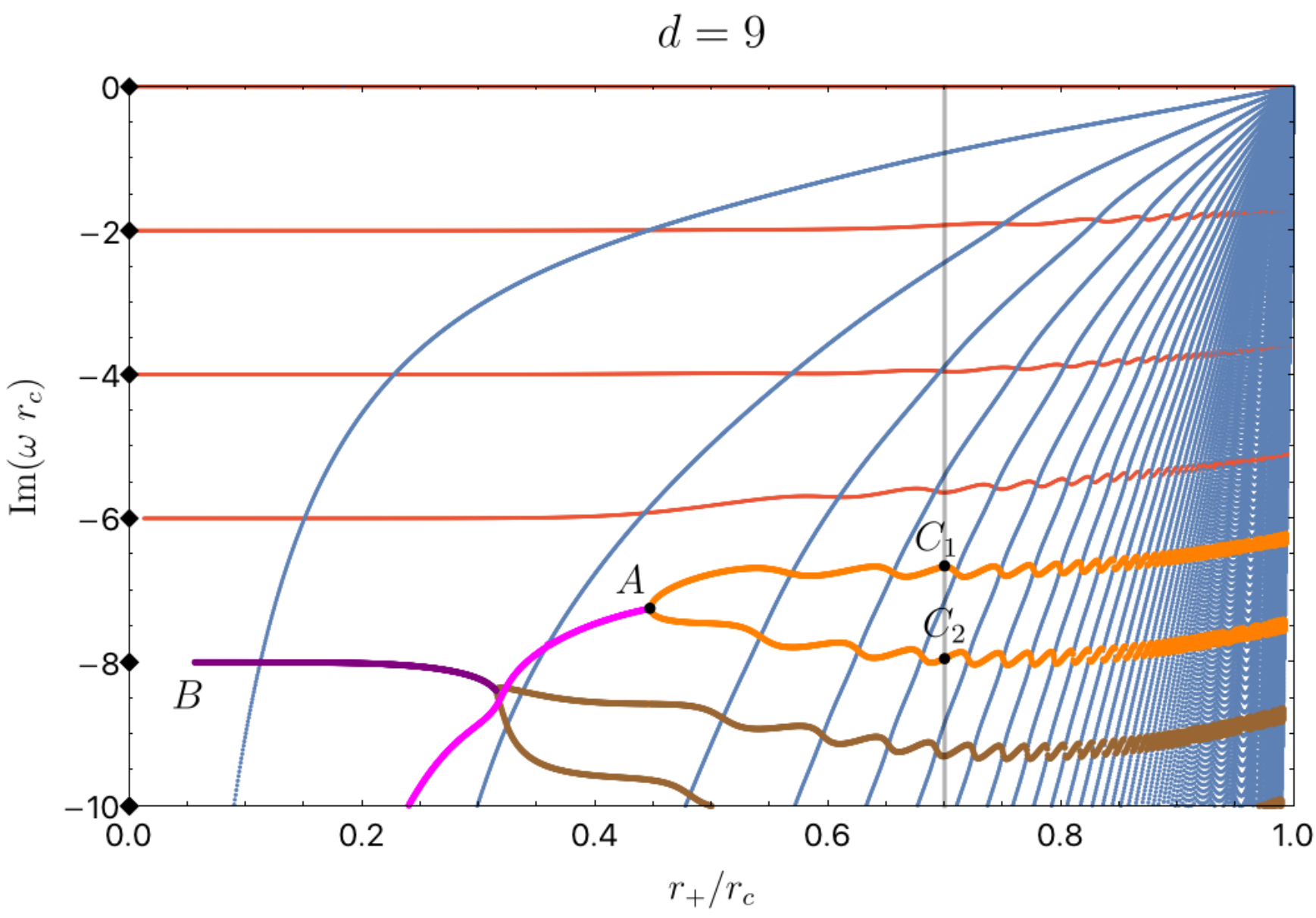}
  \end{subfigure}
  \caption{QNM spectrum of Schwarzschild-dS in $d = 6, 7, 8, 9$ dimensions, with $l=m = 0$. Red modes are purely imaginary dS modes and the blue modes are complex PS modes. The orange/brown modes are purely imaginary but not connected directly to the dS limit when $r_+\to 0$. The purple/magenta modes are PS modes with complex frequency that do not vanish in the Nariai limit $r_+ \to r_c$. The pure dS frequencies at $r_{+} = 0$ are indicated by black diamonds. These are exceptional modes in the sense that we do not observe similar modes for $d<9$ at least in the first few radial overtones.}
  \label{fig:QNM_SdS_higher_dimensions}

  \includegraphics[width=0.7\linewidth]{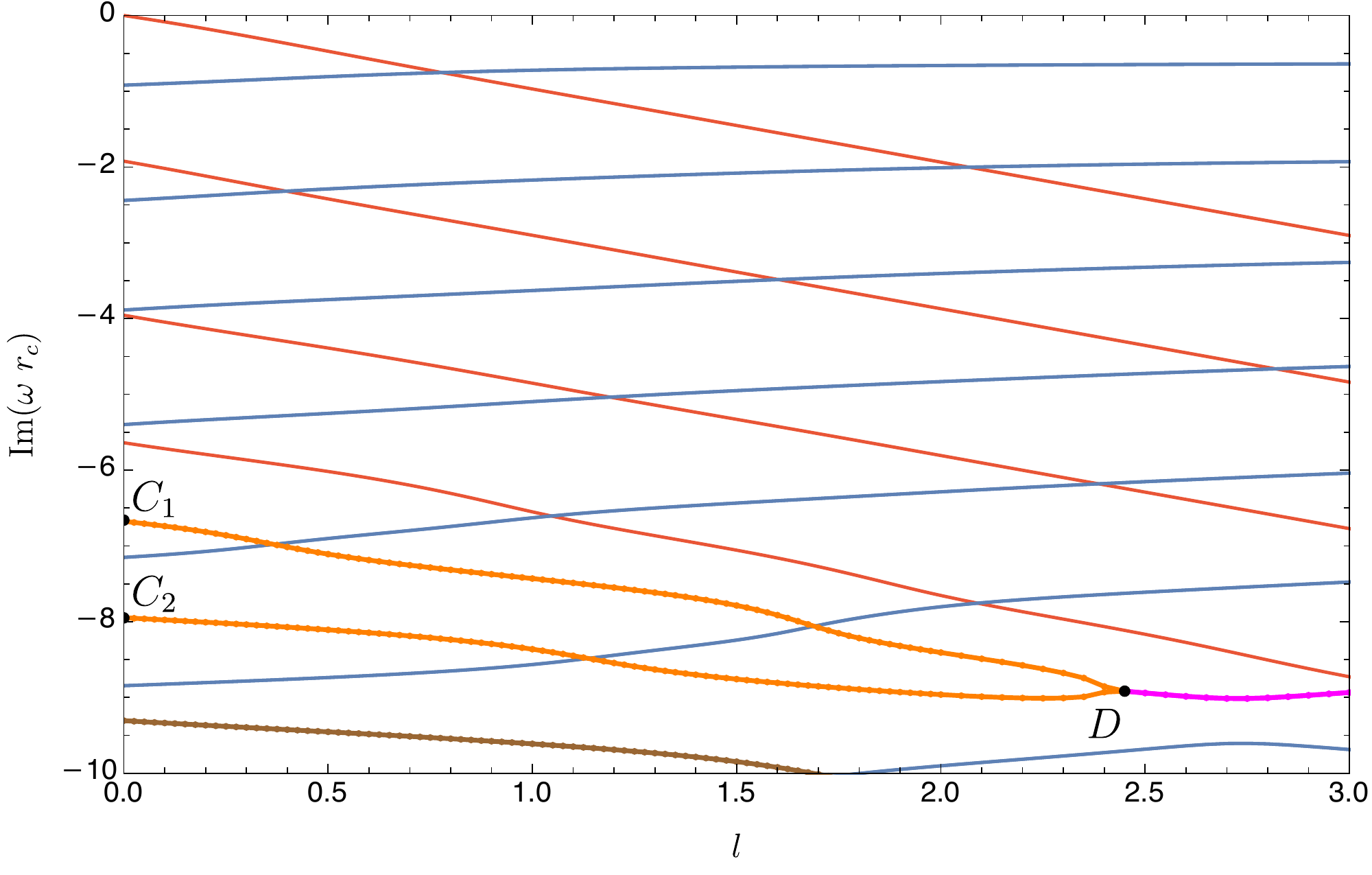}
  \caption{Marching the QNM spectra of Fig.~\ref{fig:QNM_SdS_higher_dimensions} from  $m = l=0$ all the way up to  $m = l=3$ for fixed \(y_{+} = 0.7\). The points $C_1$ and $C_2$ are those already present in the bottom-right panel of Fig.~\ref{fig:QNM_SdS_higher_dimensions} (see vertical grey line). 
The red modes are the usual purely imaginary dS modes and the blue modes  are PS modes with complex frequency (we have tracked them up to $m=l=20$ to compare with the eikonal limit of PS modes).
The orange curves starting at $C_1$ and $C_2$ have purely imaginary frequency and join at point $D$. For higher $m$, the magenta curve starting at $D$ has complex frequencies and it is a PS mode in the sense that for larger $l=m$ it agrees well with the eikonal approximation.
\label{fig:QNM_SdS_m_march}}
\end{figure}

Let us now consider what happens when $d > 5$. In Fig.~\ref{fig:QNM_SdS_higher_dimensions} we plot the QNM  spectra of Schwarzschild-dS in dimensions $d = 6, 7, 8, 9$ for $l=m=0$. The several overtones of the PS curves (blue squares) have a qualitative behaviour similar to that found in $d=4$ and $d=5$ of  Fig.~\ref{fig:QNM_schwarzschild_5d}. Moreover, the even-dimensional results are qualitatively similar to \(d = 4\) (left panel of Fig.~\ref{fig:QNM_schwarzschild_5d}) for both the PS and dS modes. In \(d = 7\), the degenerate dS modes still split as \(r_{+}/r_c\) increases above zero and tend to develop wavy shape, however they do not have the same intertwining behaviour observed in \(d = 5\) (right panel of Fig.~\ref{fig:QNM_schwarzschild_5d}). In fact, we do not observe intertwining behaviour between two dS curves in any other dimension other than $d=5$, at least up to \(d = 11\). The behaviour of the dS modes in $d=5$ is thus very unique.

In \(d = 9\) and higher, we find some exceptional modes which do not have the standard characteristics of dS or PS modes, as can be seen in the \(d = 9\) plot in the bottom-right panel of Fig.~\ref{fig:QNM_SdS_higher_dimensions}. The red and blue curves still describe the same families as before:  purely imaginary dS modes and complex frequency PS modes, respectively. However, the purple/magenta and orange/brown modes do not fit the standard classification. The purple/magenta modes are  PS modes with complex frequency (see discussion of Fig.~\ref{fig:QNM_SdS_m_march} below), but unlike the other PS modes, they do not have a standard PS behaviour at small $y_+$ or a vanishing imaginary part in the Nariai limit, $r_+/r_c\to 1$. Instead, these frequencies behave like those of pure dS space as $r_+/r_c$ grows large. For example, the magenta curve splits at \(A\) into a pair of imaginary modes (orange) which behave much like dS modes, with a non-vanishing \(\omega\) in the Nariai limit.  The magenta and purple modes appear to have different behaviours in the  limit \(r_{+} \to 0\). The magenta mode is suppressed (\emph{i.e.} $|{\rm Im}(\omega r_c)|$ becomes large as $y_+$ decreases), while the purple mode appears to tend towards the pure dS frequency, with \(\operatorname{Im}(\omega \, r_{c}) = -8i\). However, this region is difficult to resolve to the left of point \(B\). To further clarify the properties of the exceptional families that pass though points $A,C_1$ and $A, C_2$ in Fig.~\ref{fig:QNM_SdS_m_march} we do the following exercise. We fix $y_+=0.7$, which is described by the vertical grey line in  bottom-right panel of Fig.~\ref{fig:QNM_SdS_higher_dimensions} when $l=m=0$. In particular, in the exceptional orange curves this selects points $C_1$ and $C_2$ with $l=m=0$. Then, in Fig.~\ref{fig:QNM_SdS_m_march}, we run a code that marches over $m=l$  from $m=l=0$ (where $C_1$ and $C_2$ lay) all the way up to $l=m=3$. We see that in this path, the two orange curves starting at $C_1$ and $C_2$ merge at point $D$ into a single magenta curve that then extends to $m=l=3$ and beyond (not shown). Extending this plot even further to, say, $m=l=20$ we can compare it with the eikonal limit of the PS modes and conclude that the magenta curve is a high overtone PS mode (like the blue curves in Fig.~\ref{fig:QNM_SdS_higher_dimensions} and Fig.~\ref{fig:QNM_SdS_m_march}). So as $l$ increases the curves $AC_1$ and $AC_2$ of Fig.~\ref{fig:QNM_SdS_higher_dimensions} are connected to PS modes, although for $l=m=0$ they do not have the standard ${\rm Im}(\omega)\to 0$ behaviour as $y_+\to 0$. This illustrates how intricate the QNM spectra of Schwarzschild-dS can become  for higher overtones, especially for large $d$.

The quasinormal mode spectrum of other asymptotically dS spacetimes (e.g MP-dS or Reissner-Nordst\"om$-$dS) will similarly contain modes which defy the standard classification in higher dimensions, as confirmed in the  Reissner-Nordst\"om$-$dS study of \cite{diasOriginReissnerNordstrOm2020}. One might wonder why this is not visible in studies of \(\beta\) for higher-dimensional RN-dS~\cite{liuStrongCosmicCensorship2019} (although it is present in \cite{diasOriginReissnerNordstrOm2020}). In the context of Strong Cosmic Censorship, the \(n = 1\) mode is the dominant dS mode. This is the only mode which is not paired, and it has a much weaker dependence on the black hole parameters. Indeed, all of the effects we have described are present only for subdominant modes, and so are not relevant for SCC.

%===========================================================
\section{Near-horizon QNMs: explicit expressions in $d = 2N + 3$ dimensions}\label{sec:explicit_N}
%===========================================================
In Section~\ref{sec:NH_modes} we employed a matched asymptotic expansion procedure to find an analytic expression for the frequencies $\omega_{\hbox{\tiny NH}}$ of the near-horizon (NH) modes which very well approximate the associated numerical frequencies of the system near-extremality.  This $\omega_{\hbox{\tiny NH}}$ is given by \eqref{eqn:NH_modes_MP-dS_abstract}-\eqref{eqn:NHparameters} and in this appendix we provide the explicit expression for the 
quantities  $\Omega_{(1)}$ and $\kappa_{(1)}$ that appear in \eqref{eqn:NHparameters}. Note that, for all $N$, we can express the black hole parameters $(a, M, L)$ in terms of the dimensionless horizon radii $(y_{+}, \,y_{-})$ by
\begin{align}
  \frac{a^{2}}{r_{c}^{2}} &= \frac{y_-^2 y_+^2}{(1-y_-^2) (1-y_+^2) (1-y_-^{2 N+2}) (1-y_+^{2 N+2}) (y_+^2-y_-^2) (y_+^{2 N+2}-y_-^{2 N+2})}  \nonumber \\
 & \times \biggr\{(1-y_+^2) ((y_-^{4 N+4}+1) y_+^{2 N} (y_+^2-y_-^2)-(1-y_-^2) (y_-^{4 N+4}+y_+^{4 N+4})) \nonumber \\
                        &\quad +y_-^{2 N} (-y_-^4 (y_+^{2 N+2}+1)^2+2 y_-^4 (y_+^4+1) y_+^{2 N}+(1-y_+^{2 N+2})^2 (y_-^2-(1-y_-^2) y_+^2))\biggr\},\nonumber \\
  \frac{M}{r_{c}^{2 N}} &= \frac{(1-y_{-}^{2})(1-y_{+}^{2})(y_{+}^{2}-y_{-}^{2})(1-y_{-}^{2N+2})(1-y_{+}^{2N+2})(y_{+}^{2N+2}-y_{-}^{2N+2})}{2(y_{+}^{2} - y_{-}^{2} - (1-y_{-}^{2})y_{+}^{2N+4}+(1-y_{+}^{2})y_{-}^{2N+4})^{2}}, \nonumber\\
  \frac{L^{2}}{r_{c}^{2}} &= \frac{y_{-}^{2} - y_{+}^{2} + (1-y_{-}^{2})y_{+}^{2N+4}-(1-y_{+}^{2})y_{-}^{2N+4}}{y_{-}^{2}-y_{+}^{2}+(1-y_{-}^{2})y_{+}^{2N+2}-(1-y_{+}^{2})y_{-}^{2N+2}}.
  \label{eqn:explicit_horizon_radii}
\end{align}
The near-horizon modes, as previously derived in Section~\ref{sec:NH_modes} and written in~\eqref{eqn:NH_modes_MP-dS_abstract}, can be written for all $N$ as
\begin{equation}
  \omega_{\hbox{\tiny NH}} = m \, \Omega(r_{+})|_{\rm ext}+ \left\{m \Omega_{(1)} - \frac{i}{2} \left( 1 + 2n + 2 i q_{\hbox{\tiny AdS}} + \sqrt{1 + 4 {\mu_{\rm eff}}^{2} {L_{\hbox{\tiny AdS}}}^2} \right) \kappa_{(1)} \right\} \sigma,
\end{equation}
where $q_{\hbox{\tiny AdS}}$, $\mu_{\rm eff}$ and $L_{\hbox{\tiny AdS}}$ are given by~\eqref{eqn:L_AdS_explicit}-\eqref{eqn:mu_eff_explicit}. The terms $\Omega_{(1)}$ and $\kappa_{(1)}$  are the first-order coefficients of the Taylor expansions of $\Omega(r_{+})$ and $\kappa_{+}$ in the near-extremal parameter $\sigma = 1 - y_{-}/y_{+}$, as previously defined in~\eqref{eqn:MAE_variables}. Explicitly, these are given by
\begin{align}\label{eqn:kappaOmegaNH}
  \kappa_{(1)} &\equiv \frac{d\kappa_{+}}{d\sigma} \bigg|_{\sigma = 0} = \frac{1}{r_c y_+}\frac{\sqrt{1+N} (2 y_+^2 (-1+y_+^{2 N})-N (-1+y_+^2) (1+y_+^{2+2 N}))}{ \sqrt{(1-y_+^2) (1-y_+^{2+2 N})} \sqrt{1+y_+^{2+2 N} (-2+y_+^2+N (-1+y_+^2))}}, \\
  \Omega_{(1)} &\equiv \frac{d\Omega(r_{+})}{d\sigma} \bigg|_{\sigma = 0} = \frac{1}{r_{c}} \frac{2 y_{+}^{2}(1 - y_{+}^{2N}) - N(1-y_{+}^{2})(1+y_{+}^{2+2N})}{2y_{+}(1-y_{+}^{2})(1-y_{+}^{2+2N})} \nonumber \\
              &\times \frac{1+y_{+}^{4+4N}(3+2N-2(1+N)y_{+}^{2}) - y_{+}^{2+2N}(3 - (N - (1+N) y_{+}^{2})^{2})}{(1-y_{+}^{2+2N}(2+N-(1+N)y_{+}^{2}))^{3/2}\sqrt{(1+N)(N(1-y_{+}^{2})-y_{+}^{2}(1-y_{+}^{2N}))}},
\end{align}
and $\Omega(r_{+})$ at extremality is given by
\begin{equation}
  \Omega(r_{+})|_{\rm ext} = \frac{1}{r_c y_+}\sqrt{\frac{N (1-y_+^2)-y_+^2 (1-y_+^{2 N})}{(1+N) (1+y_+^{2+2 N} (-2+y_+^2+N (-1+y_+^2)))}}.
\end{equation}

%===========================================================
%===========================================================
\section{Numerical convergence tests}\label{sec:convergence_tests}\label{sec:convergence}
%===========================================================
%===========================================================

\begin{figure}
  \centering
  \includegraphics[width=0.8\linewidth]{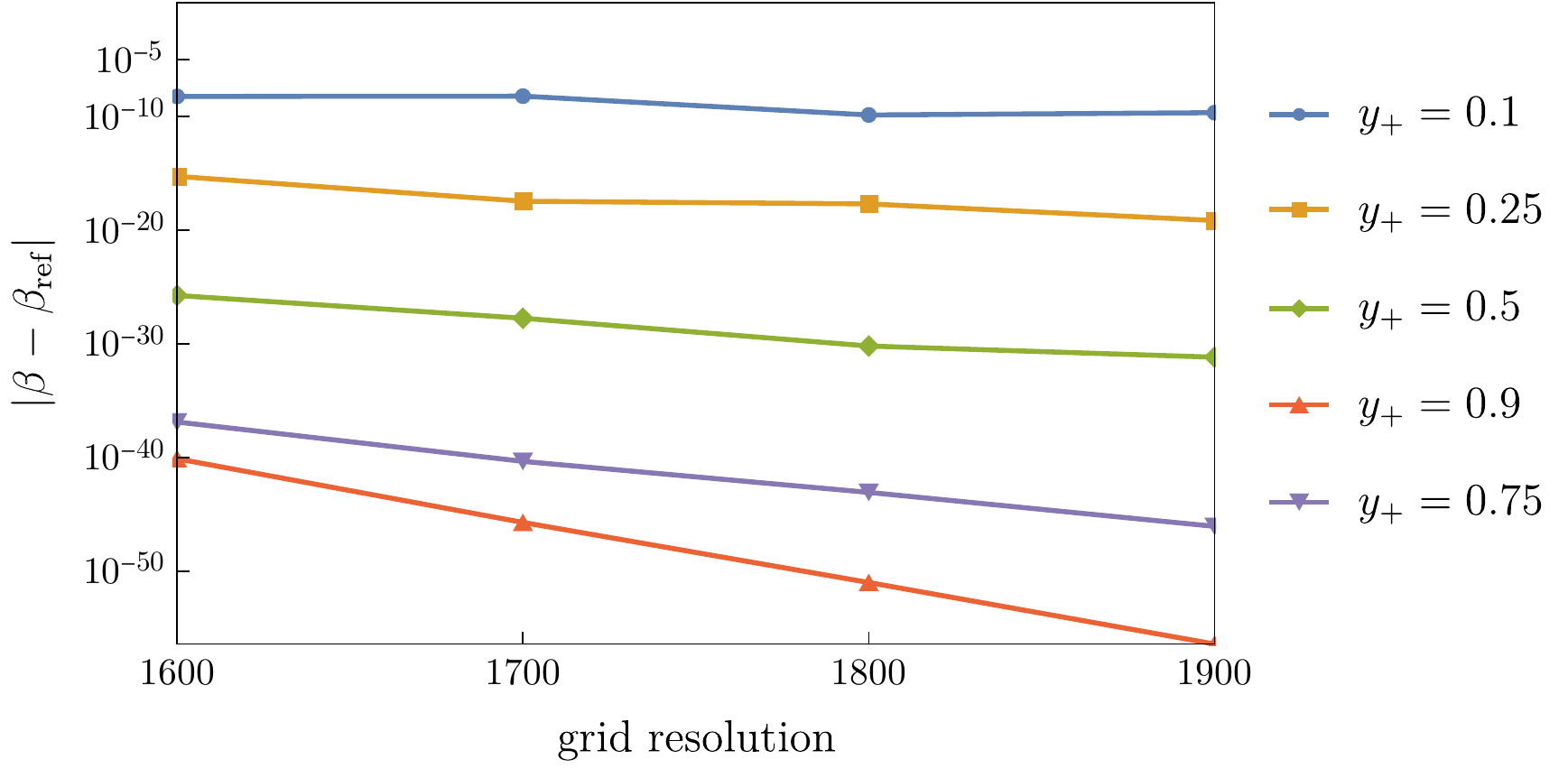}
  \caption{Convergence of $\beta$ for $d=5$ MP-dS with $m = l = 10$ and $r_{-} = 0.9995 \, r_{+}$, as displayed in Fig.~\ref{fig:beta_near_extremality_all_N}. These modes were computed at $(\textrm{resolution, precision}) = (1600, 500)$. The precision is scaled linearly with the grid resolution, \emph{i.e.} a maximum $(\textrm{resolution, precision}) = (2000, 625)$ which is used to compute the reference value $\beta_{\rm ref}$.\label{fig:convergence_near_extremal_Nd4}}
\end{figure}

We use pseudospectral collocation methods to generate our numerical data, and thus our numerical results should (and do) have exponential convergence as the number of points used to discretise the numerical grid increases (see e.g \cite{diasNumericalMethodsFinding2016}).
All of our numerical results have converged with an error that is not higher than $10^{-8}$. To illustrate our numerical error analysis, a convergence test is given in Fig.~\ref{fig:convergence_near_extremal_Nd4} for $d = 5$ MP-dS at $r_- = 0.9995 \, r_+$, with $m = l = 10$. To test convergence, we recompute these modes with increasing grid resolution and precision (increasing the precision proportionally to the resolution), up to $(\textrm{resolution}, \textrm{precision}) = (2000, 625)$. This maximum value is used to compute $\beta_{\rm ref}$. The maximum error $|\beta - \beta_{\rm ref}|$ is $\sim 10^{-8.2}$, as expected.

% Bibliography
\bibliographystyle{JHEP}
\bibliography{refs_QNM_SCC}

\end{document}